\documentclass[traditabstract]{aa}  

\usepackage{psfig}

\begin{document}

\title{Unveiling the nature of {\it INTEGRAL} objects through optical 
spectroscopy\thanks{Based on observations collected at the following 
observatories: Cerro Tololo Interamerican Observatory (Chile); 
Observatorio del Roque de los Muchachos of the Instituto de 
Astrof\'{\i}sica de Canarias (Canary Islands, Spain); ESO (La Silla, 
Chile) under programme 083.D-0110(A); Astronomical Observatory of 
Bologna in Loiano (Italy); Astronomical Observatory of Asiago (Italy);
Observatorio Astron\'omico Nacional (San Pedro M\'artir, M\'exico); 
South African Astronomical Observatory (Sutherland, South Africa);  
Anglo-Australian Observatory (Siding Spring, Australia); Apache 
Point Observatory (New Mexico, USA).}}

\subtitle{VIII. Identification of 44 newly detected hard X--ray sources}

\author{N. Masetti\inst{1},
P. Parisi\inst{1,2},
E. Palazzi\inst{1},
E. Jim\'enez-Bail\'on\inst{3},
V. Chavushyan\inst{4},
L. Bassani\inst{1},
A. Bazzano\inst{5}, 
A.J. Bird\inst{6},
A.J. Dean\inst{6},
P.A. Charles\inst{6,7},
G. Galaz\inst{8},
R. Landi\inst{1},
A. Malizia\inst{1},
E. Mason\inst{9},
V.A. McBride\inst{6},
D. Minniti\inst{8,10},
L. Morelli\inst{11},
F. Schiavone\inst{1},
J.B. Stephen\inst{1} and
P. Ubertini\inst{5}
}

\institute{
INAF -- Istituto di Astrofisica Spaziale e Fisica Cosmica di 
Bologna, Via Gobetti 101, I-40129 Bologna, Italy
\and
Dipartimento di Astronomia, Universit\`a di Bologna,
Via Ranzani 1, I-40127 Bologna, Italy
\and
Instituto de Astronom\'{\i}a, Universidad Nacional Aut\'onoma de M\'exico,
Apartado Postal 70-264, 04510 M\'exico D.F., M\'exico
\and
Instituto Nacional de Astrof\'{i}sica, \'Optica y Electr\'onica,
Apartado Postal 51-216, 72000 Puebla, M\'exico
\and
INAF -- Istituto di Astrofisica Spaziale e Fisica Cosmica di
Roma, Via Fosso del Cavaliere 100, I-00133 Rome, Italy
\and
School of Physics \& Astronomy, University of Southampton, Southampton, 
Hampshire, SO17 1BJ, United Kingdom  
\and
South African Astronomical Observatory, PO Box 9, Observatory 7935, South Africa
\and
Departamento de Astronom\'{i}a y Astrof\'{i}sica, Pontificia Universidad 
Cat\'olica de Chile, Casilla 306, Santiago 22, Chile
\and
European Southern Observatory, Alonso de Cordova 3107, Vitacura,
Santiago, Chile
\and
Specola Vaticana, V-00120 Citt\`a del Vaticano
\and
Dipartimento di Astronomia, Universit\`a di Padova,
Vicolo dell'Osservatorio 3, I-35122 Padua, Italy
}

\offprints{N. Masetti (\texttt{masetti@iasfbo.inaf.it)}}
\date{Received 23 April 2010; accepted 20 May 2010}

\abstract{Hard X--ray surveys performed by the {\it INTEGRAL} satellite
have discovered a conspicuous fraction (up to 30\%) of unidentified 
objects among the detected sources. Here we continue our program of 
identification of these objects by (i) selecting probable optical 
candidates by means of positional cross-correlation of the {\it INTEGRAL} 
detections with soft X--ray, radio, and/or optical archives and (ii) 
performing optical spectroscopy on them.
As a result, we pinpointed and identified, or more accurately 
characterized, 44 definite or likely counterparts of {\it INTEGRAL} sources.
Among them, 32 are active galactic nuclei (AGNs; 18 with broad emission 
lines, 13 with narrow emission lines only, and one X--ray bright, optically 
normal galaxy) with redshift 0.019 $< z <$ 0.6058, 6 cataclysmic variables 
(CVs), 5 high-mass X--ray binaries (2 of which in the Small Magellanic 
Cloud), and 1 low-mass X--ray binary. This was achieved by using 7 
telescopes of various sizes and archival data from two online 
spectroscopic surveys.
The main physical parameters of these hard X--ray sources were also 
determined using the multiwavelength information available in the 
literature. In general, AGNs are the most abundant population among 
hard X--ray objects, and our results confirm the tendency of finding AGNs 
more frequently than any other type of hard X--ray emitting object among 
unidentified {\it INTEGRAL} sources when optical spectroscopy is used as 
an identification tool.
Moreover, the deeper sensitivity of the more recent {\it INTEGRAL} surveys 
enables one to begin detecting hard X--ray emission above 20 keV from 
sources such as LINER-type AGNs and non-magnetic CVs.}


\titlerunning{The nature of 44 newly detected {\it INTEGRAL} sources}
\authorrunning{N. Masetti et al.}

\maketitle

\section{Introduction}

\begin{figure*}[th!]
\hspace{-.1cm}
\centering{\mbox{\psfig{file=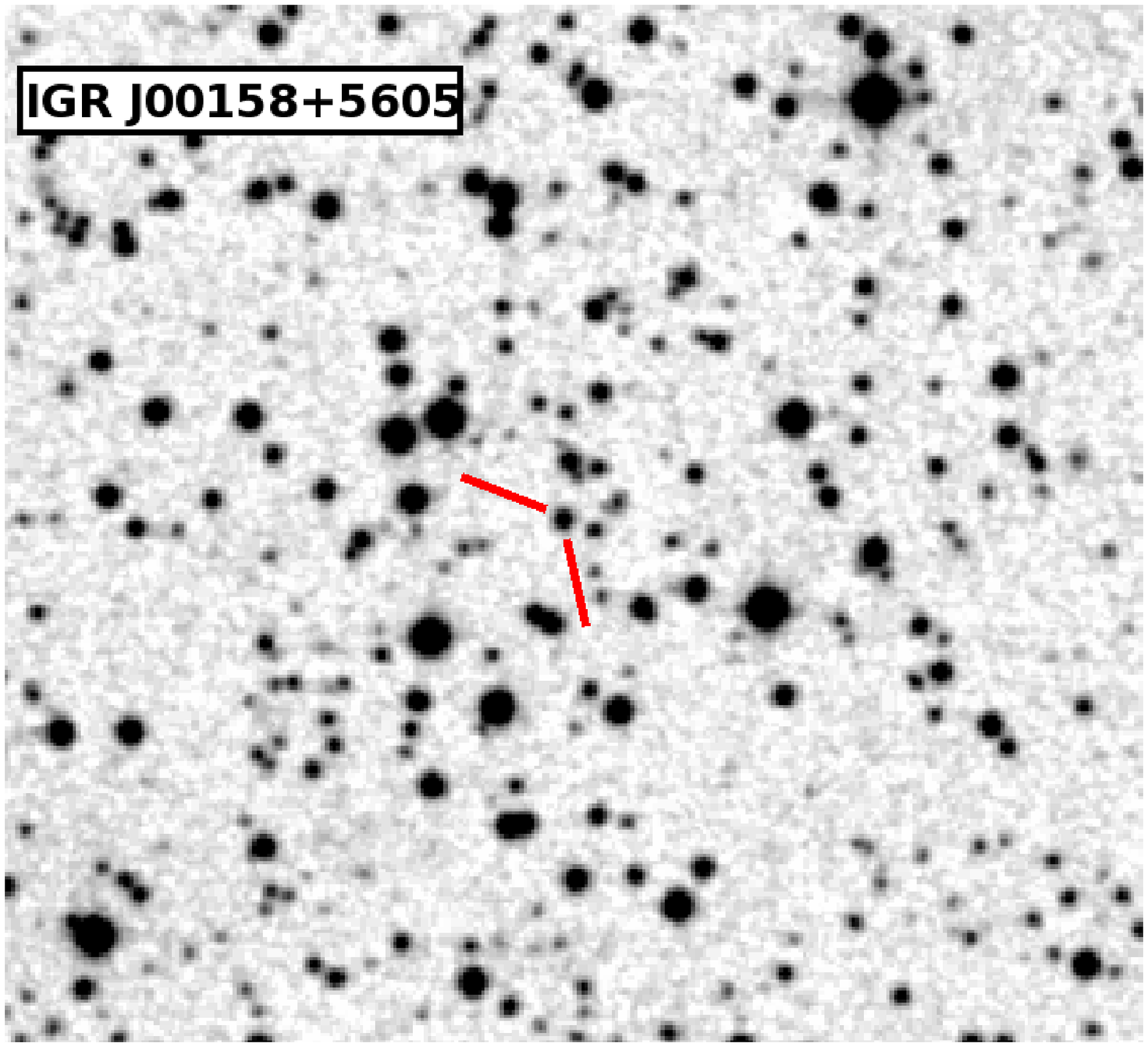,width=5.9cm}}}
\centering{\mbox{\psfig{file=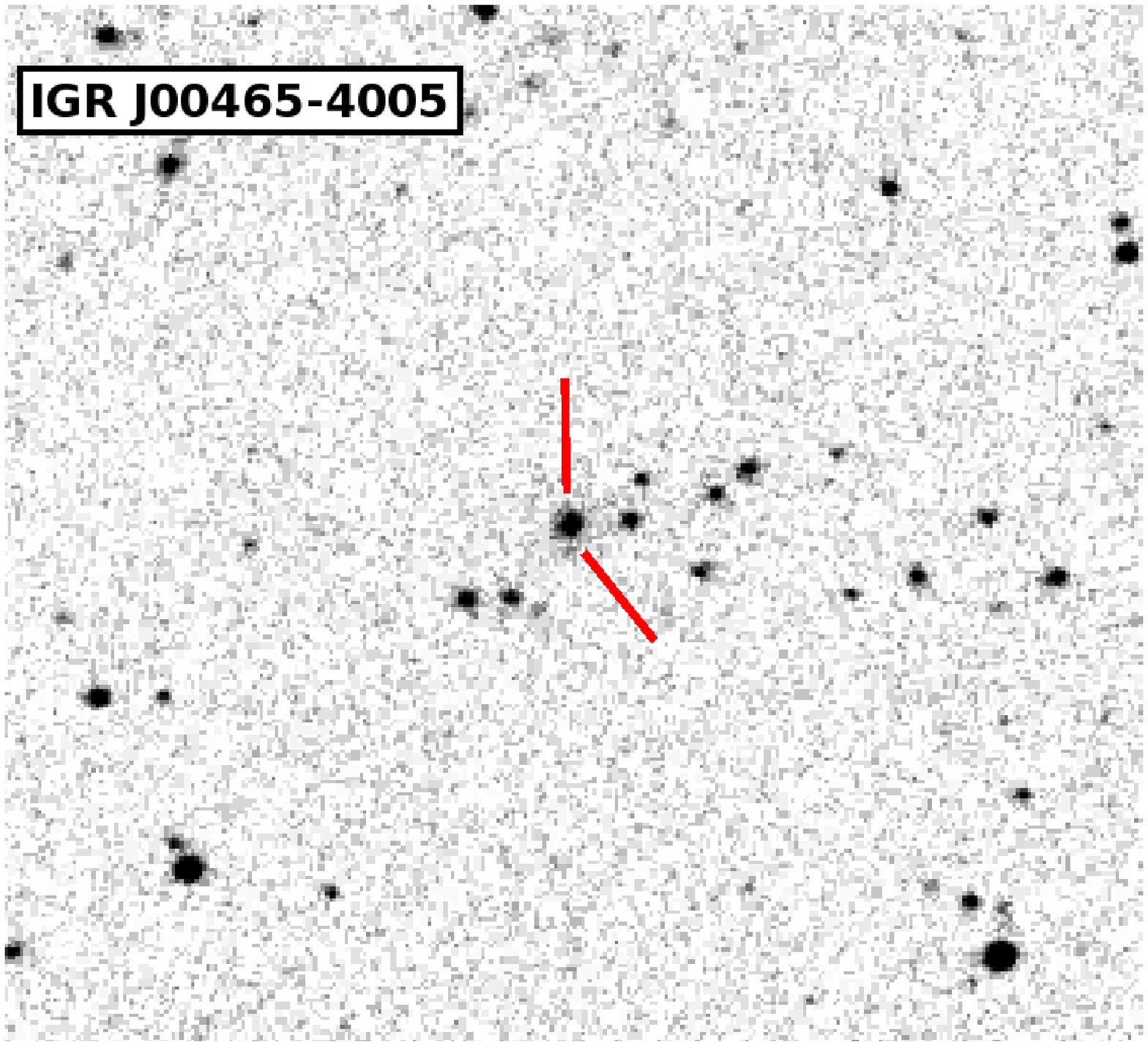,width=5.9cm}}}
\centering{\mbox{\psfig{file=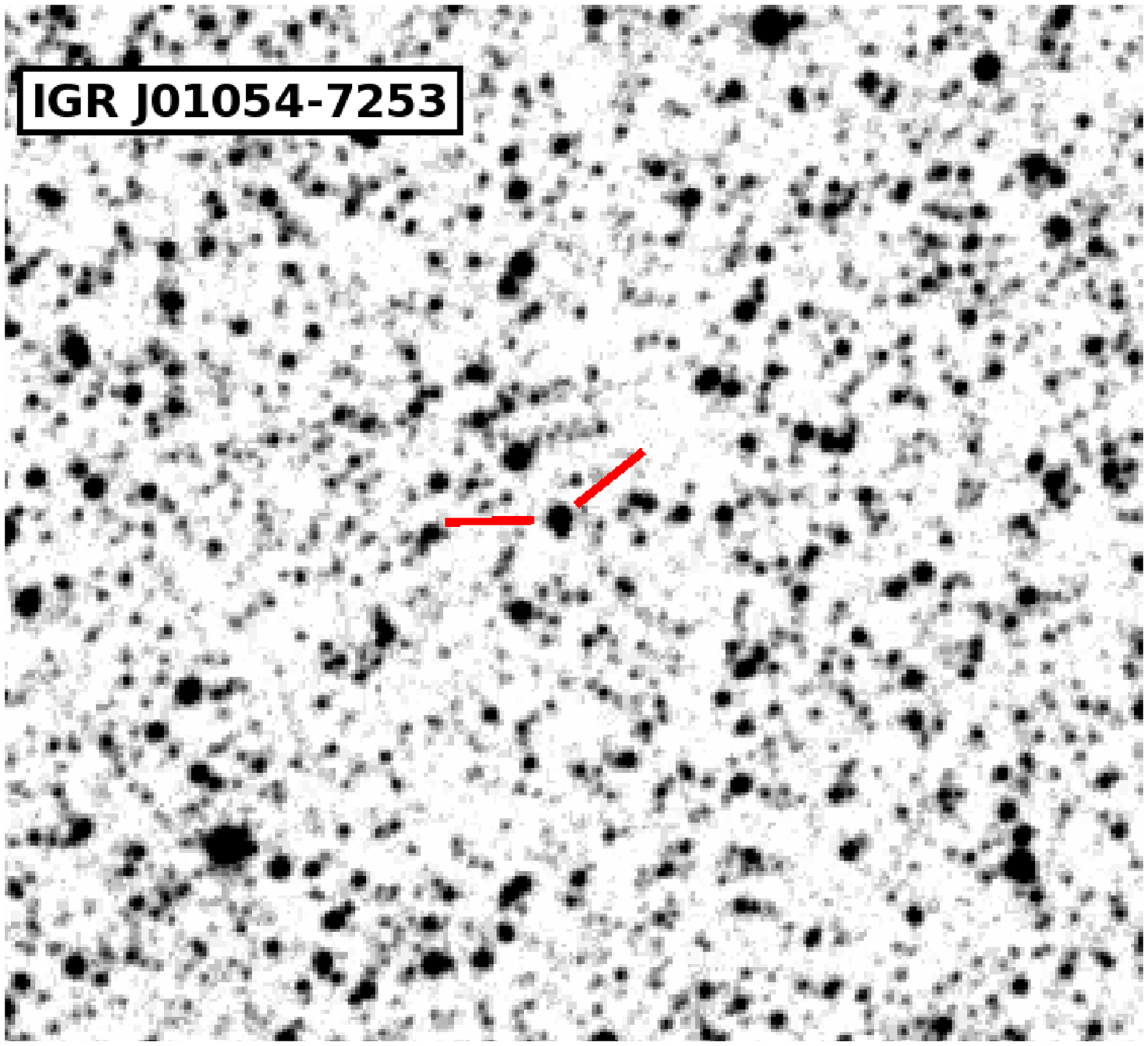,width=5.9cm}}}
\centering{\mbox{\psfig{file=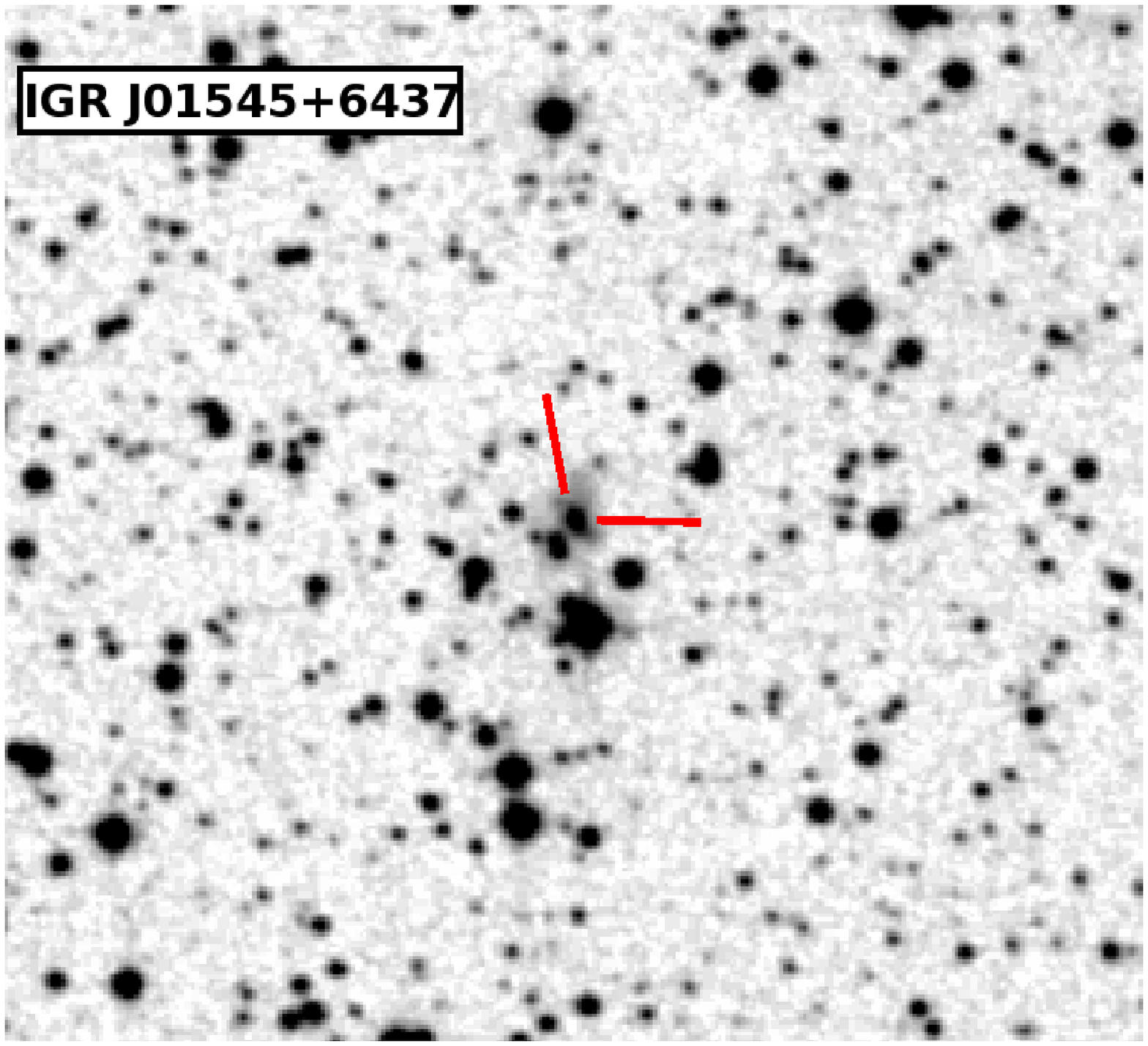,width=5.9cm}}}
\centering{\mbox{\psfig{file=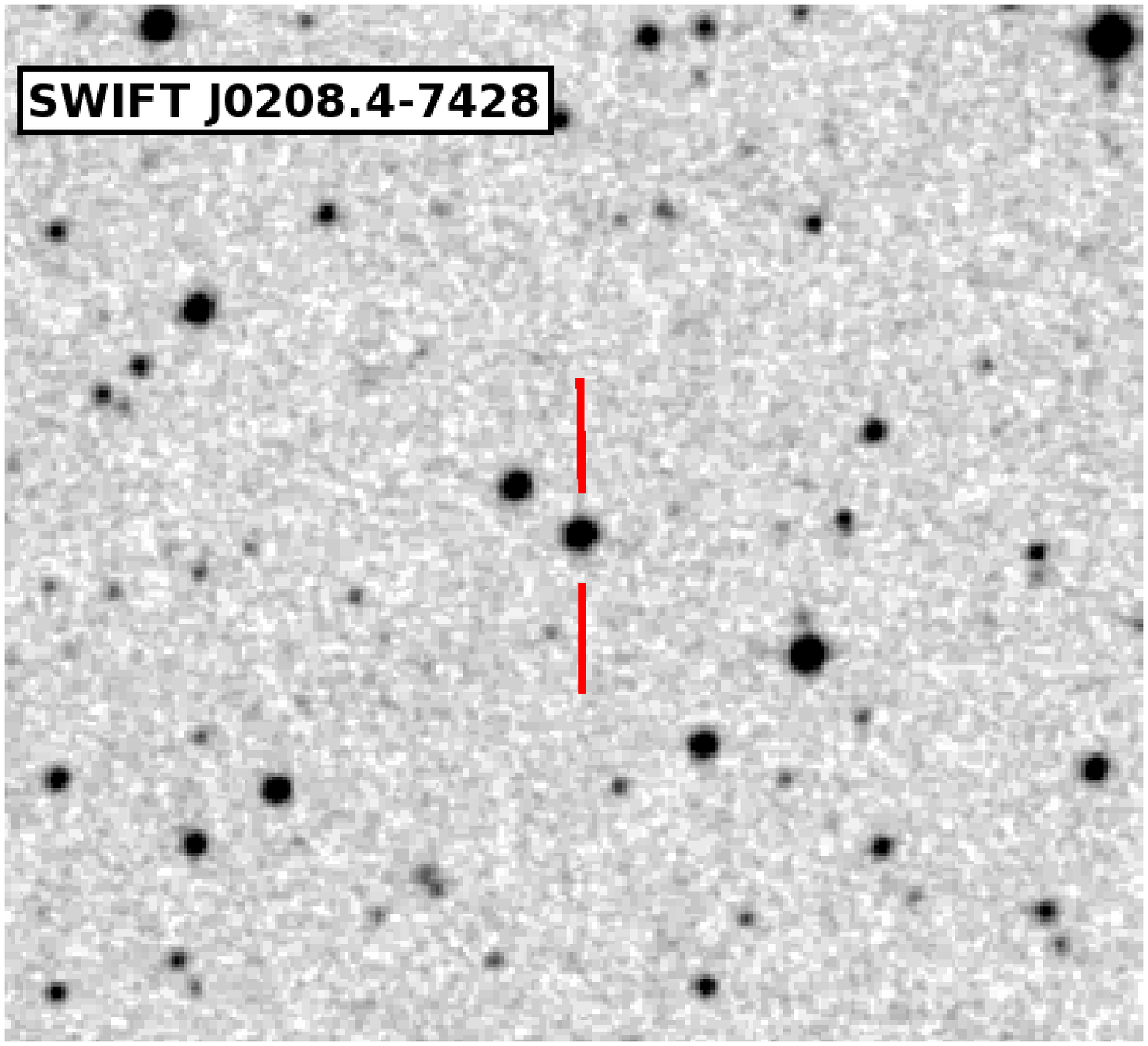,width=5.9cm}}}
\centering{\mbox{\psfig{file=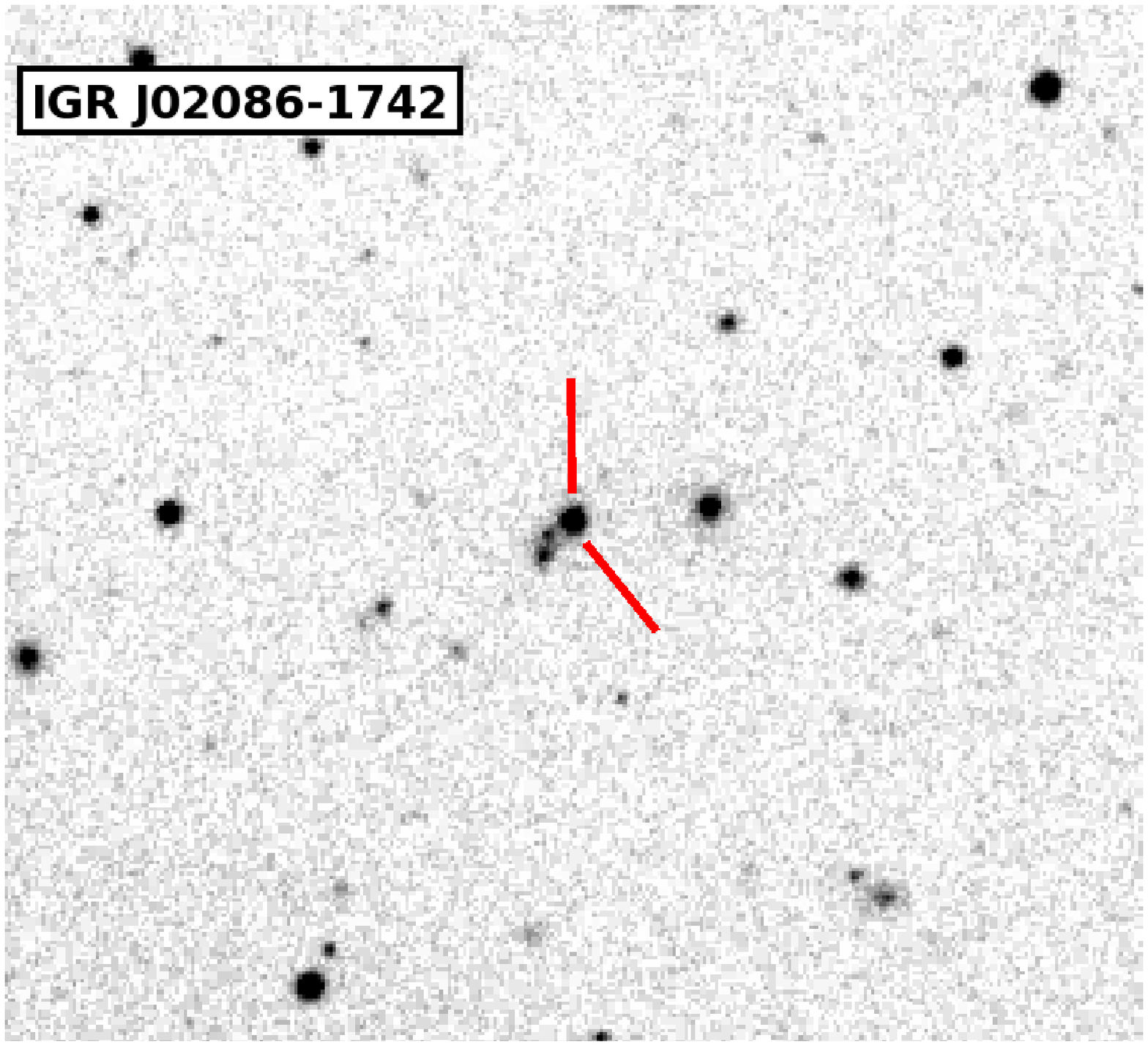,width=5.9cm}}}
\centering{\mbox{\psfig{file=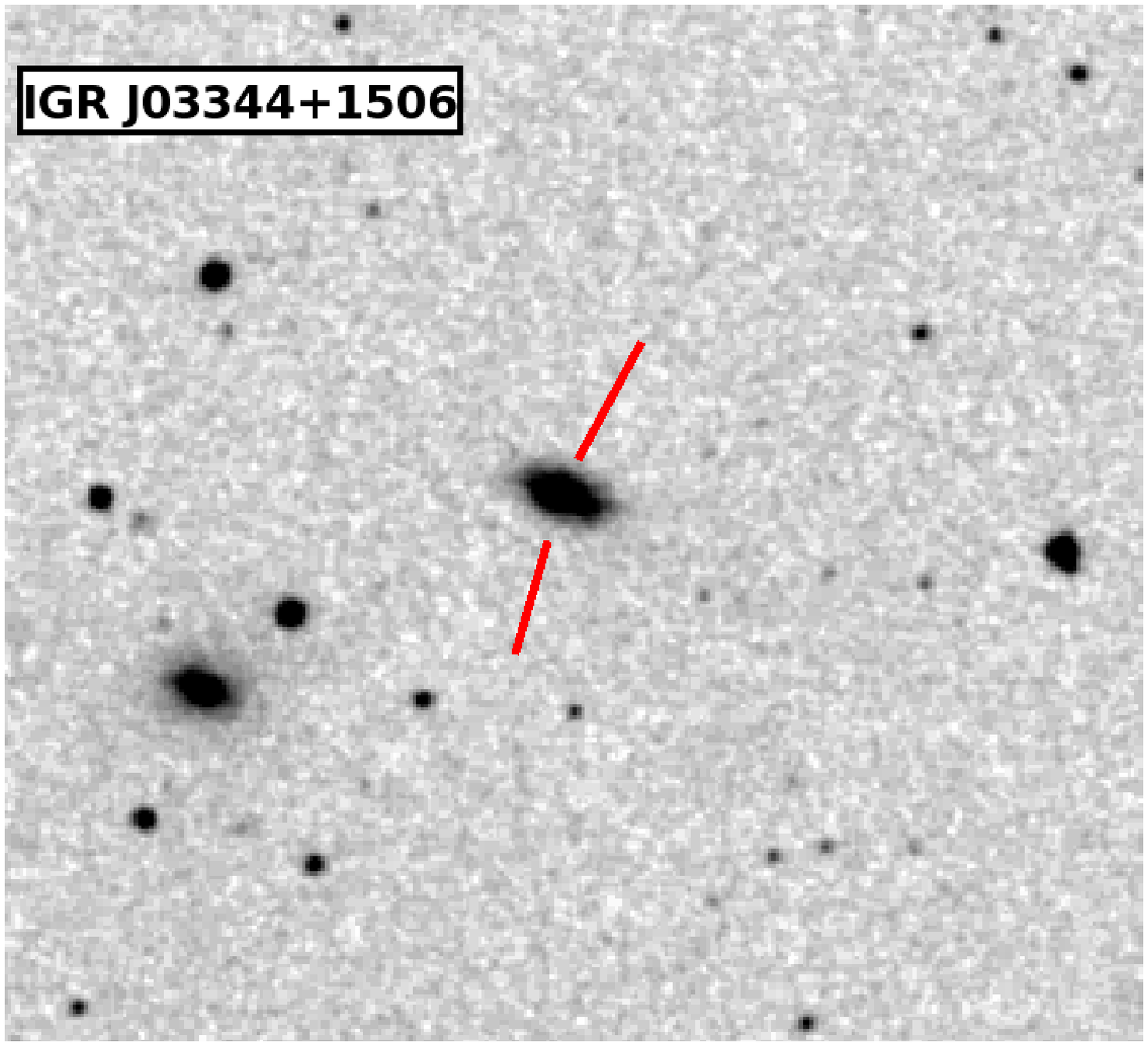,width=5.9cm}}}
\centering{\mbox{\psfig{file=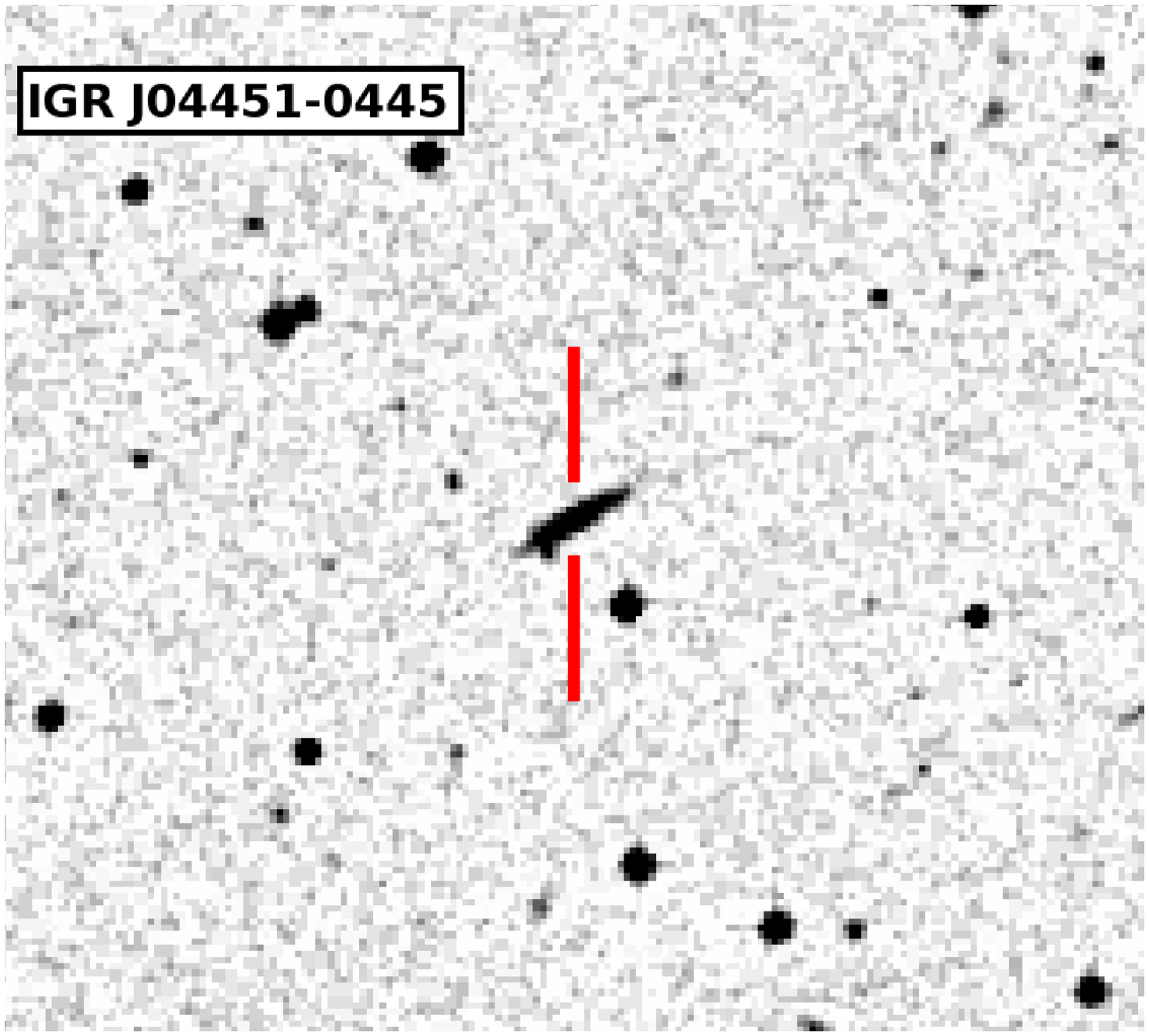,width=5.9cm}}}
\centering{\mbox{\psfig{file=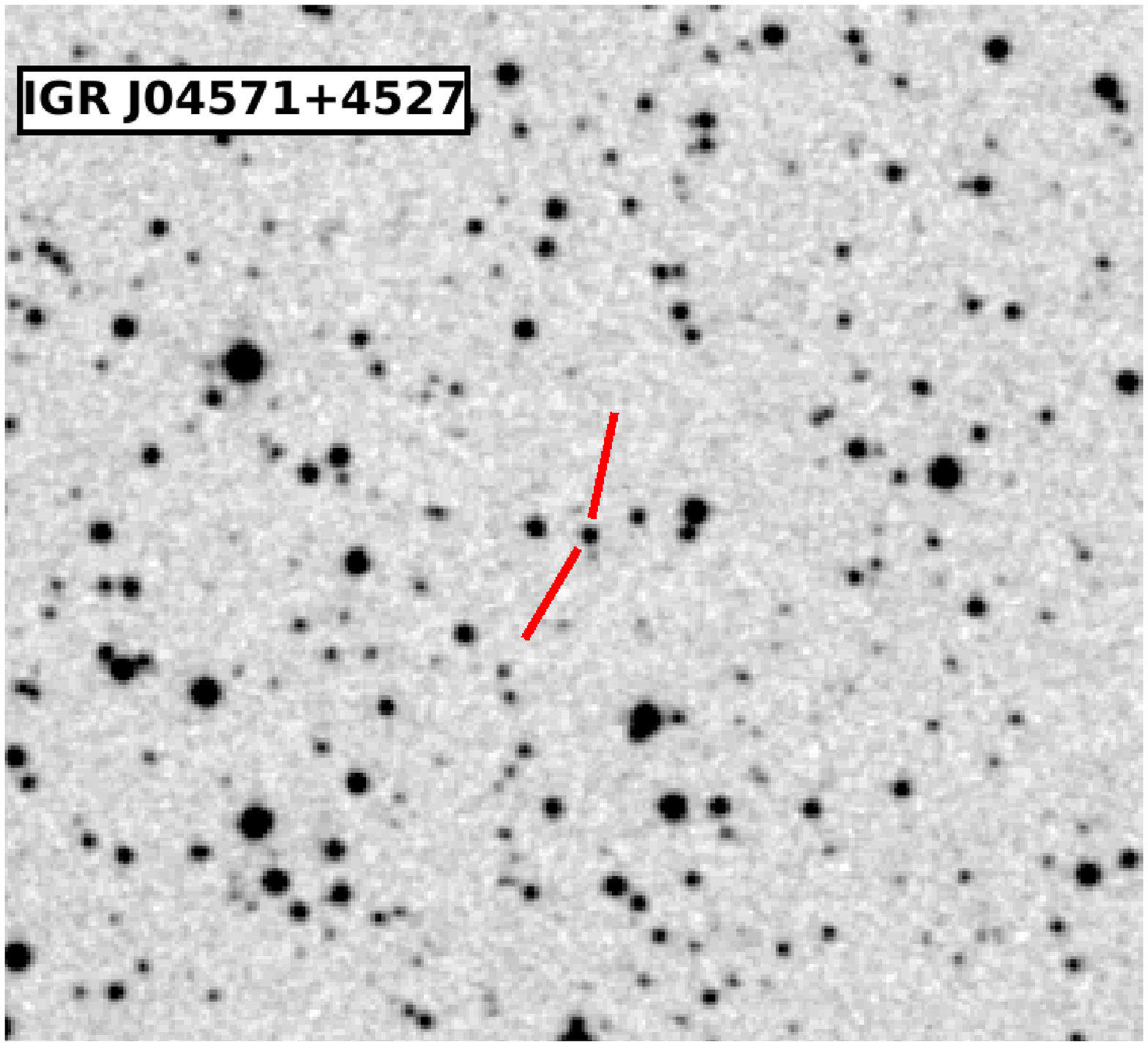,width=5.9cm}}}
\caption{Optical images of the fields of 9 of the {\it INTEGRAL} hard 
X--ray sources selected in this paper for optical spectroscopic 
follow-up (see Table 1). The object name is indicated in each panel.
The proposed optical counterparts are indicated with tick marks. Field 
sizes are 5$'$$\times$5$'$ and are extracted from the DSS-II-Red survey 
with the exception of IGR J04451$-$0445, which is extracted from the DSS-I 
survey. In all cases, north is up and east to the left.}
\end{figure*}

\begin{figure*}[th!]
\hspace{-.1cm}
\centering{\mbox{\psfig{file=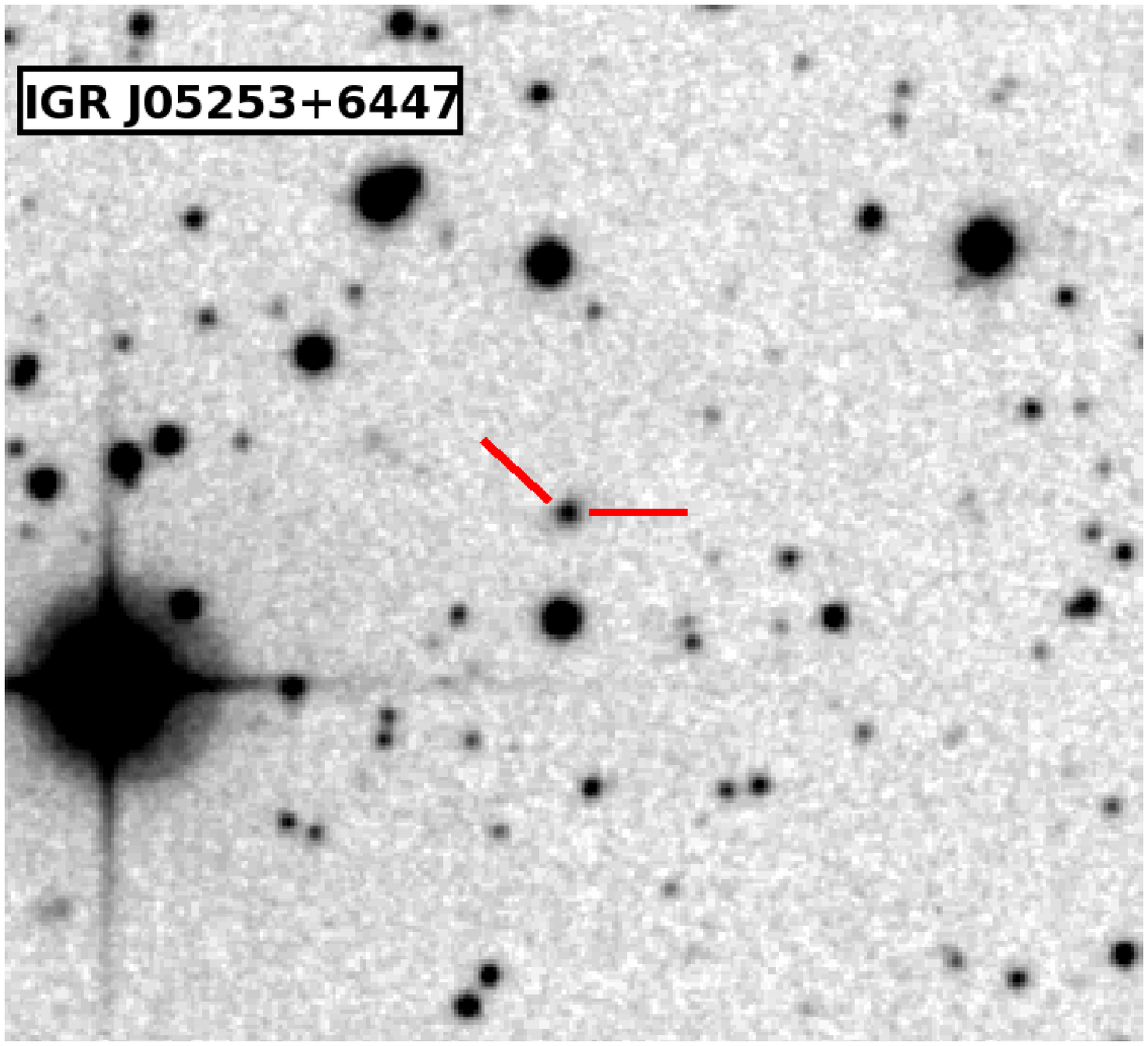,width=5.9cm}}}
\centering{\mbox{\psfig{file=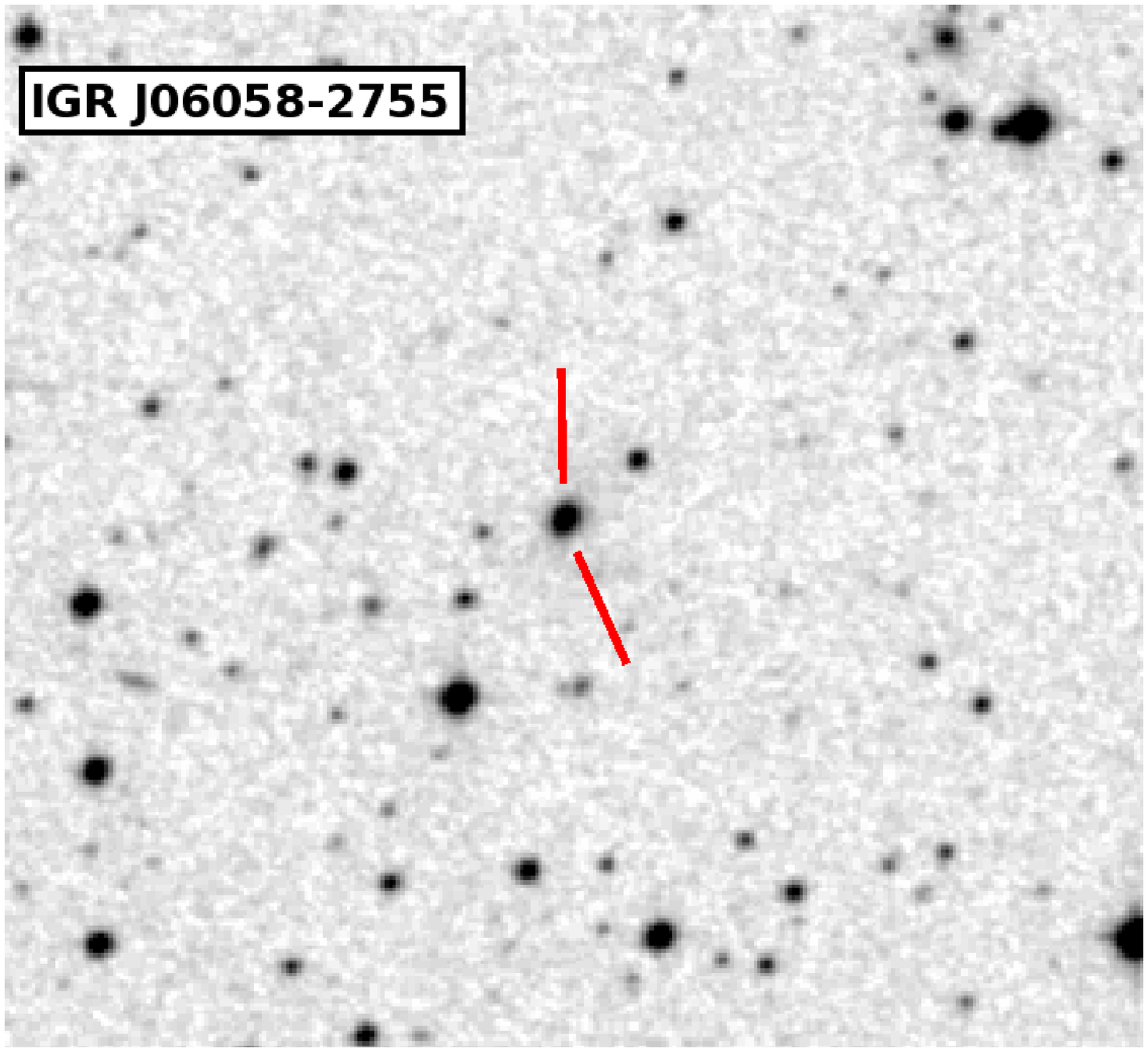,width=5.9cm}}}
\centering{\mbox{\psfig{file=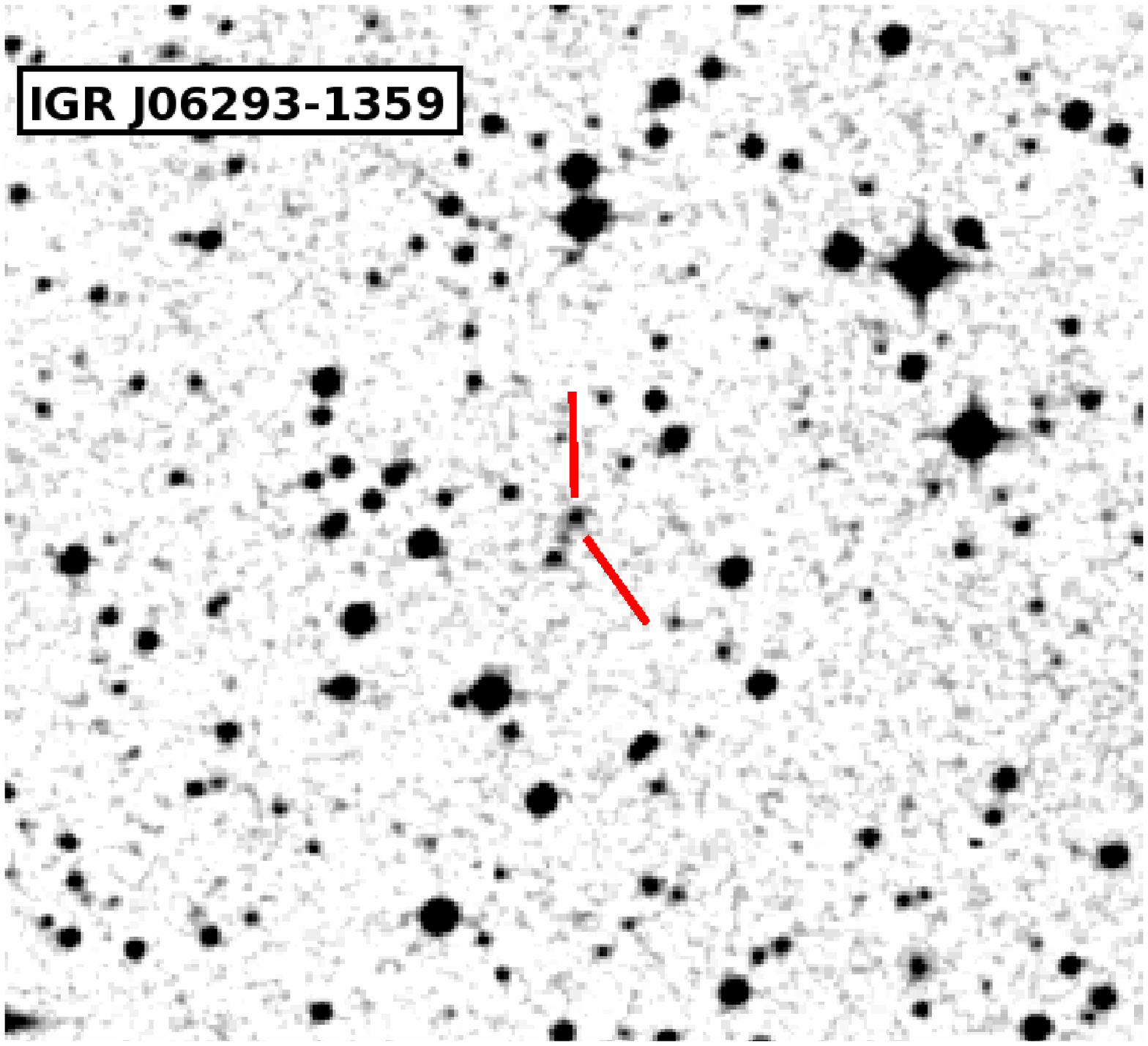,width=5.9cm}}}
\centering{\mbox{\psfig{file=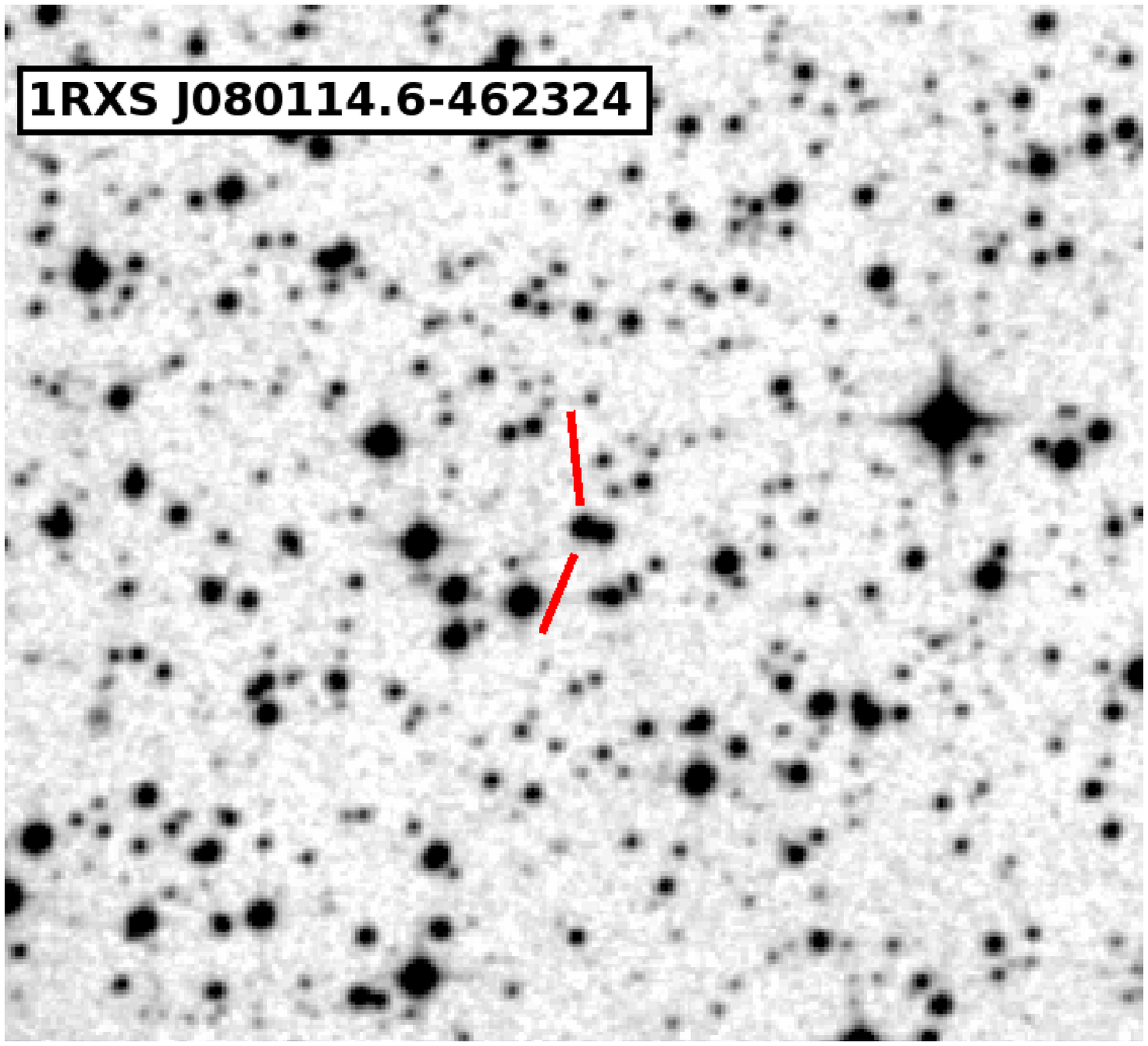,width=5.9cm}}}
\centering{\mbox{\psfig{file=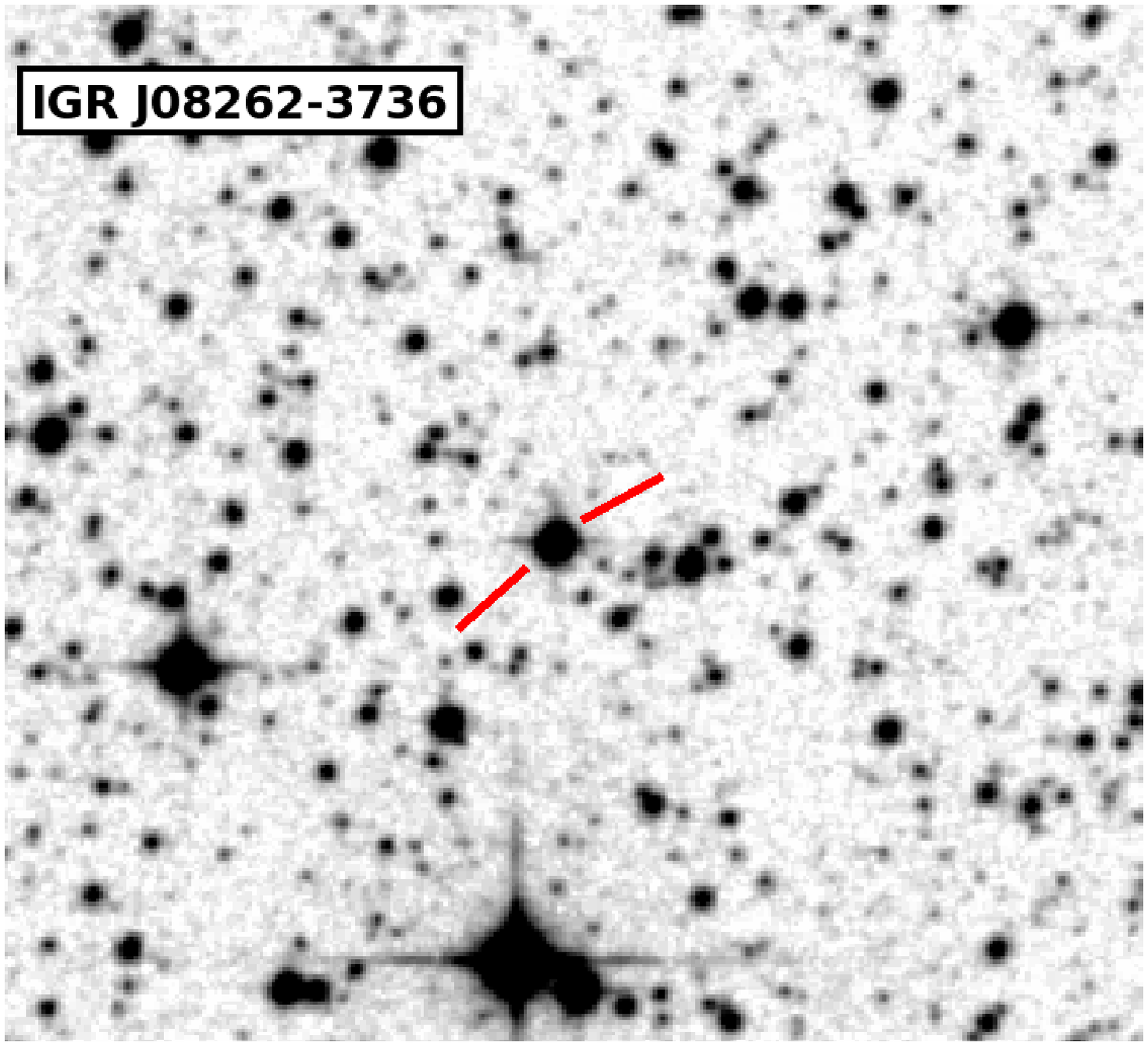,width=5.9cm}}}
\centering{\mbox{\psfig{file=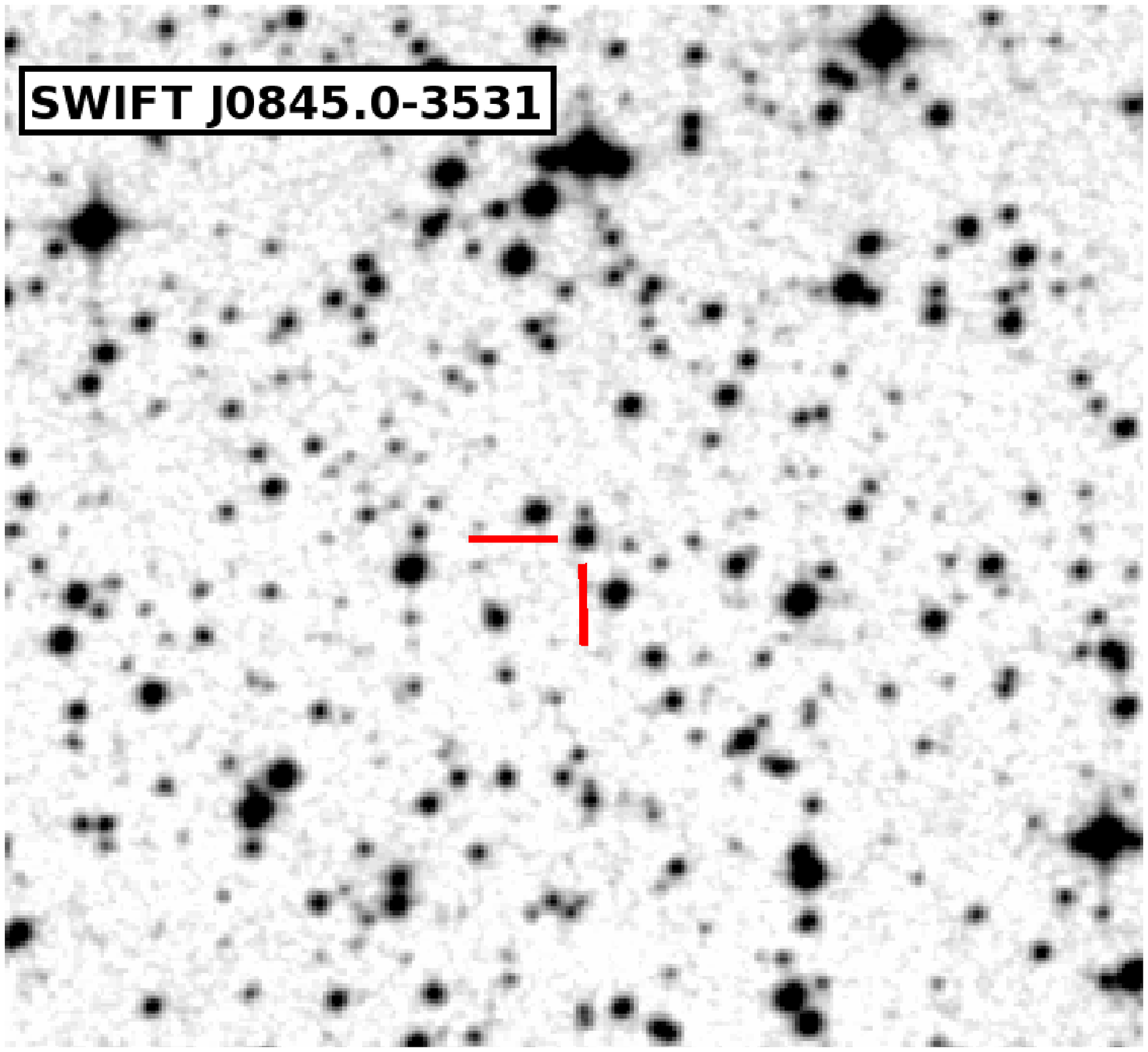,width=5.9cm}}}
\centering{\mbox{\psfig{file=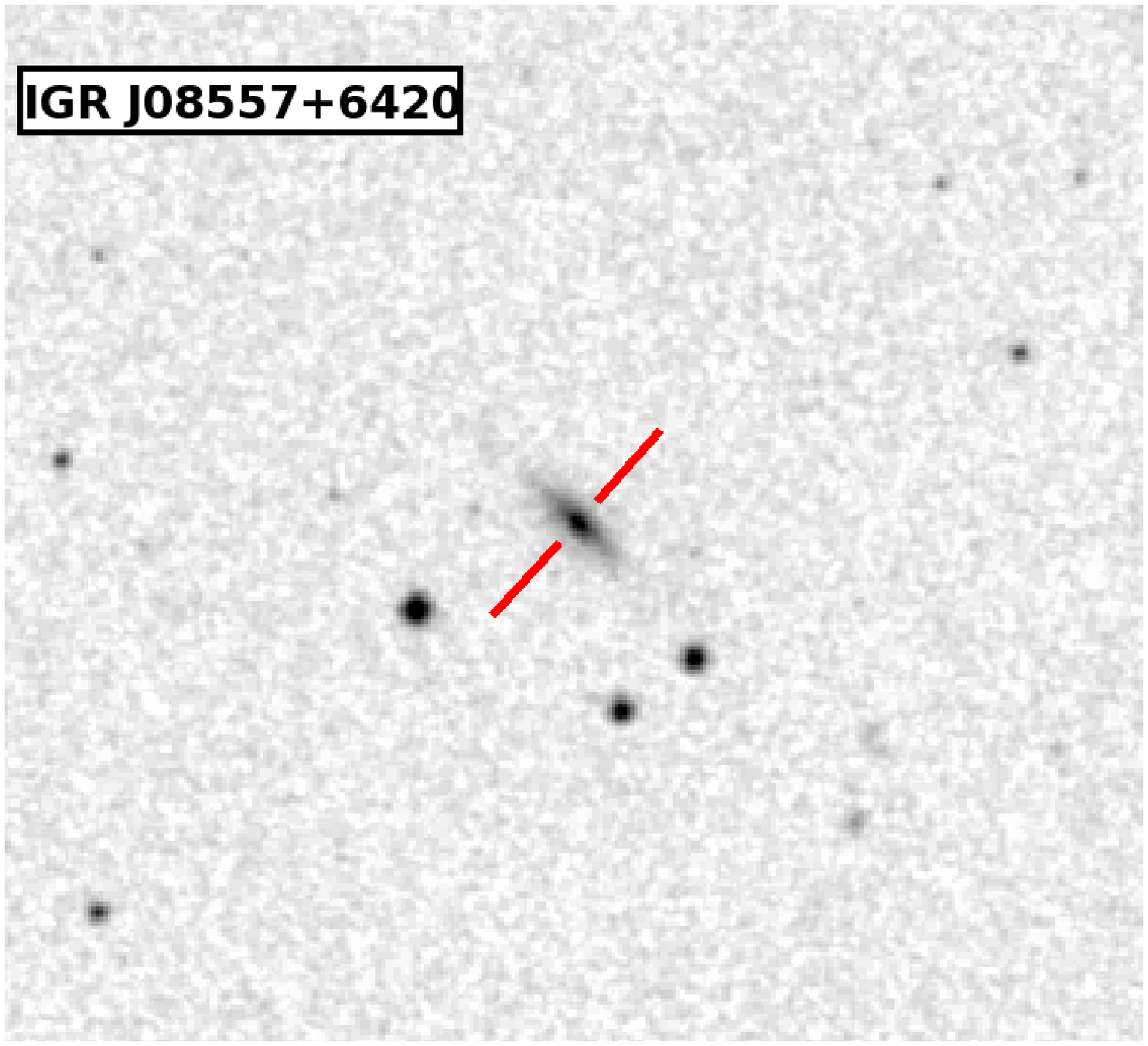,width=5.9cm}}}
\centering{\mbox{\psfig{file=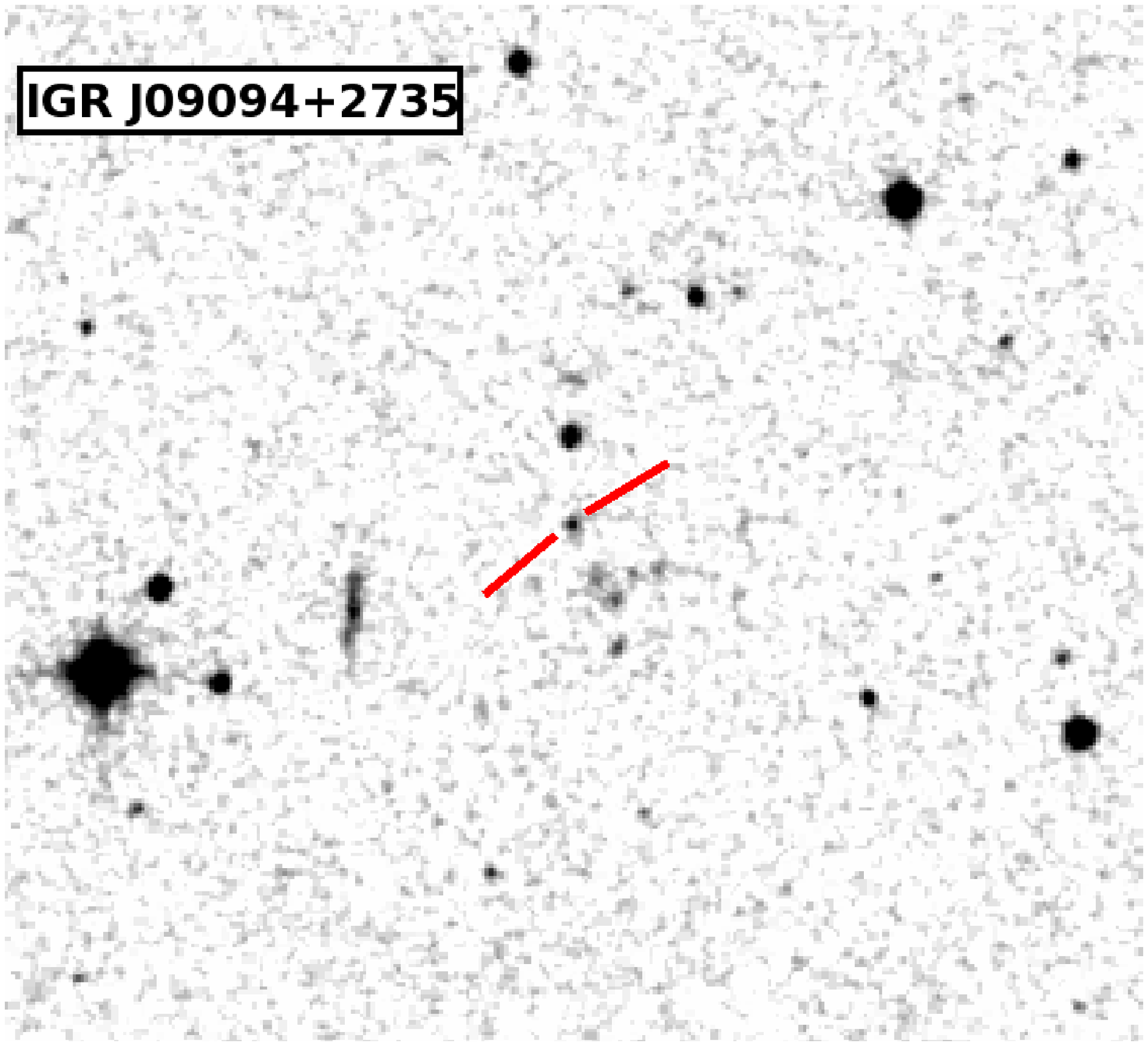,width=5.9cm}}}
\centering{\mbox{\psfig{file=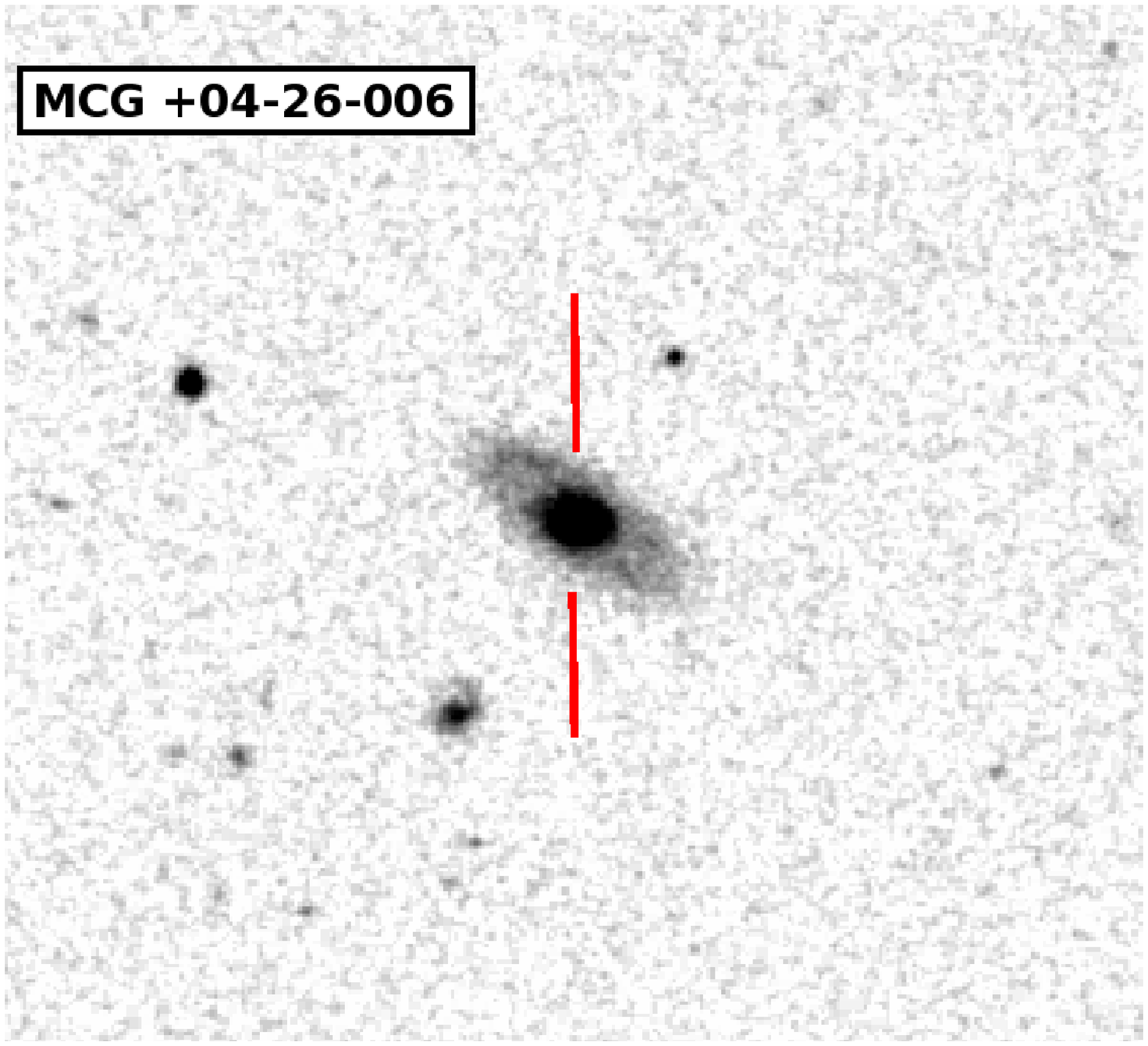,width=5.9cm}}}
\caption{As Fig. 1, but for 9 more {\it INTEGRAL} sources of our
sample (see Table 1).}
\end{figure*}

\begin{figure*}[th!]
\hspace{-.1cm}
\centering{\mbox{\psfig{file=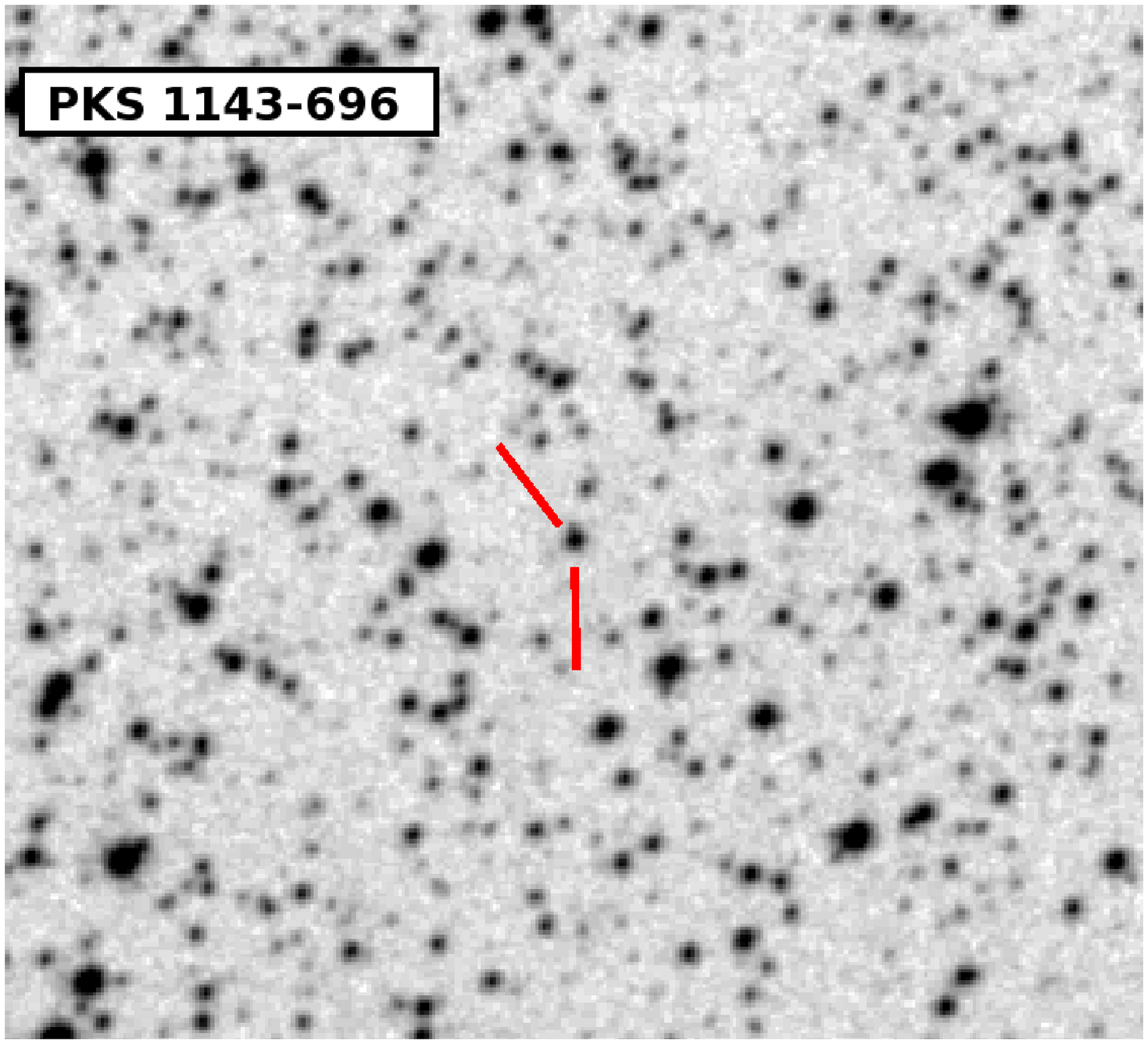,width=5.9cm}}}
\centering{\mbox{\psfig{file=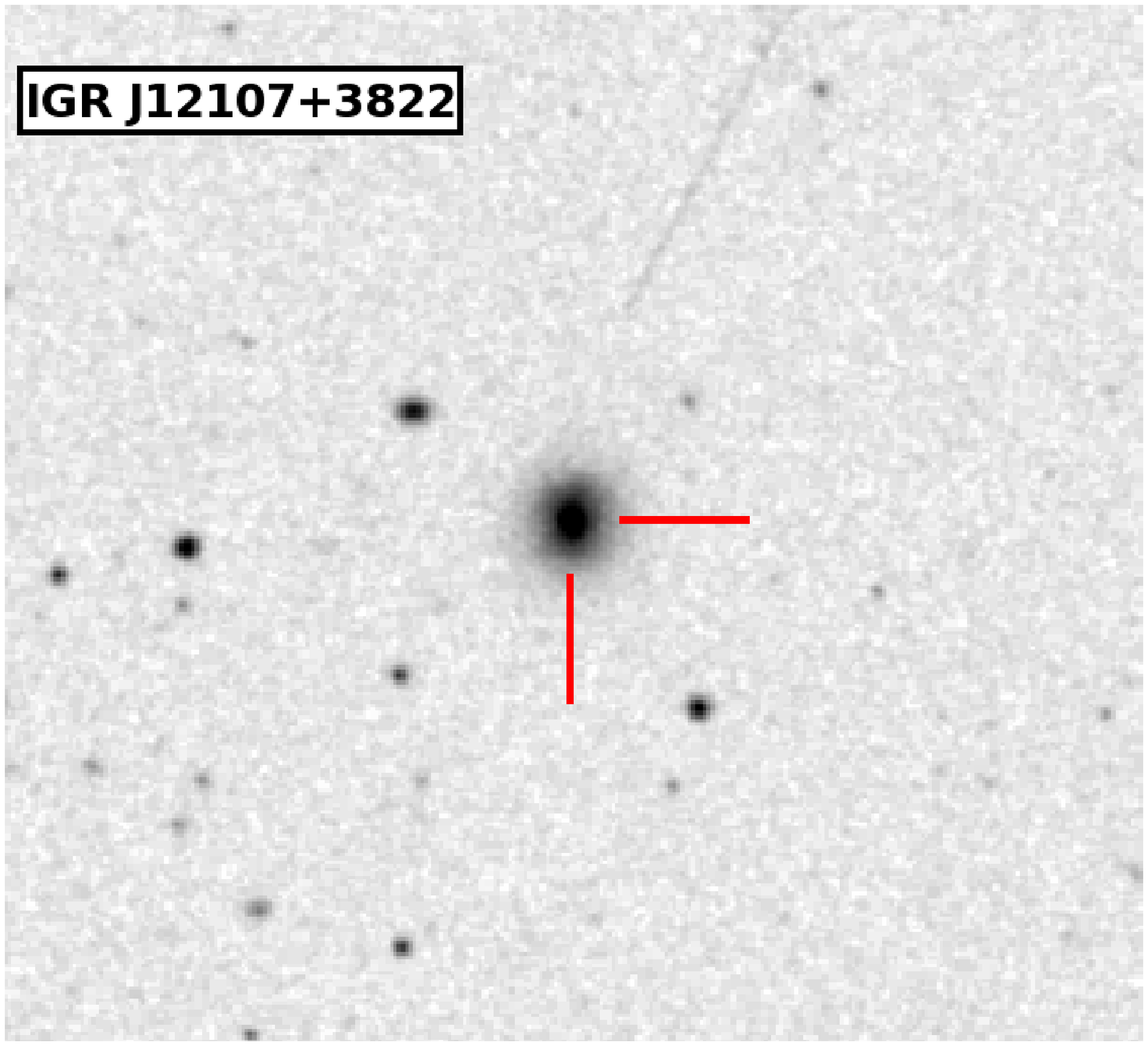,width=5.9cm}}}
\centering{\mbox{\psfig{file=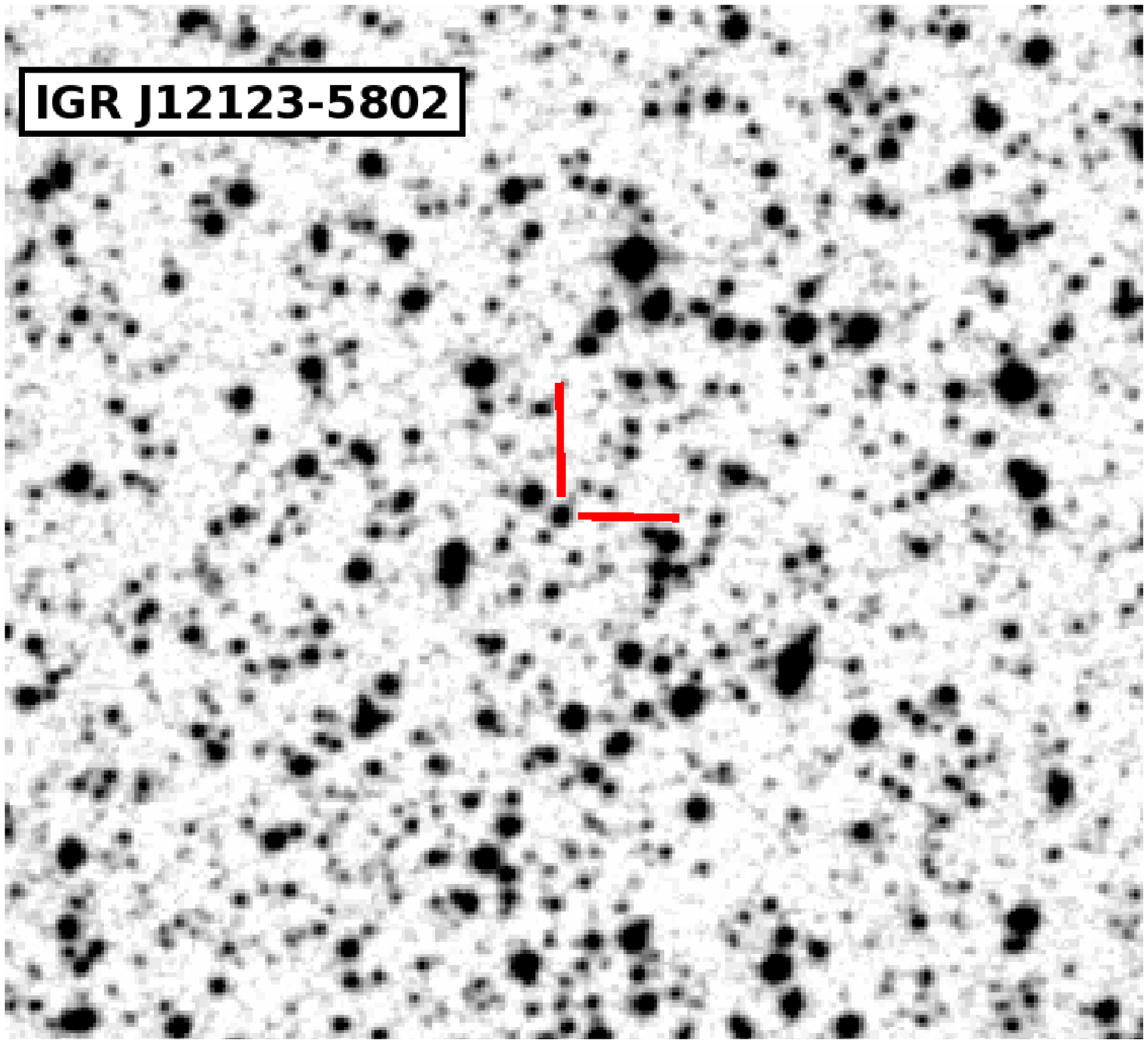,width=5.9cm}}}
\centering{\mbox{\psfig{file=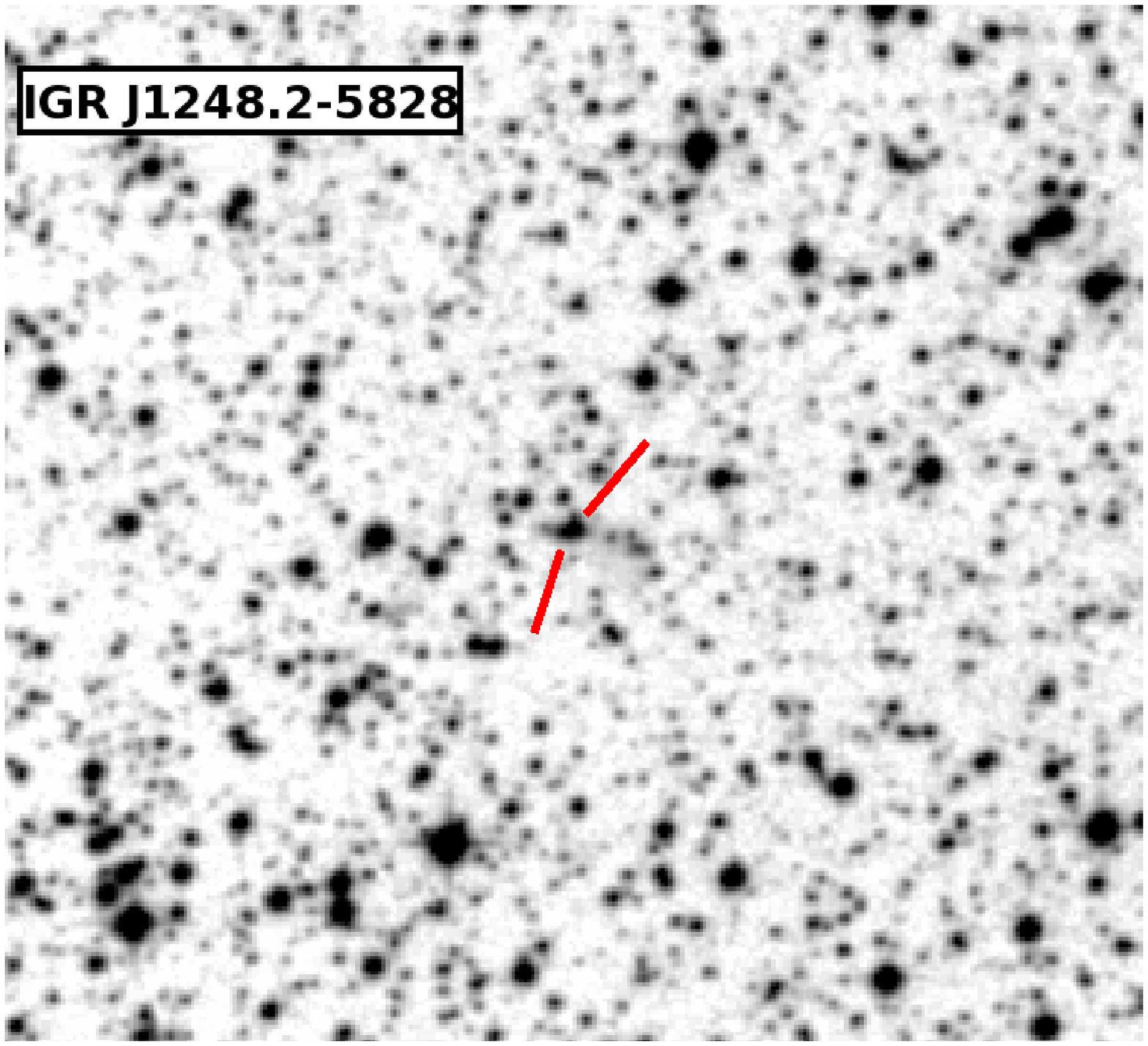,width=5.9cm}}}
\centering{\mbox{\psfig{file=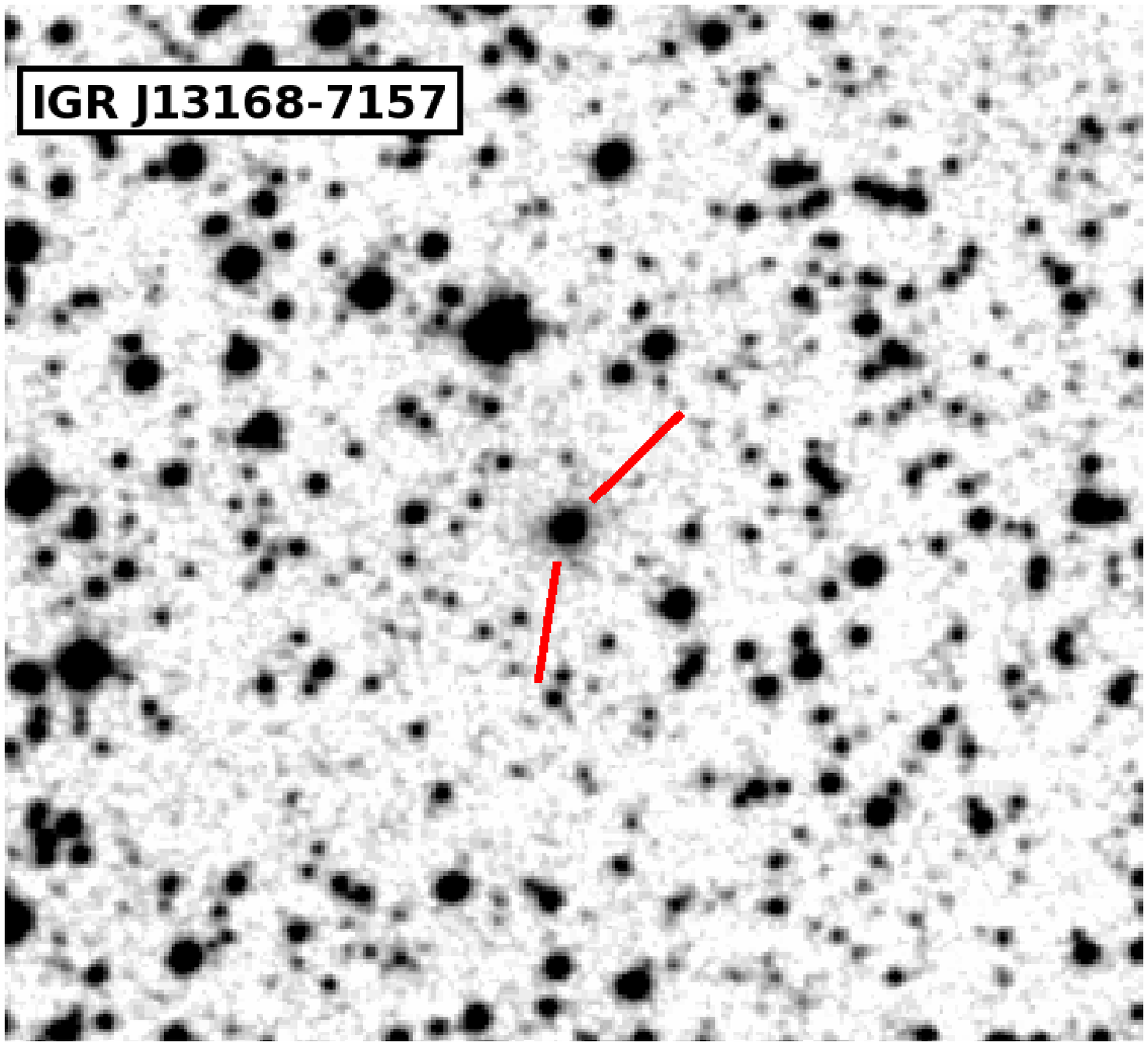,width=5.9cm}}}
\centering{\mbox{\psfig{file=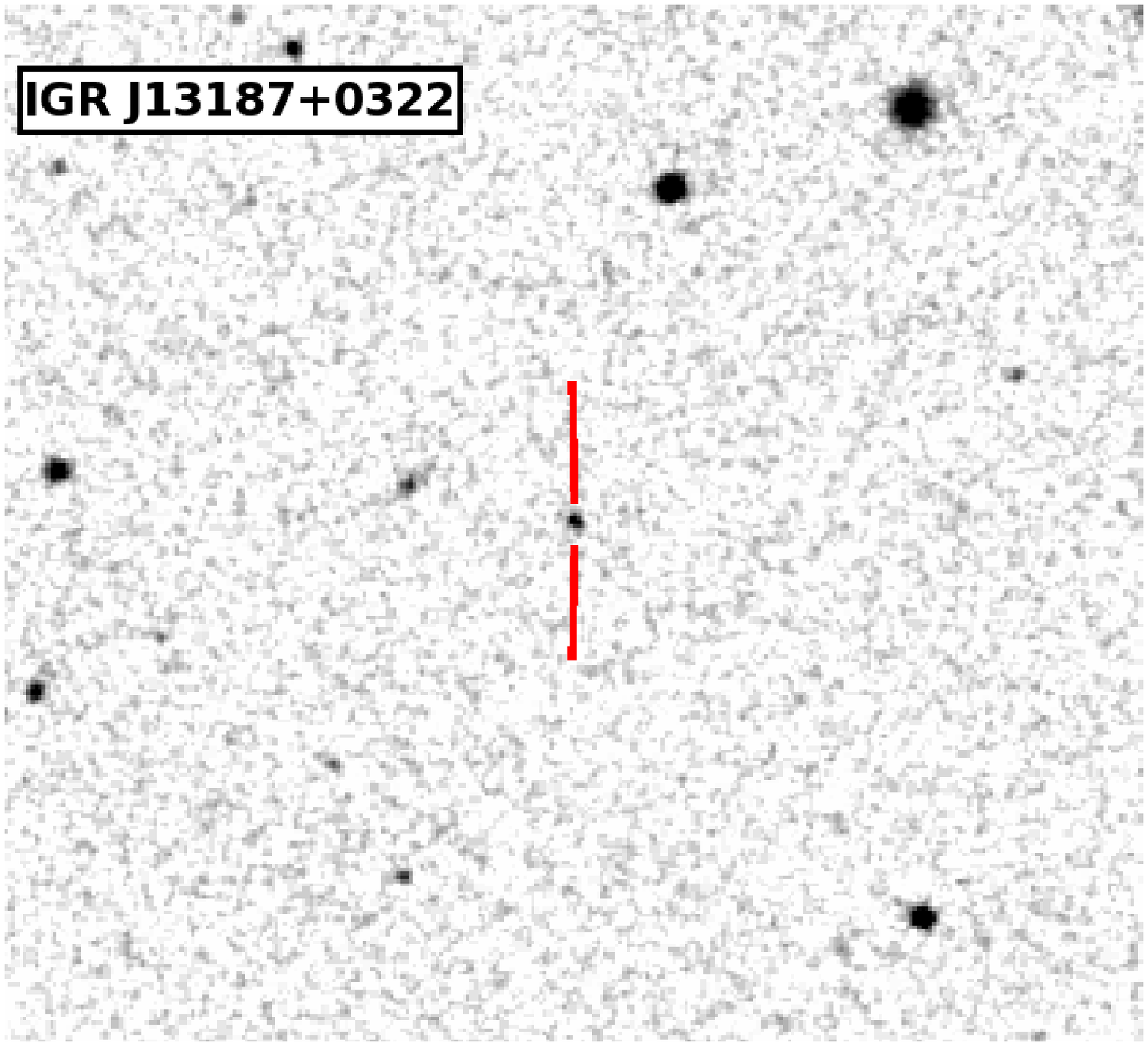,width=5.9cm}}}
\centering{\mbox{\psfig{file=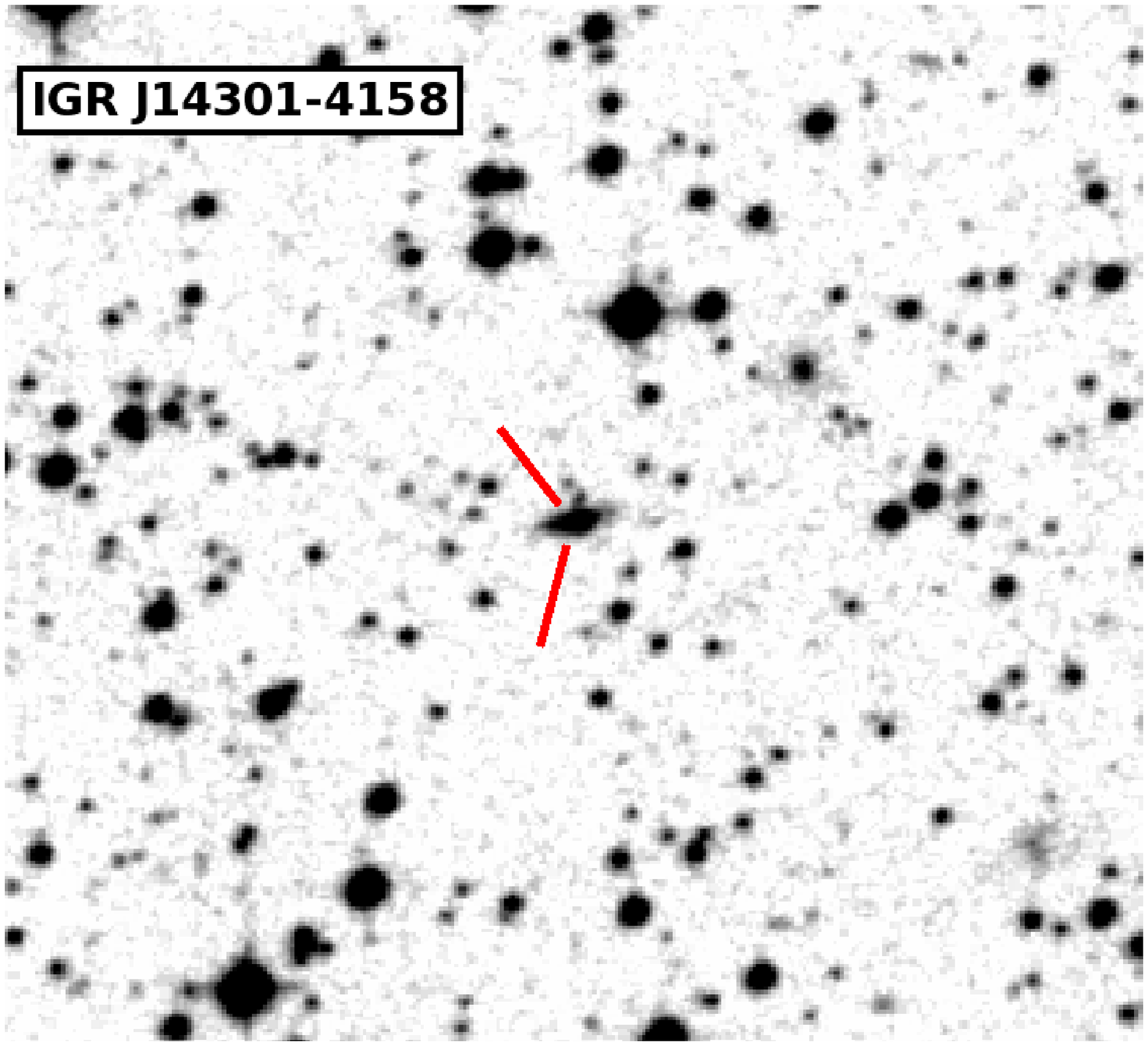,width=5.9cm}}}
\centering{\mbox{\psfig{file=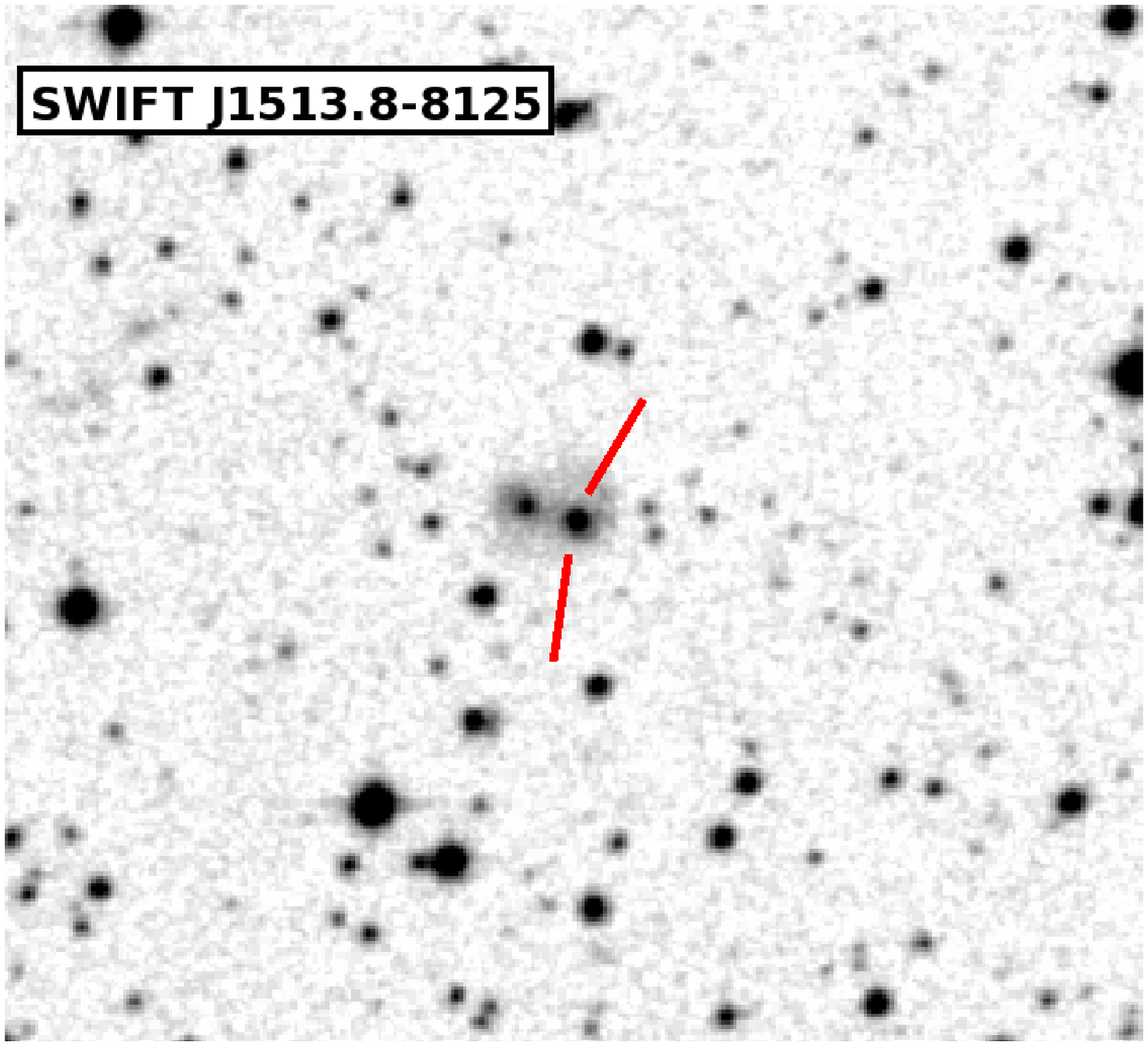,width=5.9cm}}}
\centering{\mbox{\psfig{file=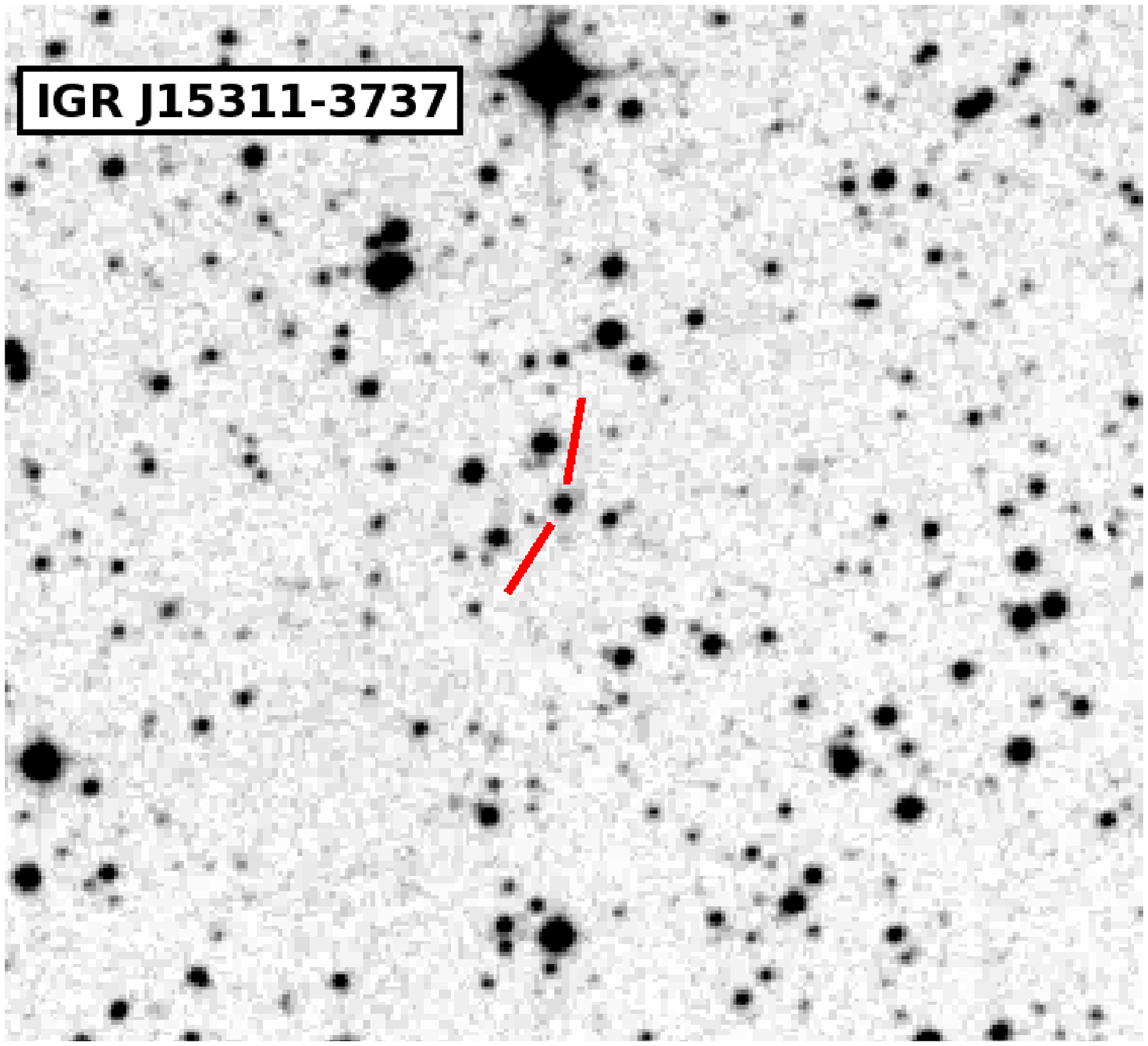,width=5.9cm}}}
\caption{As Fig. 1, but for 9 more {\it INTEGRAL} sources of our 
sample (see Table 1).}
\end{figure*}

\begin{figure*}[th!]
\hspace{-.1cm}
\centering{\mbox{\psfig{file=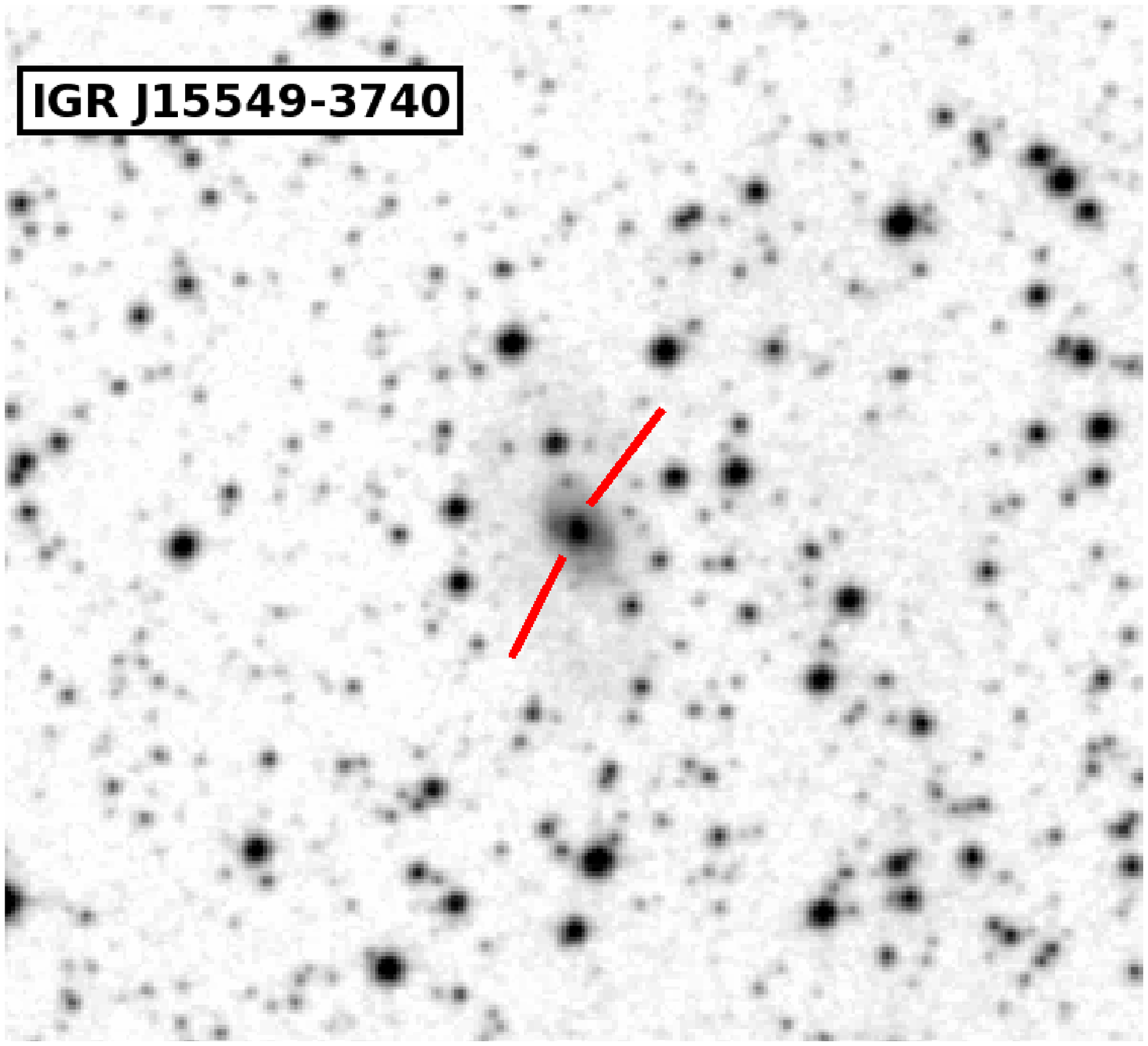,width=5.9cm}}}
\centering{\mbox{\psfig{file=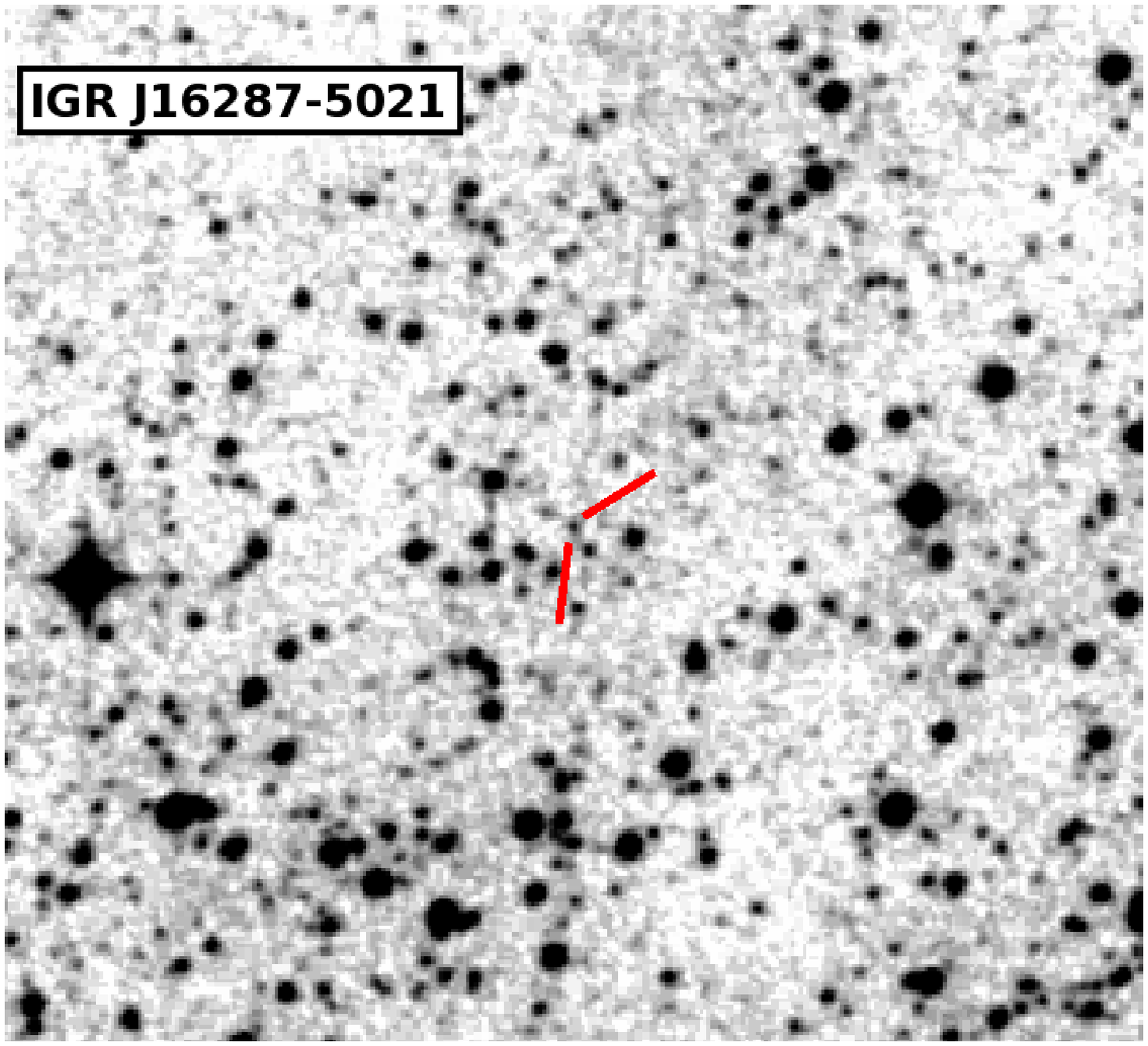,width=5.9cm}}}
\centering{\mbox{\psfig{file=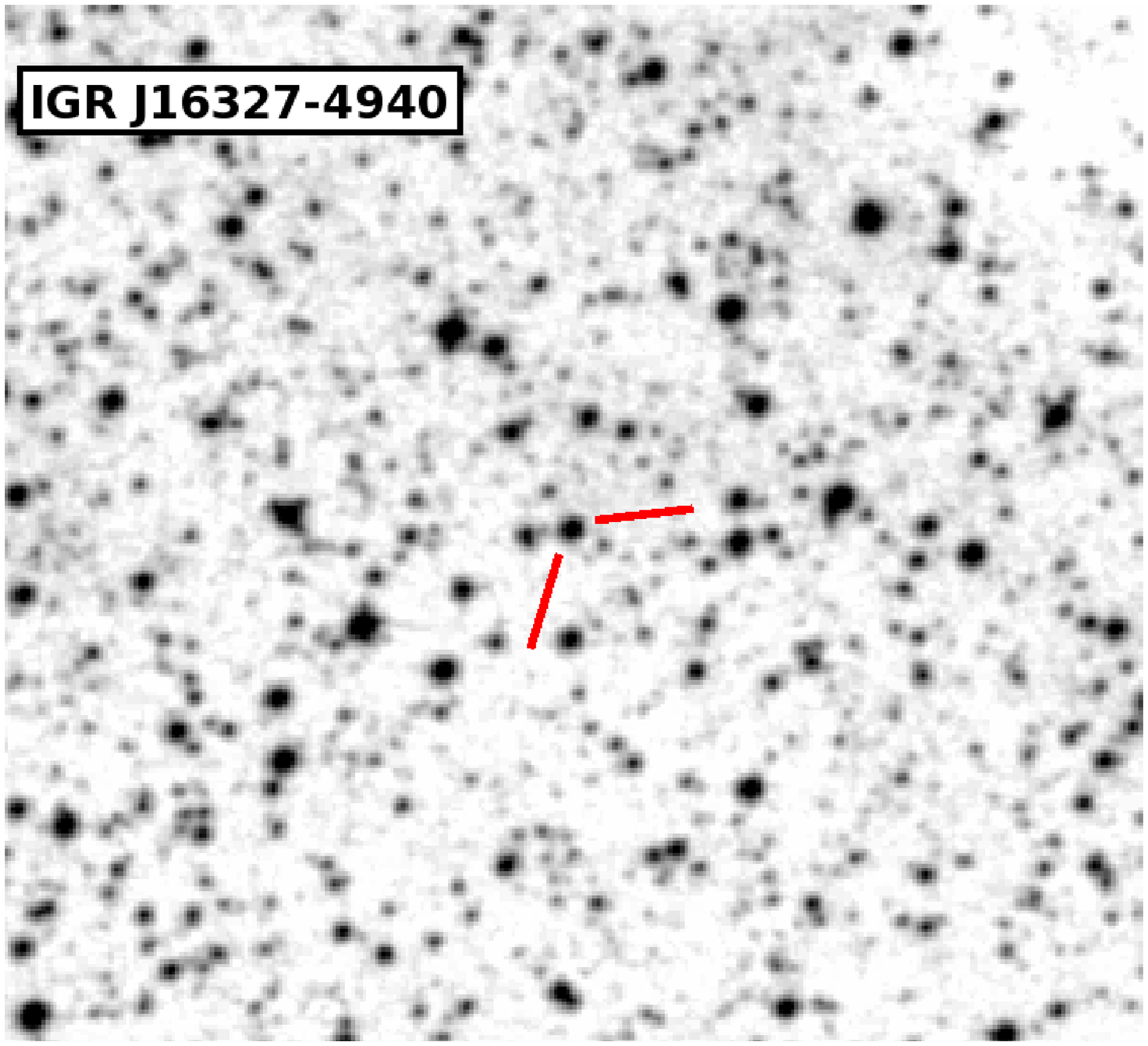,width=5.9cm}}}
\centering{\mbox{\psfig{file=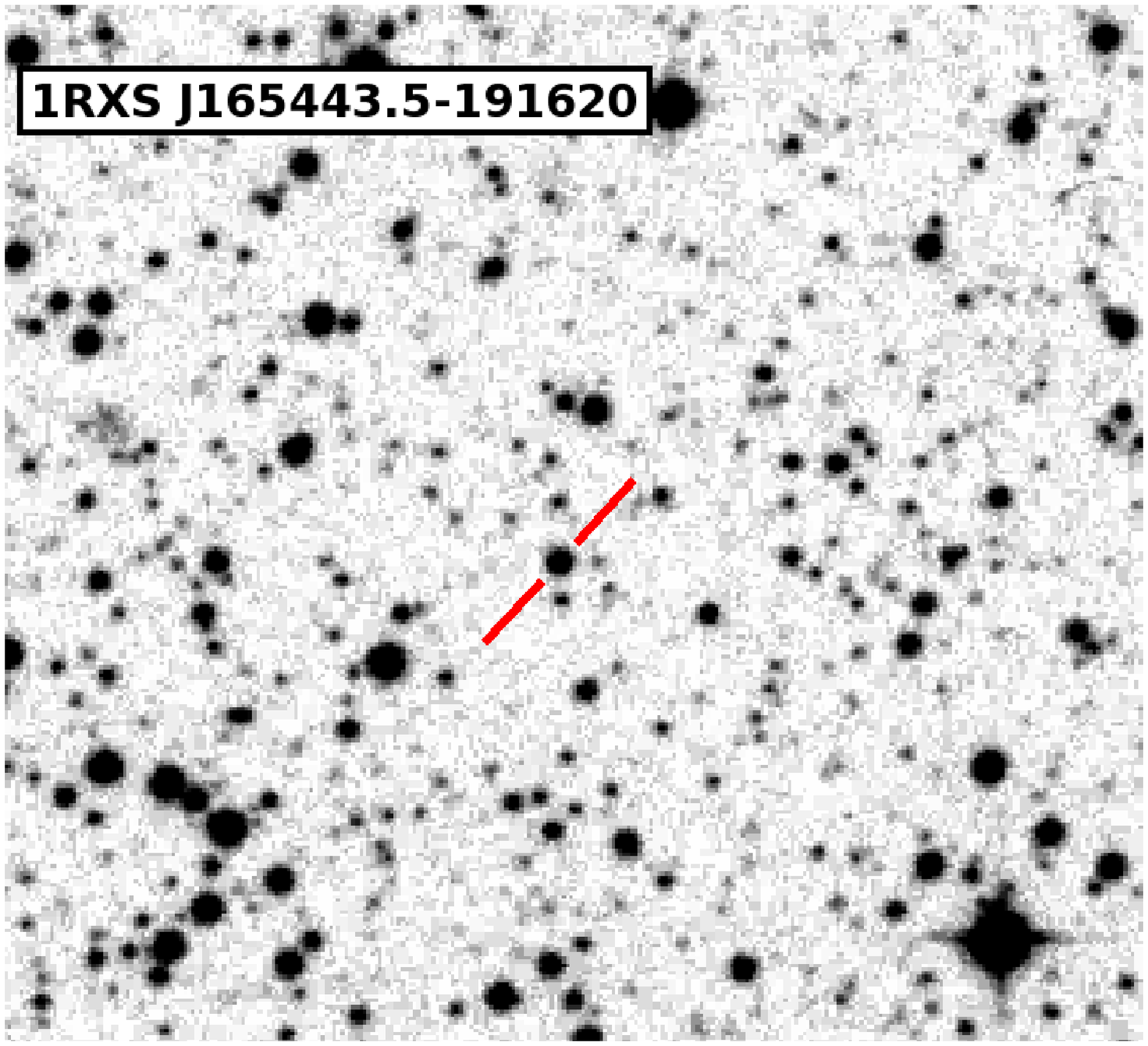,width=5.9cm}}}
\centering{\mbox{\psfig{file=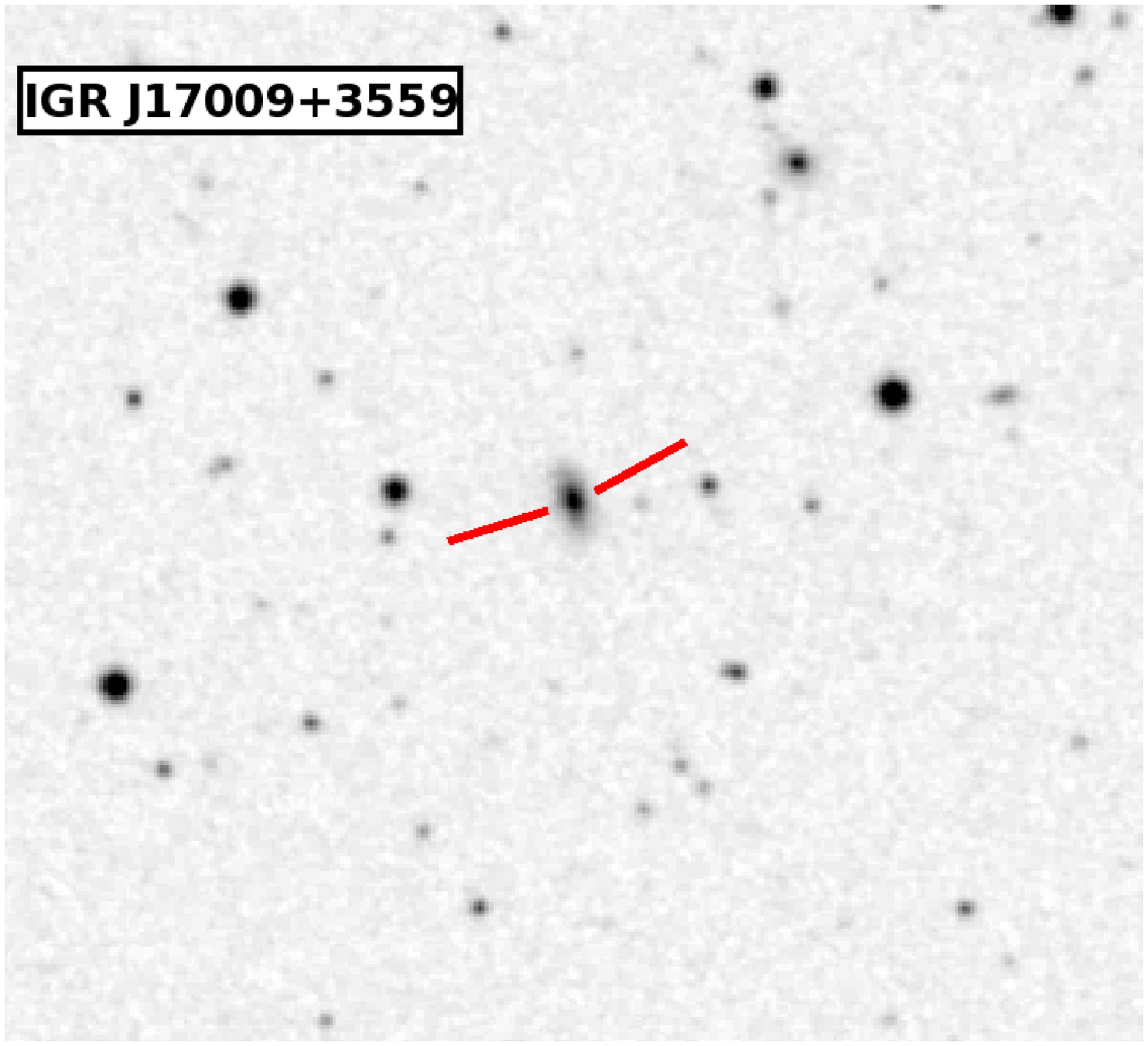,width=5.9cm}}}
\centering{\mbox{\psfig{file=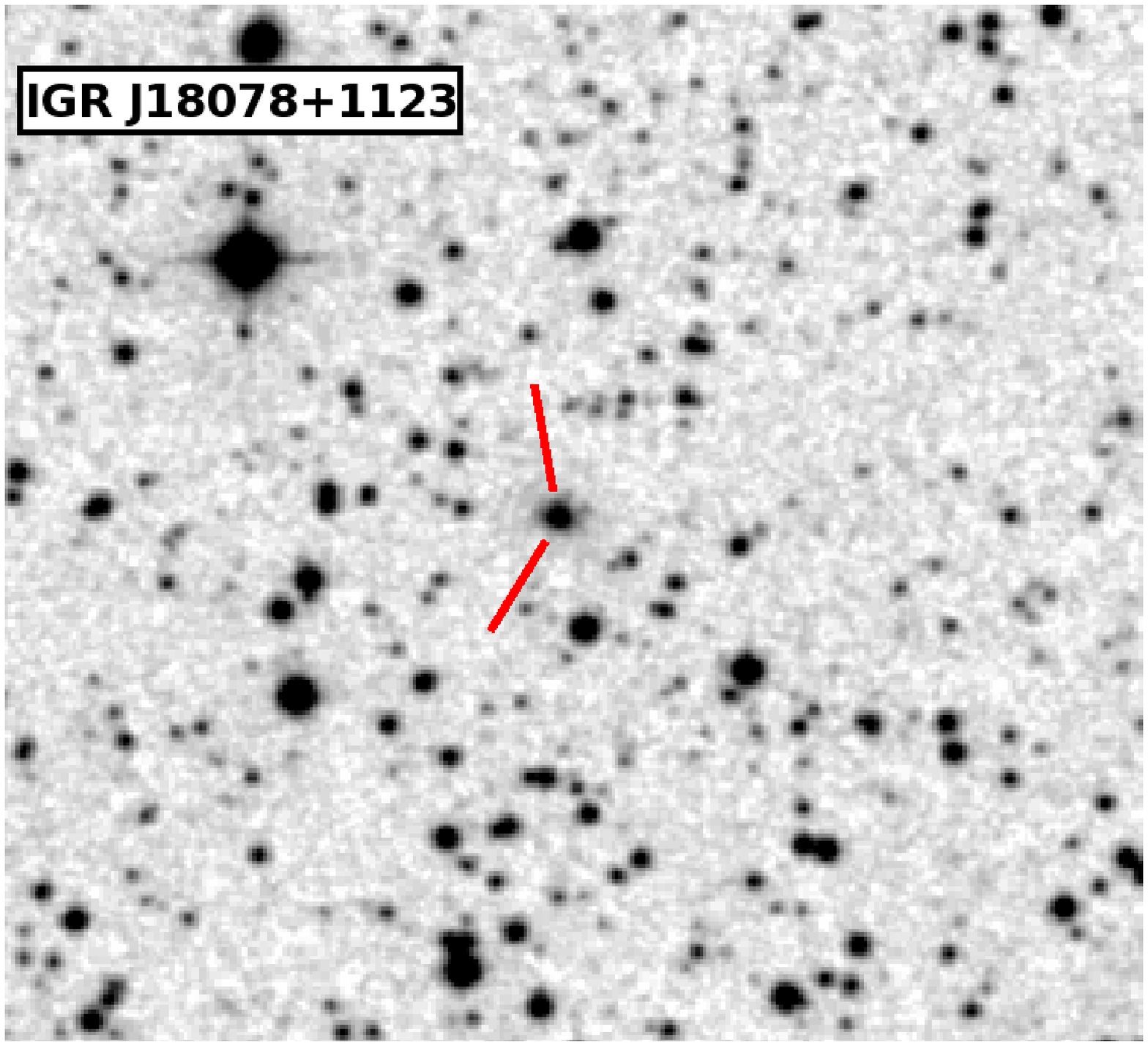,width=5.9cm}}}
\centering{\mbox{\psfig{file=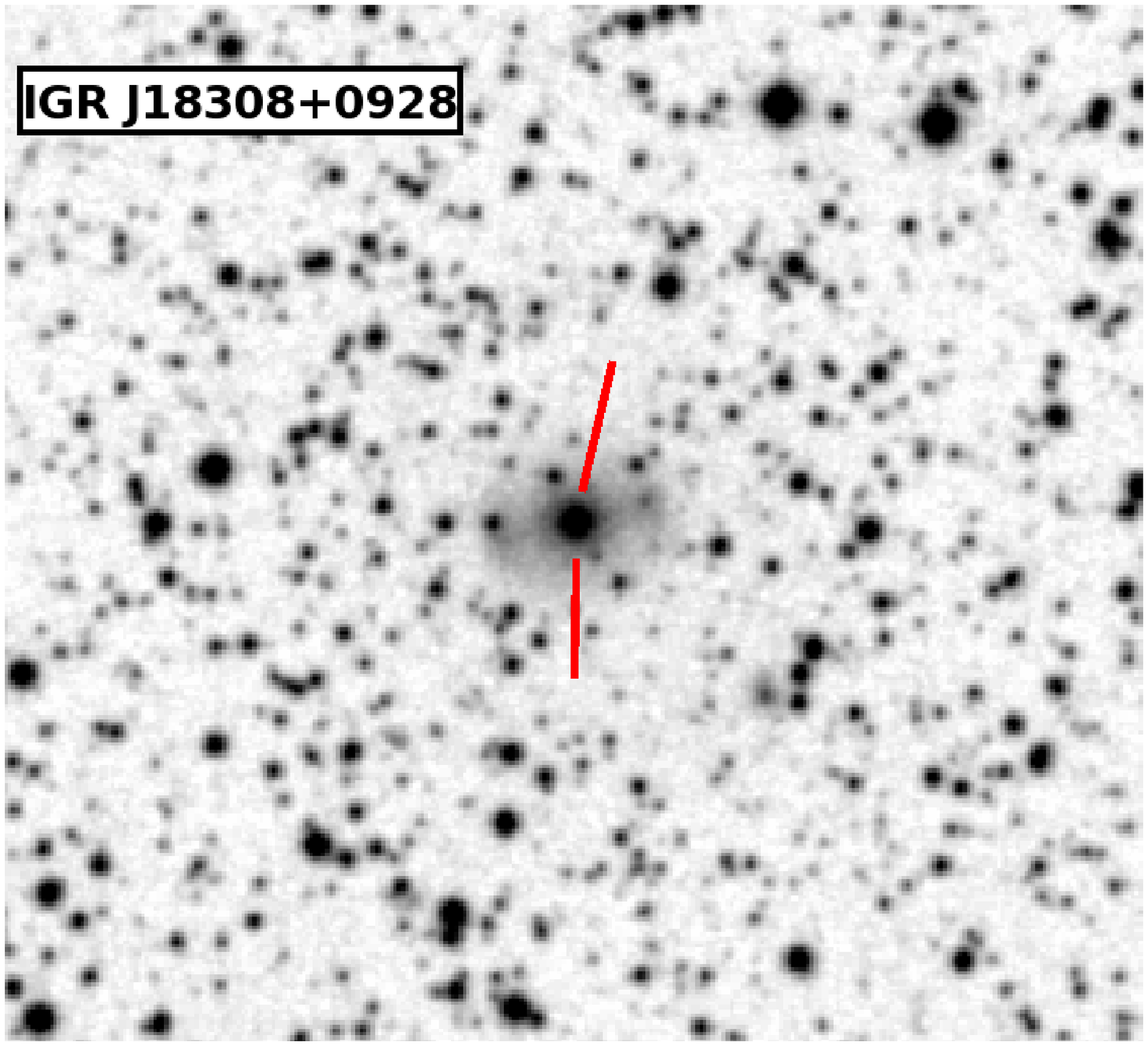,width=5.9cm}}}
\centering{\mbox{\psfig{file=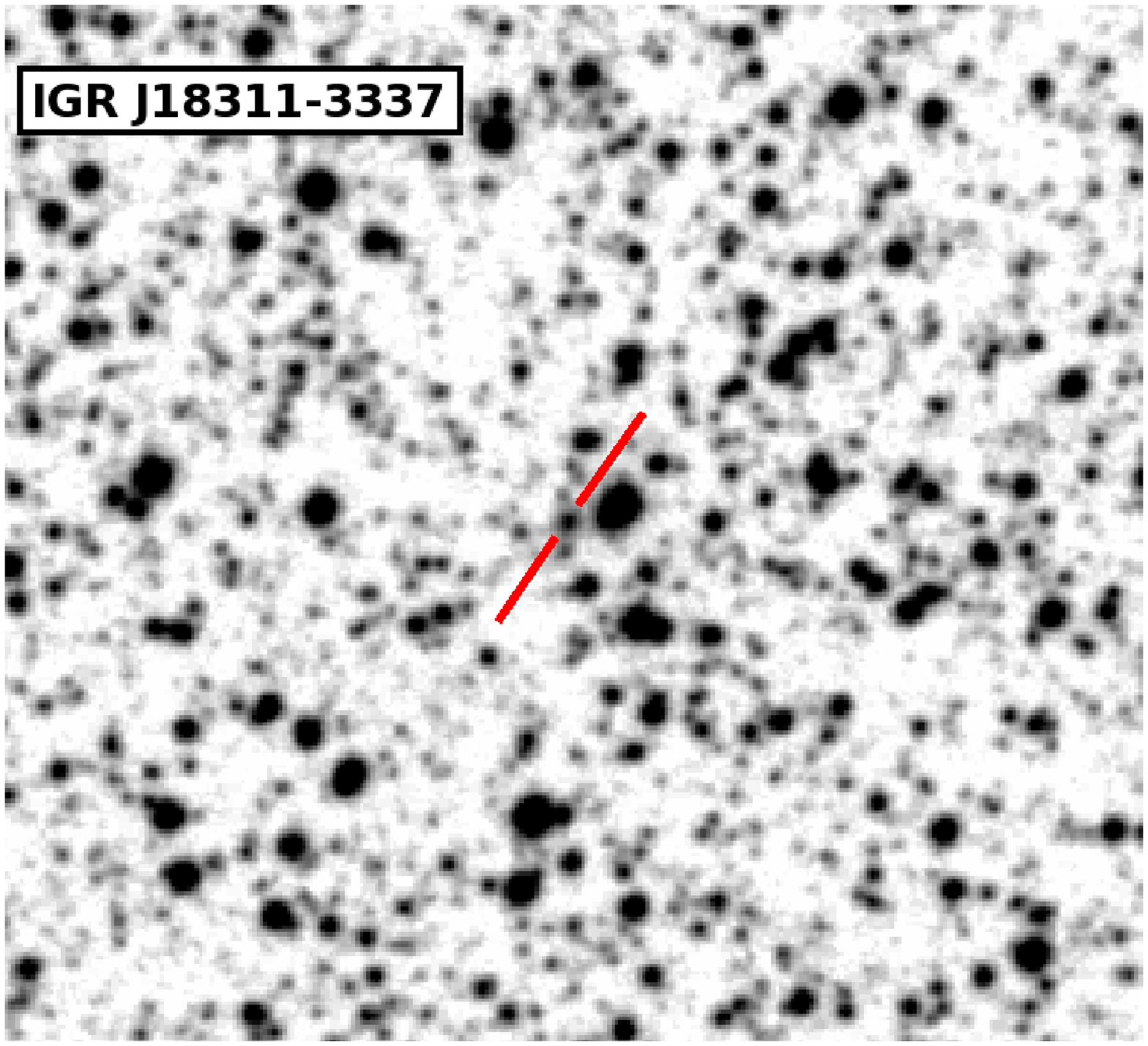,width=5.9cm}}}
\centering{\mbox{\psfig{file=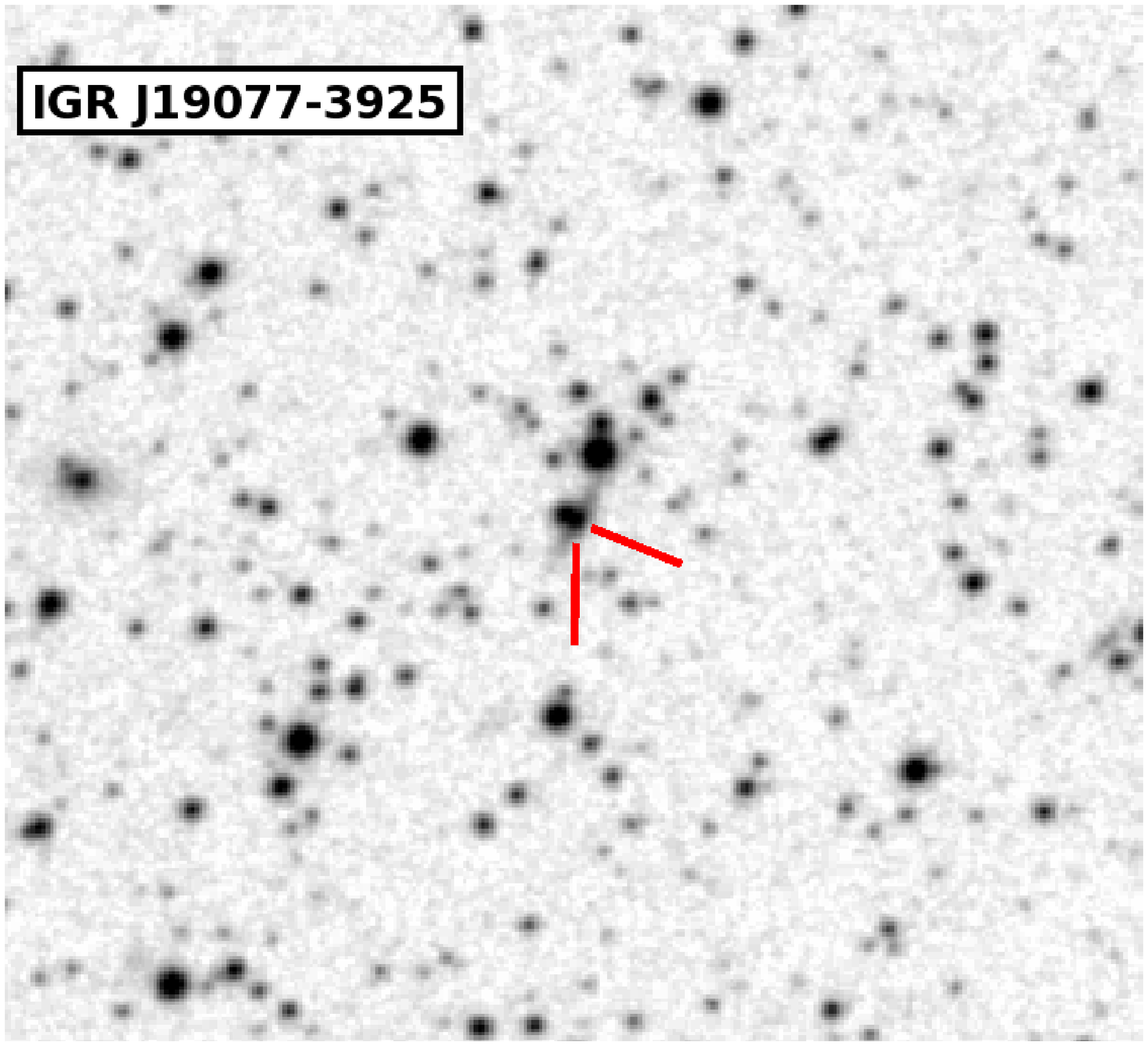,width=5.9cm}}}
\caption{As Fig. 1, but for 9 more {\it INTEGRAL} sources of our
sample (see Table 1).}
\end{figure*}

\begin{figure*}[th!]
\hspace{-.1cm}
\centering{\mbox{\psfig{file=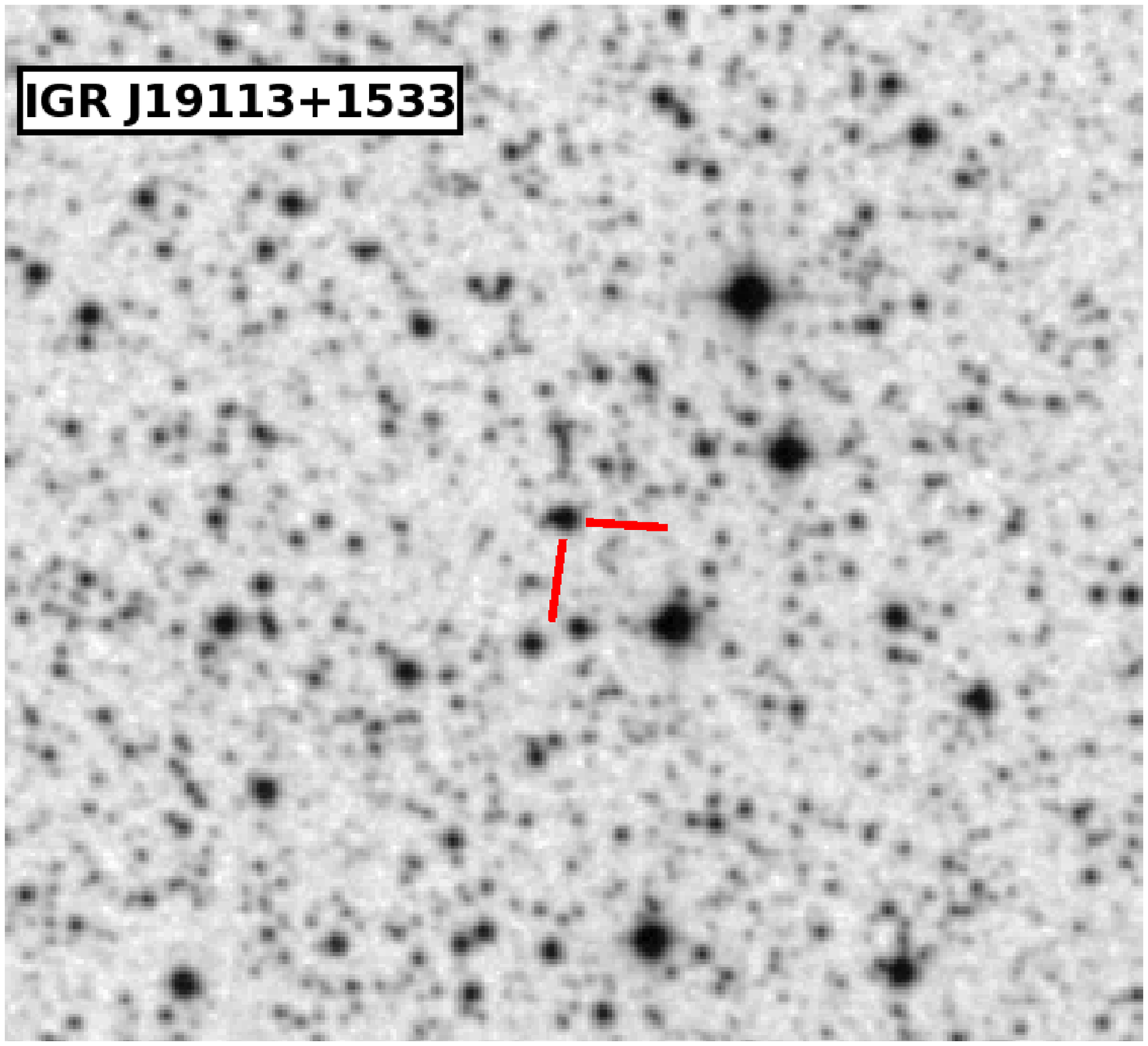,width=5.9cm}}}
\centering{\mbox{\psfig{file=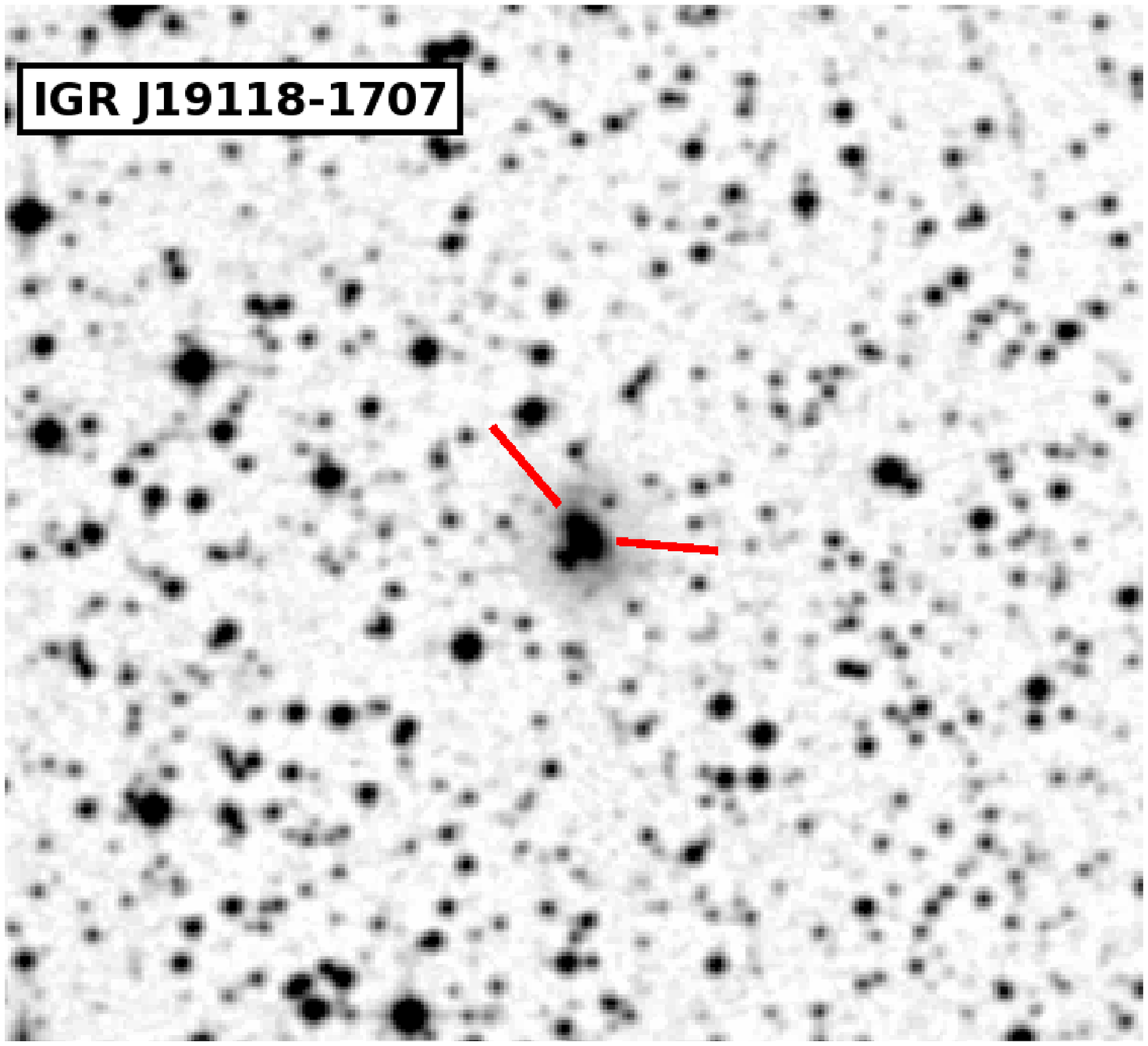,width=5.9cm}}}
\centering{\mbox{\psfig{file=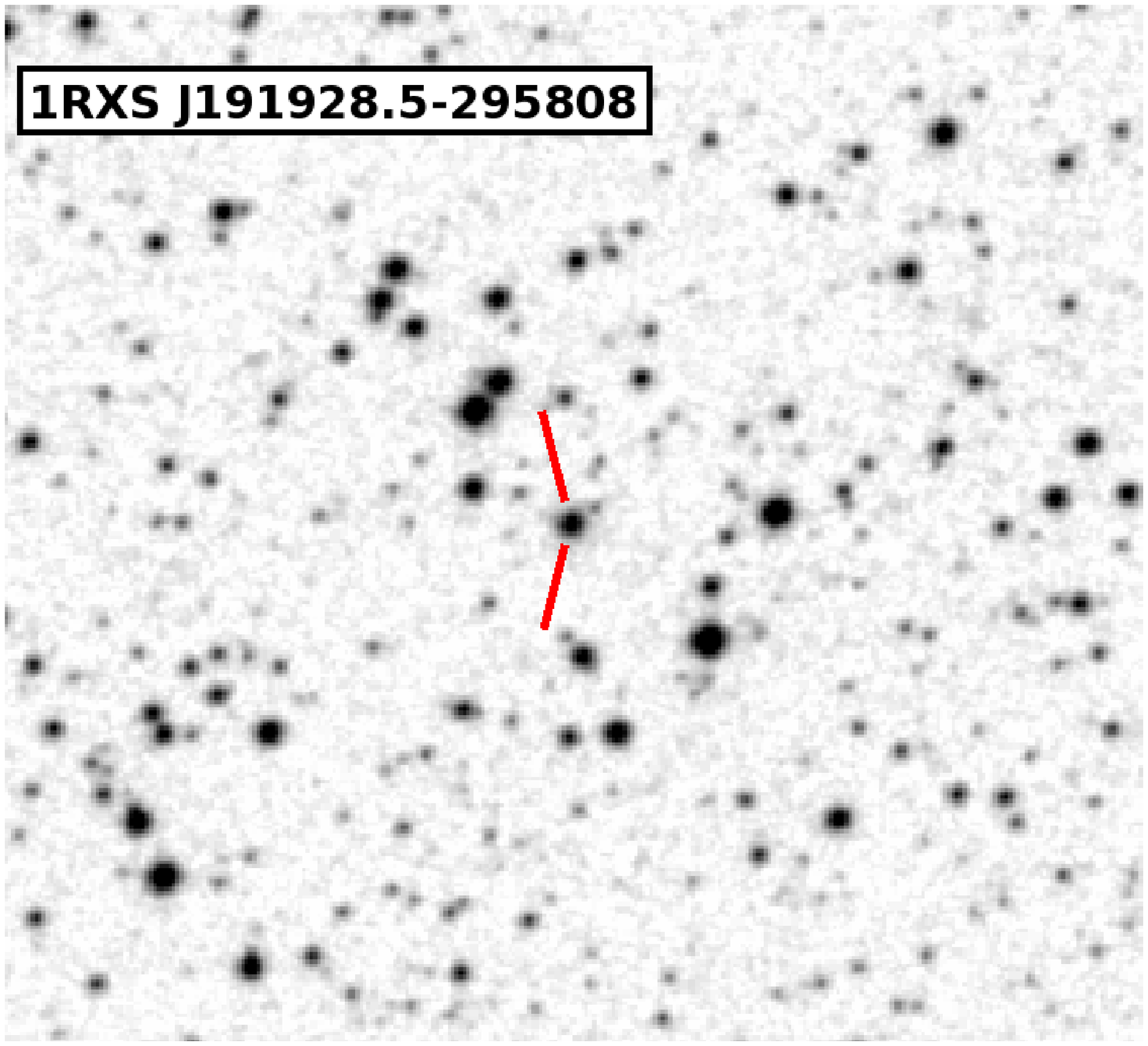,width=5.9cm}}}
\centering{\mbox{\psfig{file=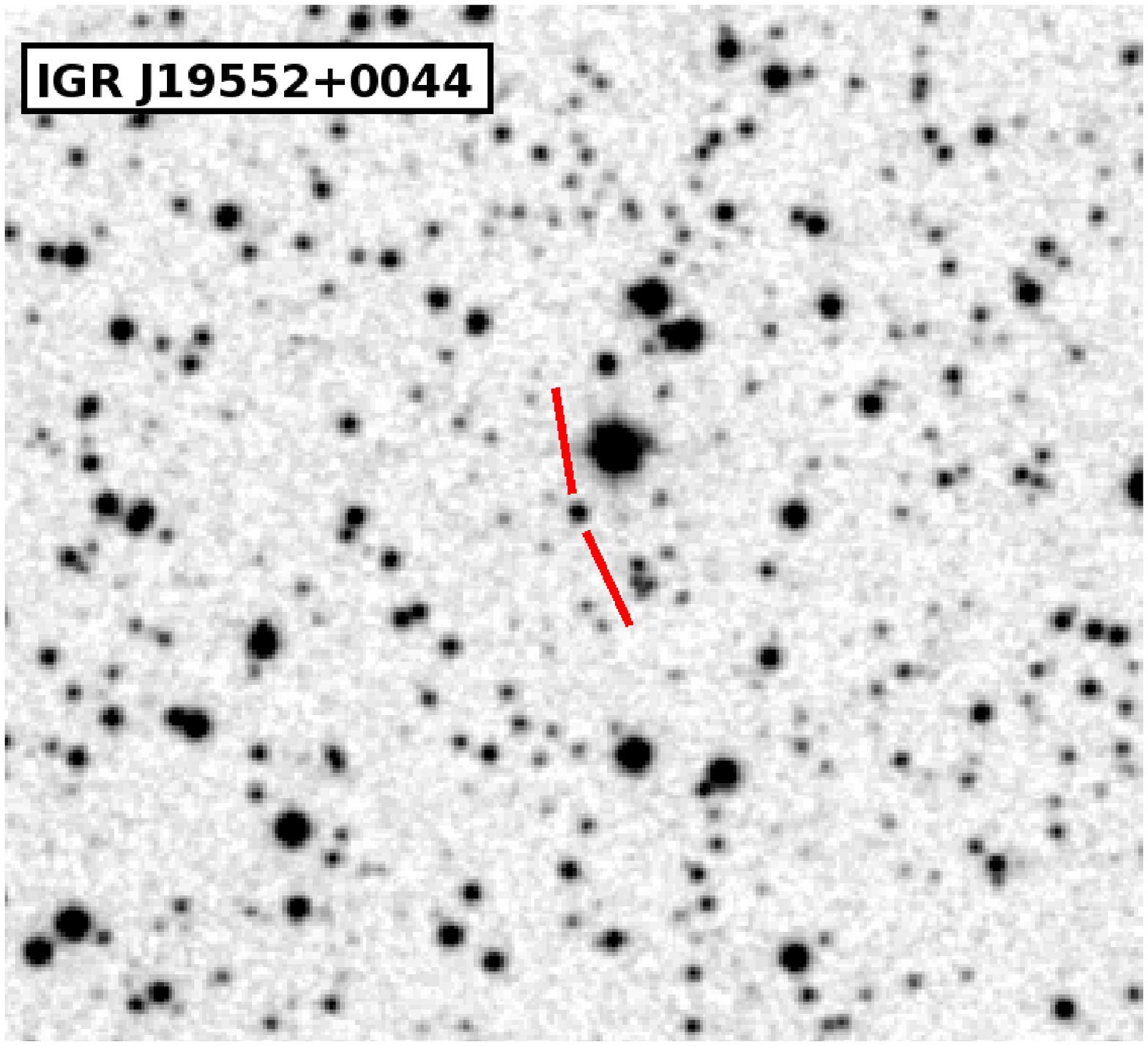,width=5.9cm}}}
\centering{\mbox{\psfig{file=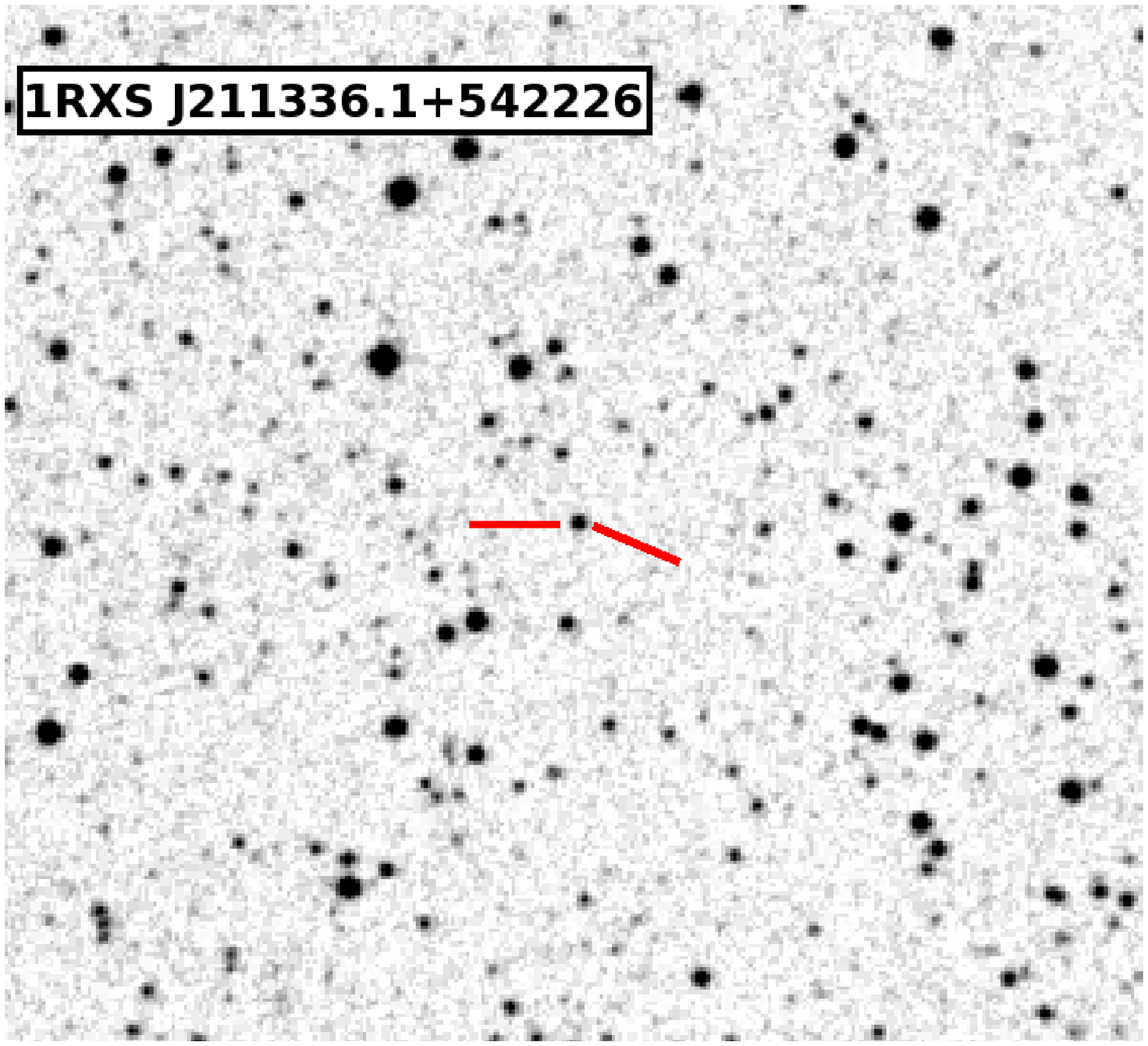,width=5.9cm}}}
\centering{\mbox{\psfig{file=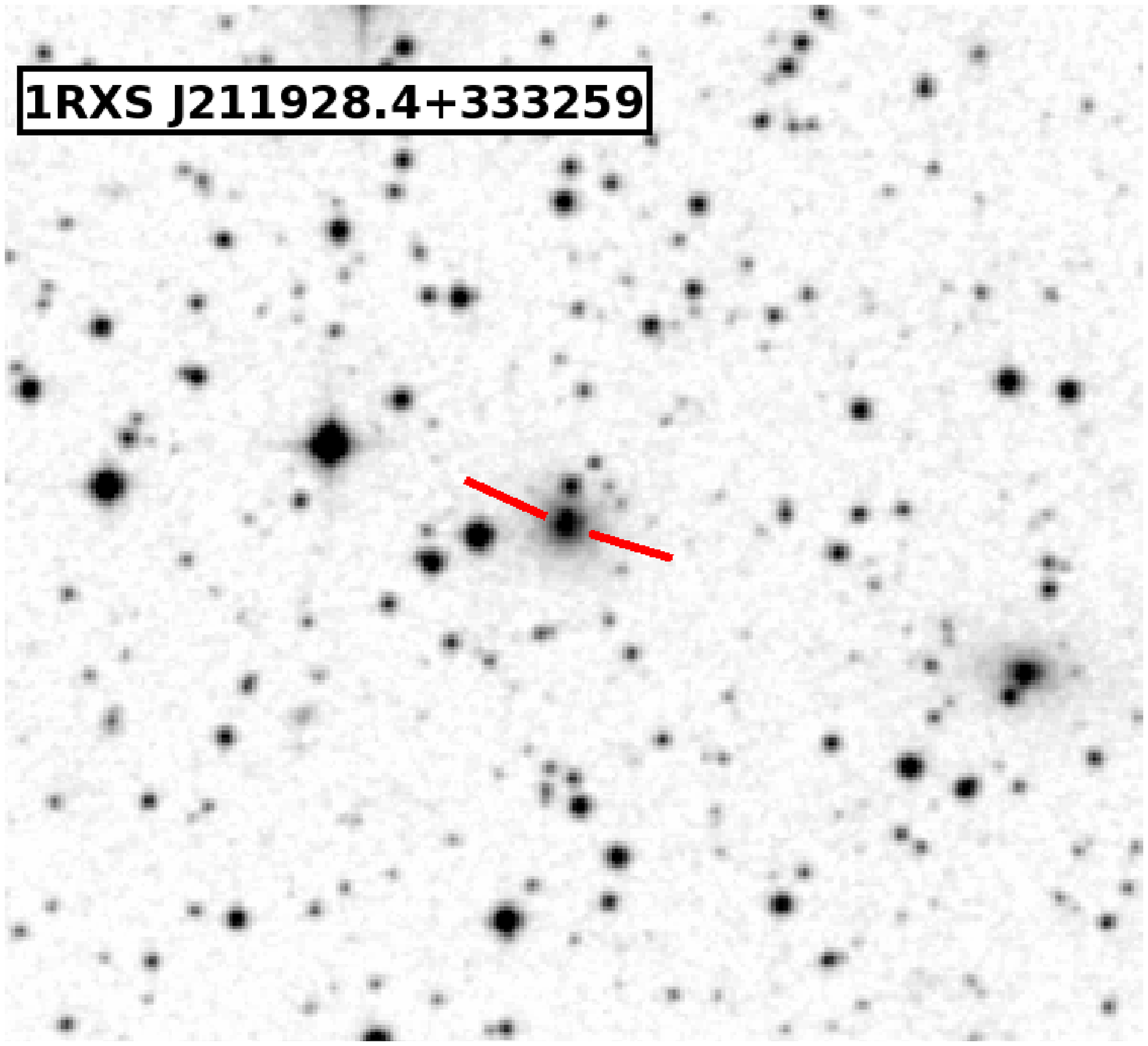,width=5.9cm}}}
\parbox{6cm}{
\psfig{file=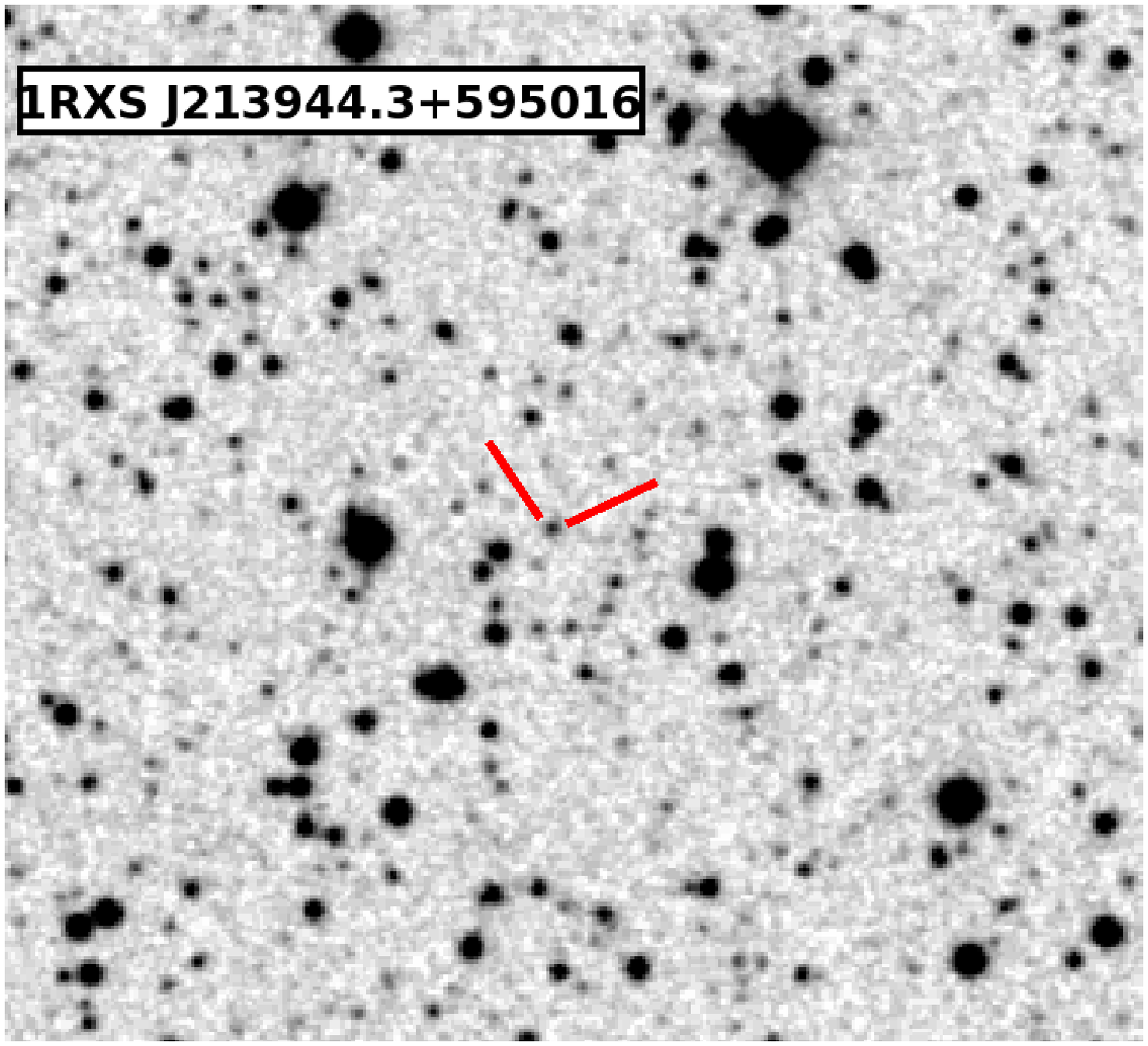,width=5.9cm}
}
\hspace{0.8cm}
\parbox{11cm}{
\vspace{-.5cm}
\caption{As Fig. 1, but for 7 more {\it INTEGRAL} sources of our
sample (see Table 1).}}
\end{figure*}

The fourth survey (Bird et al. 2010) performed by the IBIS instrument 
(Ubertini et al. 2003) onboard the {\it INTEGRAL} satellite (Winkler et 
al. 2003) has allowed the detection of more than 700 hard X--ray sources 
over the whole sky in the 20--100 keV band down to an average flux level 
of about 1 mCrab and with positional accuracies ranging between 0.2 and 
$\sim$5 arcmin.

In this survey, IBIS found mostly active galactic nuclei (AGNs, 35\% of 
the total number of detected objects) followed by known Galactic X--ray 
binaries (26\%) and cataclysmic variables (CVs, 5\%). A large 
number of the remaining objects (29\% of all detections) 
has no obvious counterpart at other wavelengths and therefore cannot be 
associated with any known class of high-energy emitting objects. 
Therefore, a multiwavelength observational campaign on these unidentified 
sources is crucial to determine their nature, and is especially 
important given that they constitute one third of the whole set of IBIS 
detections.

X--ray analysis methods have helped in identifying the nature of new 
{\it INTEGRAL} sources, such as for example 
X--ray timing (by the detection of pulsations or orbital periods, e.g., 
Walter et al. 2006; Corbet et al. 2010; La Parola et al. 2010) or, in 
general, X--ray spectroscopy and imaging (Tomsick et al. 2009 and 
references therein). Alternatively, and also quite effectively, 
cross-correlation with soft X--ray catalogs and optical spectroscopy on 
thereby selected candidates allows the determination of the nature and 
main multiwavelength characteristics of unidentified or poorly studied
hard X--ray objects.

We thus continue the identification work on {\it INTEGRAL} sources 
started in 2004, which has allowed us to identify up to now more than 100 
sources by means of optical spectroscopy (see Masetti et al. 2004, 
2006a,b,c,d, 2008a, 2009 --- hereafter Papers I-VII, respectively --- and 
Masetti et al. 2007, 2008b), by presenting optical spectra of the firm or 
likely counterparts of 42 unidentified, unclassified or poorly studied 
sources belonging to the 4$^{\rm th}$ IBIS catalog (Bird et al. 2010) or 
the online {\it INTEGRAL} All-Sky Survey Source Catalog (Krivonos et 
al. 2010\footnote{A preliminary version of this catalog can be found at \\
{\tt http://hea.iki.rssi.ru/rsdc/catalog/index.php}}); we moreover included 
two additional sources in our sample, that is, IGR J01054$-$7253 (Bozzo et 
al. 2009; Coe et al. 2010) and Swift J0208.4$-$7428 (McBride et al. 2010) 
observed in the Small Magellanic Cloud (SMC), thus reaching a total of 44 
objects. Optical spectroscopy for all of them was acquired using 7 
different telescopes and 2 public spectroscopic archives.

The paper is structured as follows: in Sect. 2, we explain the criteria 
used to select the sample of {\it INTEGRAL} and optical objects considered 
in this work.  In Sect. 3, a brief description of the observations is 
given. Section 4 reports and discusses the results, divided into three 
broad classes of sources (CVs, X--ray binaries, and AGNs), and an update 
of the statistical outline of the identifications of {\it INTEGRAL} 
sources obtained until now. Conclusions are drawn in Sect. 5.

The main results of this paper, along with the information about the {\it 
INTEGRAL} sources that have been identified (by us or by other groups) 
using optical or near-infrared (NIR) observations, are also collected in a 
web page\footnote{{\tt http://www.iasfbo.inaf.it/extras/IGR/main.html}} 
that we maintain as a service to the scientific community (Masetti \& 
Schiavone 2008). In this work, if not otherwise stated, errors and limits 
are reported at 1$\sigma$ and 3$\sigma$ confidence levels, respectively. 
We also note that this work supersedes the results presented in the 
preliminary analysis of Masetti et al. (2010) in which the identifications 
for a subsample of 25 sources were reported.

\section{Sample selection}

In a similar way to Papers I-VII, and this time using the two largest 
and most recently available IBIS surveys (Bird et al. 2010; Krivonos et 
al. 2010), we selected unidentified or unclassified hard X--ray 
sources that contain, within the 
IBIS 90\% confidence level error box, a single bright soft X--ray object 
detected either in the {\it ROSAT} all-sky surveys (Voges et al. 1999, 
2000), or with {\it Swift}/XRT (from Baumgartner et al. 2008,
Landi et al. 2009, 2010, Mescheryakov et al. 2009, Krivonos et al. 
2009, Coe et al. 2009, Kniazev et al. 2010, McBride et al. 2010, and 
--- in some cases independently --- from the 
XRT archive\footnote{XRT archival data are freely available at \\ {\tt 
http://www.asdc.asi.it/}}), or in the {\it XMM-Newton} Slew Survey (Saxton 
et al. 2008), or with {\it Chandra} (Tomsick et al. 2009). This approach 
was proven by Stephen et al. (2006) to be very effective in associating, 
with a high degree of probability, IBIS sources with a softer 
X--ray counterpart and in turn drastically reducing their positional 
error circles to a few arcsec in radius, thus shrinking the search 
area by a factor of $\sim$10$^4$.

After this first selection, we chose among these objects those that had, 
within their refined 90\% confidence level soft X--ray error 
boxes\footnote{When needed, for the cases in which the soft X--ray 
positional error is given at 1$\sigma$ confidence level (basically, the
{\it ROSAT} and {\it XMM-Newton} Slew Survey sources) we rescaled it
to the corresponding 90\% confidence level assuming a Gaussian probability 
distribution.}, a single or a few (3 at most) possible optical counterparts 
with magnitudes $R \la$ 19 in the DSS-II-Red survey\footnote{Available at 
{\tt http://archive.eso.org/dss/dss}}, for which optical spectroscopy 
could be obtained with reasonable signal-to-noise ratio (S/N) at 
telescopes with diameter smaller than 4 metres.

This allowed us to pinpoint 31 sources, to which we added two hard X--ray
objects detected with {\it INTEGRAL} in the SMC, as mentioned above (Bozzo 
et al. 2009; Coe et al. 2009, 2010; McBride et al. 2010).

To enlarge the sample, we performed a similar procedure by
cross-correlating the above IBIS surveys with radio catalogs such as the 
NVSS (Condon et al. 1998), SUMSS (Mauch et al. 2003), and MGPS (Murphy et 
al. 2007) surveys when a soft X--ray observation of the hard X--ray source 
field was not available. This step provided 5 more {\it INTEGRAL} sources 
with likely optical counterparts. Finally, we also considered IBIS objects
that contain within their error box a conspicuous galaxy belonging 
to the 2MASX archive (Skrutskie et al. 2006), or an emission-line object 
present in the SIMBAD database\footnote{Available at {\tt 
http://simbad.u-strasbg.fr}}. This identified 6 additional IBIS sources to 
be included in our sample.

Although selected by means of cross-correlation with optical catalogs 
only, the position of the possible counterpart to IGR J08262$-$3736 is
within the 2$\sigma$ error circle of a {\it ROSAT} all-sky faint 
survey source (Voges et al. 2000); however, given the large extent of its 
{\it ROSAT} error circle (31$''$ in radius at 1$\sigma$ confidence 
level), we consider the association between the optical and the {\it ROSAT} 
sources too loose to be assumed as firm.

Thus, in total we gathered a sample of 44 {\it INTEGRAL} objects with
possible optical counterparts, which we explored by means of optical 
spectroscopy. Their names and accurate coordinates (to 1$''$ or less) 
are reported in Table 1, while their optical finding charts are shown in 
Figs. 1-5, with the corresponding putative counterparts indicated 
with tick marks. The finding chart for source IGR J22292+6647 is not 
reported here as it was already published in Landi et al. (2009).

For the source naming in Table 1, we simply adopted the names as they 
appear in the relevant catalogs (Bird et al. 2010; Krivonos 2010) or 
papers (Bozzo et al. 2009; Coe et al. 2010; McBride et al. 2010), and the 
``IGR" alias when available.
However, we note that for one of the hard X--ray objects selected 
by means of cross-correlation with soft X--ray catalogs, i.e., 1RXS 
J191928.5$-$295808, we chose to use its {\it ROSAT} name (thus associated 
with its soft X--ray emission) rather than the denomination reported in 
the 4$^{\rm th}$ IBIS Survey (PKS 1916$-$300; Bird et al. 2010) because 
the latter refers to a radio source that is slightly but significantly 
offset from the optical and soft X--ray positions of the possible 
counterpart to this {\it INTEGRAL} hard X--ray object.

In any case, we wish to point out that the 11 sources with no 
associated soft X--ray counterpart should be considered to have only a 
tentative, albeit likely, counterpart: this is indicated with an asterisk 
in Table 1. This caution, although to a much lesser extent, should also be 
applied to those IBIS sources (IGR J03344+1506, IGR J09094+2735, and 1RXS 
J211336.1+542226) associated only with an object belonging to the {\it 
ROSAT} all-sky faint survey (Voges et al. 2000) as pointed out by Stephen 
et al. (2006). The reader is nevertheless referred to Paper III for the 
caveats and the shortcomings of choosing, within an IBIS error box, 
``peculiar" sources that are not straightforwardly linked to an 
arcsec-sized soft X--ray position.

We also note that, for the high-energy sources in our sample with more 
than one optical candidate in the corresponding arcsec-sized error box, 
all objects with $R\la$ 19 were spectroscopically observed. However, we 
only indicate their firm or likely optical counterpart, identified 
on the basis of peculiar spectral features (that is, emission 
lines). All other candidates are considered no further because their 
spectra do not exhibit any peculiarity.

To summarize, we emphasize that in our final sample there are 4 {\it 
INTEGRAL} sources (IGR J12107+3822, Mescheryakov et al. 2009; IGR 
J13168$-$7157, Kniazev et al. 2010; IGR J17009+3559, Krivonos et al. 2009; 
and IGR J22292+6647, Butler et al. 2009) that, although already 
identified by these authors, have incomplete information at longer 
wavelengths or were independently observed by us before their 
identification was published. Our observations are thus presented here to 
confirm the nature of these objects and improve their classification 
and the amount of information known about them.

\begin{table*}[th!]
\caption[]{Log of the spectroscopic observations presented in this paper
(see text for details). If not indicated otherwise, source coordinates
were extracted from the 2MASS catalog and have an accuracy better than 
0$\farcs$1.}
\scriptsize
\begin{center}
\begin{tabular}{llllcccr}
\noalign{\smallskip}
\hline
\hline
\noalign{\smallskip}
\multicolumn{1}{c}{{\it (1)}} & \multicolumn{1}{c}{{\it (2)}} & \multicolumn{1}{c}{{\it (3)}} & \multicolumn{1}{c}{{\it (4)}} & 
{\it (5)} & {\it (6)} & {\it (7)} & \multicolumn{1}{c}{{\it (8)}} \\
\multicolumn{1}{c}{Object} & \multicolumn{1}{c}{RA} & \multicolumn{1}{c}{Dec} & 
\multicolumn{1}{c}{Telescope+instrument} & $\lambda$ range & Disp. & \multicolumn{1}{c}{UT Date \& Time}  & Exposure \\
 & \multicolumn{1}{c}{(J2000)} & \multicolumn{1}{c}{(J2000)} & & (\AA) & (\AA/pix) & 
\multicolumn{1}{c}{at mid-exposure} & time (s)  \\

\noalign{\smallskip}
\hline
\noalign{\smallskip}

IGR J00158+5605         & 00:15:54.19 &   +56:02:57.5 & Cassini+BFOSC       & 3500-8700  & 4.0 & 16 Nov 2009, 19:46 & 2$\times$1800 \\ 
IGR J00465$-$4005       & 00:46:20.68 & $-$40:05:49.1 & CTIO 1.5m+RC Spec.  & 3300-10500 & 5.7 & 12 Sep 2009, 04:22 & 2$\times$1800 \\ 
IGR J01054$-$7253       & 01:04:42.28 & $-$72:54:03.7 & Radcliffe+Gr. Spec. & 3850-7200  & 2.3 & 16 Aug 2009, 03:17 & 2$\times$1800 \\ 
IGR J01545+6437$^*$     & 01:54:35.29 &   +64:37:57.5 & Copernicus+AFOSC    & 3500-7800  & 4.2 & 07 Dec 2008, 22:51 & 2$\times$1800 \\ 
Swift J0208.4$-$7428    & 02:06:45.17 & $-$74:27:47.7 & CTIO 1.5m+RC Spec.  & 3300-10500 & 5.7 & 27 Dec 2009, 01:13 & 2$\times$1200 \\ 
IGR J02086$-$1742       & 02:08:34.95 & $-$17:39:34.8 & SPM 2.1m+B\&C Spec. & 3500-7800  & 4.0 & 03 Dec 2008, 06:28 & 2$\times$1800 \\ 
IGR J03344+1506         & 03:34:32.77 &   +15:08:01.2 & SPM 2.1m+B\&C Spec. & 3500-7800  & 4.0 & 19 Sep 2009, 10:31 & 2$\times$1800 \\ 
IGR J04451$-$0445$^*$   & 04:44:52.93 & $-$04:46:39.5 & SPM 2.1m+B\&C Spec. & 3500-7800  & 4.0 & 19 Sep 2009, 11:49 & 2$\times$1800 \\ 
IGR J04571+4527         & 04:57:06.98 &   +45:27:48.5 & SPM 2.1m+B\&C Spec. & 3500-7800  & 4.0 & 29 Jan 2009, 07:01 & 2$\times$1800 \\ 
IGR J05253+6447         & 05:24:28.61 &   +64:44:43.7 & Cassini+BFOSC       & 3500-8700  & 4.0 & 09 Dec 2009, 22:35 & 2$\times$1800 \\ 
IGR J06058$-$2755       & 06:05:48.96 & $-$27:54:40.1 & CTIO 1.5m+RC Spec.  & 3300-10500 & 5.7 & 30 Nov 2009, 02:51 & 2$\times$1000 \\ 
IGR J06293$-$1359$^*$   & 06:29:09.3$^\ddagger$ & $-$14:04:49$^\ddagger$ & SPM 2.1m+B\&C Spec. & 3500-7800  & 4.0 & 05 Dec 2008, 05:38 & 2$\times$1800 \\ 
1RXS J080114.6$-$462324 & 08:01:17.03 & $-$46:23:27.5 & NTT+EFOSC2          & 3650-9300  & 5.5 & 01 Jun 2009, 00:31 &          1200 \\ 
IGR J08262$-$3736$^*$   & 08:26:13.65 & $-$37:37:11.9 & CTIO 1.5m+RC Spec.  & 3300-10500 & 5.7 & 30 Nov 2009, 08:18 &  2$\times$900 \\ 
Swift J0845.0$-$3531    & 08:45:21.38 & $-$35:30:24.2 & CTIO 1.5m+RC Spec.  & 3300-10500 & 5.7 & 19 Dec 2009, 06:10 & 2$\times$1800 \\ 
IGR J08557+6420         & 08:55:12.54 &   +64:23:45.5 & SPM 2.1m+B\&C Spec. & 3500-7800  & 4.0 & 30 Jan 2009, 07:17 & 2$\times$1800 \\ 
IGR J09094+2735         & 09:09:18.75 &   +27:37:33.7 & SDSS+CCD Spec.      & 3800-9200  & 1.0 & 17 Jan 2005, 06:22 &          2700 \\
MCG +04$-$26$-$006      & 10:46:42.51 &   +25:55:53.9 & SPM 2.1m+B\&C Spec. & 3500-7800  & 4.0 & 03 Dec 2008, 11:51 & 2$\times$1800 \\ 
PKS 1143$-$696          & 11:45:53.62 & $-$69:54:01.8 & CTIO 1.5m+RC Spec.  & 3300-10500 & 5.7 & 31 Dec 2009, 06:19 & 2$\times$1800 \\ 
IGR J12107+3822         & 12:10:44.28 &   +38:20:10.2 & SDSS+CCD Spec.      & 3800-9200  & 1.0 & 13 Apr 2005, 07:12 &          2300 \\
IGR J12123$-$5802       & 12:12:26.24 & $-$58:00:20.5 & CTIO 1.5m+RC Spec.  & 3300-10500 & 5.7 & 31 Dec 2009, 07:22 & 2$\times$1800 \\ 
IGR J1248.2$-$5828      & 12:47:57.84 & $-$58:30:00.2 & Radcliffe+Gr. Spec. & 3850-7200  & 2.3 & 13 Aug 2009, 17:56 &          1800 \\ 
IGR J13168$-$7157       & 13:16:54.28 & $-$71:55:27.1 & CTIO 1.5m+RC Spec.  & 3300-10500 & 5.7 & 28 Jan 2010, 05:15 & 2$\times$1200 \\ 
IGR J13187+0322$^*$     & 13:18:31.24 &   +03:19:48.9 & SDSS+CCD Spec.      & 3800-9200  & 1.0 & 22 Jan 2002, 13:07 &          4800 \\
IGR J14301$-$4158$^*$   & 14:30:12.17 & $-$41:58:31.4 & AAT+6dF             & 3900-7600  & 1.6 & 29 Jun 2003, 12:00 &      600+1200 \\ 
Swift J1513.8$-$8125    & 15:14:41.92 & $-$81:23:38.0 & Radcliffe+Gr. Spec. & 3850-7200  & 2.3 & 13 Aug 2009, 20:37 & 2$\times$1800 \\ 
IGR J15311$-$3737       & 15:30:51.79 & $-$37:34:57.3 & Radcliffe+Gr. Spec. & 3850-7200  & 2.3 & 15 Aug 2009, 19:36 & 2$\times$1800 \\ 
IGR J15549$-$3740$^*$   & 15:54:46.76 & $-$37:38:19.1 & AAT+6dF             & 3900-7600  & 1.6 & 12 May 2002, 15:38 &      1200+600 \\ 
IGR J16287$-$5021       & 16:28:27.36 & $-$50:22:42.9 & NTT+EFOSC2          & 3650-9300  & 5.5 & 31 May 2009, 01:41 &          1200 \\ 
IGR J16327$-$4940$^*$   & 16:32:39.95 & $-$49:42:13.8 & Radcliffe+Gr. Spec. & 3850-7200  & 2.3 & 10 Aug 2009, 20:44 & 2$\times$1800 \\ 
1RXS J165443.5$-$191620 & 16:54:43.74 & $-$19:16:31.1 & SPM 2.1m+B\&C Spec. & 3500-7800  & 4.0 & 21 Jun 2009, 06:13 & 2$\times$1800 \\ 
IGR J17009+3559         & 17:00:53.00 &   +35:59:56.2 & Copernicus+AFOSC    & 3500-7800  & 4.2 & 16 Oct 2009, 18:04 & 2$\times$1800 \\ 
IGR J18078+1123         & 18:07:49.91 &   +11:20:49.1 & SPM 2.1m+B\&C Spec. & 3500-7800  & 4.0 & 21 Jun 2009, 18:49 & 2$\times$1800 \\ 
IGR J18308+0928         & 18:30:50.64 &   +09:28:41.7 & Cassini+BFOSC       & 3500-8700  & 4.0 & 18 May 2009, 23:24 & 2$\times$1200 \\ 
IGR J18311$-$3337$^*$   & 18:31:14.75 & $-$33:36:08.5 & AAT+6dF             & 3900-7600  & 1.6 & 27 Jul 2003, 12:48 &      1200+600 \\ 
IGR J19077$-$3925       & 19:07:50.36 & $-$39:23:31.9 & AAT+6dF             & 3900-7600  & 1.6 & 26 Jun 2003, 15:25 &      1200+600 \\ 
IGR J19113+1533$^*$     & 19:11:18.83 &   +15:32:32.8 & SPM 2.1m+B\&C Spec. & 3500-7800  & 4.0 & 04 Dec 2008, 02:06 &          1200 \\ 
IGR J19118$-$1707$^*$   & 19:11:42.64 & $-$17:10:05.1 & AAT+6dF             & 3900-7600  & 1.6 & 21 Apr 2004, 18:34 &      1200+600 \\ 
1RXS J191928.5$-$295808 & 19:19:28.04 & $-$29:58:08.0 & AAT+6dF             & 3900-7600  & 1.6 & 10 Sep 2002, 10:10 &      1200+600 \\ 
IGR J19552+0044         & 19:55:12.47 &   +00:45:36.6 & Cassini+BFOSC       & 3500-8700  & 4.0 & 19 May 2009, 01:58 & 2$\times$1800 \\ 
1RXS J211336.1+542226   & 21:13:35.38$^\dagger$ & +54:22:32.8$^\dagger$ & SPM 2.1m+B\&C Spec. & 3500-7800  & 4.0 & 22 Jun 2009, 10:21 & 3$\times$1800 \\ 
1RXS J211928.4+333259   & 21:19:29.13 &   +33:32:57.0 & SPM 2.1m+B\&C Spec. & 3500-7800  & 4.0 & 04 Dec 2008, 03:12 & 2$\times$1800 \\ 
1RXS J213944.3+595016   & 21:39:45.1$\ddagger$ & +59:50:14$\ddagger$ & SPM 2.1m+B\&C Spec. & 3500-7800  & 4.0 & 24 Jun 2009, 10:44 & 2$\times$1800 \\ 
IGR J22292+6647         & 22:29:13.84 &   +66:46:51.5 & TNG+DOLoReS         & 3800-8000  & 2.5 & 13 Aug 2008, 01:27 &          1800 \\ 

\noalign{\smallskip}
\hline
\noalign{\smallskip}
\multicolumn{8}{l}{$^*$: tentative association (see text).}\\
\multicolumn{8}{l}{$^\dagger$: coordinates extracted from the USNO catalogs, having 
an accuracy of about 0$\farcs$2 (Deutsch 1999; Assafin et al. 2001; Monet et al. 2003).}\\
\multicolumn{8}{l}{$^\ddagger$: coordinates extracted from the DSS-II-Red frames, having an
accuracy of $\sim$1$''$.} \\
\noalign{\smallskip}
\hline
\hline
\noalign{\smallskip}
\end{tabular}
\end{center}
\end{table*}

\section{Optical spectroscopy}

Similarly to Papers VI and VII, the data presented in this work were 
collected in the course of a campaign that lasted one year and a half 
involving observations at the following telescopes:

\begin{itemize}
\item the 1.5m at the Cerro Tololo Interamerican Observatory (CTIO), Chile;
\item the 1.52m ``Cassini'' telescope of the Astronomical Observatory of 
Bologna, in Loiano, Italy; 
\item the 1.82m ``Copernicus'' telescope of the Astronomical Observatory of
Asiago, Italy;
\item the 1.9m ``Radcliffe'' telescope of the South African Astronomical 
Observatory (SAAO) in Sutherland, South Africa;
\item the 2.1m telescope of the Observatorio Astron\'omico Nacional in San 
Pedro Martir (SPM), M\'exico;
\item the 3.58m ``Telescopio Nazionale Galileo" (TNG) at the Roque de 
Los Muchachos Observatory in La Palma, Spain; 
\item the 3.58m ``New Technology Telescope'' (NTT) at the ESO-La Silla 
Observatory, Chile.
\end{itemize}

The spectroscopic data acquired at these telescopes were optimally 
extracted (Horne 1986) and reduced following standard procedures using 
IRAF\footnote{IRAF is the Image Reduction and Analysis Facility made 
available to the astronomical community by the National Optical Astronomy 
Observatories, which are operated by AURA, Inc., under contract with the 
U.S. National Science Foundation. It is available at {\tt 
http://iraf.noao.edu/}}.  Calibration frames (flat fields and bias) were 
taken on the day preceeding or following the observing night.  The 
wavelength calibration was performed using lamp data acquired soon after 
each on-target spectroscopic acquisiton; the uncertainty in this 
calibration was $\sim$0.5~\AA~in all cases according to our checks made 
using the positions of background night sky lines. Flux calibration was 
performed using catalogued spectrophotometric standards.

Additional spectra were retrieved from two different astronomical 
archives: the Sloan Digital Sky Survey\footnote{{\tt 
http://www.sdss.org/}} (SDSS, Adelman-McCarthy et al. 2007) archive, and 
the Six-degree Field Galaxy Survey\footnote{{\tt 
http://www.aao.gov.au/local/www/6df/}} (6dFGS) archive (Jones et al. 
2004). Since the 6dFGS archive provides spectra that are not 
flux-calibrated, we used the optical photometric information in Jones 
et al. (2005) to calibrate the 6dFGS spectra presented in this work.

We report in Table 1 the detailed log of all observations. We list in 
Col. 1 the names of the observed {\it INTEGRAL} sources. In Cols. 2 
and 3, we indicate the possible optical counterpart coordinates, extracted 
from the 2MASS catalog (with an accuracy of $\leq$0$\farcs$1, according to 
Skrutskie et al. 2006), from the USNO catalogs (with uncertainties of 
about 0$\farcs$2: Deutsch 1999; Assafin et al. 2001; Monet et al. 2003), 
or from the DSS-II-Red astrometry (which has a precision of $\sim$1$''$). 
In Col. 4, we report the telescope and the instrument used for the 
observations. The characteristics of each spectrograph are presented in 
Cols. 5 and 6. Column 7 provides the observation date and the UT time at 
mid-exposure, while Col. 8 reports the exposure times and the number of 
observations for each source.

To provide additional information about the possible counterpart to IGR 
J16287$-$5021, we also analyzed an optical $R$-band frame acquired with 
the NTT plus EFOSC2 on 31 May 2009 (start time: 01:27 UT; duration: 20 s) 
in seeing conditions of 1$\farcs$8; the 2$\times$2-rebinned CCD of EFOSC2 
provides a plate scale of 0$\farcs$24/pix and a useful field of 
4$\farcm$1$\times$4$\farcm$1.
This imaging frame was corrected for both bias and flat-field using 
standard procedures and was calibrated using nearby 
USNO-A2.0\footnote{Available at \\ {\tt 
http://archive.eso.org/skycat/servers/usnoa/}} stars. Simple aperture 
photometry, within the 
MIDAS\footnote{\texttt{http://www.eso.org/projects/esomidas}} package, was 
then used to evaluate the $R$-band magnitude of the possible optical 
counterpart to IGR J16287$-$5021.

\section{Results}

We describe the identification and classification criteria for 
the optical spectra of the 44 sources belonging to the sample considered 
in this work. The optical magnitudes quoted below, if not stated otherwise,
are extracted from the USNO-A2.0 catalog.

To determinate the reddening along the line of sight to the Galactic 
sources of our sample, when possible and applicable, we considered an 
intrinsic H$_\alpha$/H$_\beta$ line ratio of 2.86 (Osterbrock 1989) and 
inferred the corresponding color excess by comparing the intrinsic line 
ratio with the measured one by applying the Galactic extinction law of 
Cardelli et al. (1989).

To determine the distances of the compact Galactic X--ray sources of our 
sample, for CVs we assumed an absolute magnitude M$_V \sim$ 9 
and an intrinsic color index $(V-R)_0 \sim$ 0 mag (Warner 1995), whereas 
for high-mass X--ray binaries (HMXBs), when applicable, we used the 
intrinsic stellar color indices and absolute magnitudes reported in 
Lang (1992) and Wegner (1994). For the single low-mass X--ray binary 
(LMXB) in our sample, we considered $(V-R)_0$ $\sim$ 0 $\sim$ M$_R$ 
(e.g., van Paradijs \& McClintock 1995). Although these methods
basically provide an order-of-magnitude value for the distance of Galactic
sources, our past experience (Papers I-VII) tells us that these estimates
are in general correct to within 50\% of the refined value subsequently 
determined with more precise approaches.

For the emission-line AGN classification, we used the criteria of Veilleux 
\& Osterbrock (1987) and the line ratio diagnostics of both Ho et al. 
(1993, 1997) and Kauffmann et al. (2003); for the subclass assignation of 
Seyfert 1 nuclei, we used the H$_\beta$/[O {\sc iii}]$\lambda$5007 line 
flux ratio criterion described in Winkler (1992), and the criteria of 
Osterbrock \& Pogge (1985) for the classification of narrow-line Seyfert 1 
galaxies.

The spectra of the galaxies shown here were not corrected for starlight 
contamination (see, e.g., Ho et al. 1993, 1997) because of the limited S/N 
and spectral resolution. In this case, we also do not expect this 
to affect any of our main results and conclusions.

In the following, we consider a cosmology with $H_{\rm 0}$ = 65 km s$^{-1}$ 
Mpc$^{-1}$, $\Omega_{\Lambda}$ = 0.7, and $\Omega_{\rm m}$ = 0.3; the 
luminosity distances of the extragalactic objects reported in this paper 
were computed for these parameters using the Cosmology Calculator of 
Wright (2006). When not explicitly stated otherwise, for our X--ray flux 
estimates we assume a Crab-like spectrum except for the {\it XMM-Newton} 
sources, for which we considered the fluxes reported in Saxton et al. (2008).

In the following subsections, we consider the object identifications by 
dividing them into three broad classes (CVs, X--ray binaries, and AGNs) 
listed according to their increasing distance from Earth.

\subsection{CVs}

\begin{figure*}[th!]
\mbox{\psfig{file=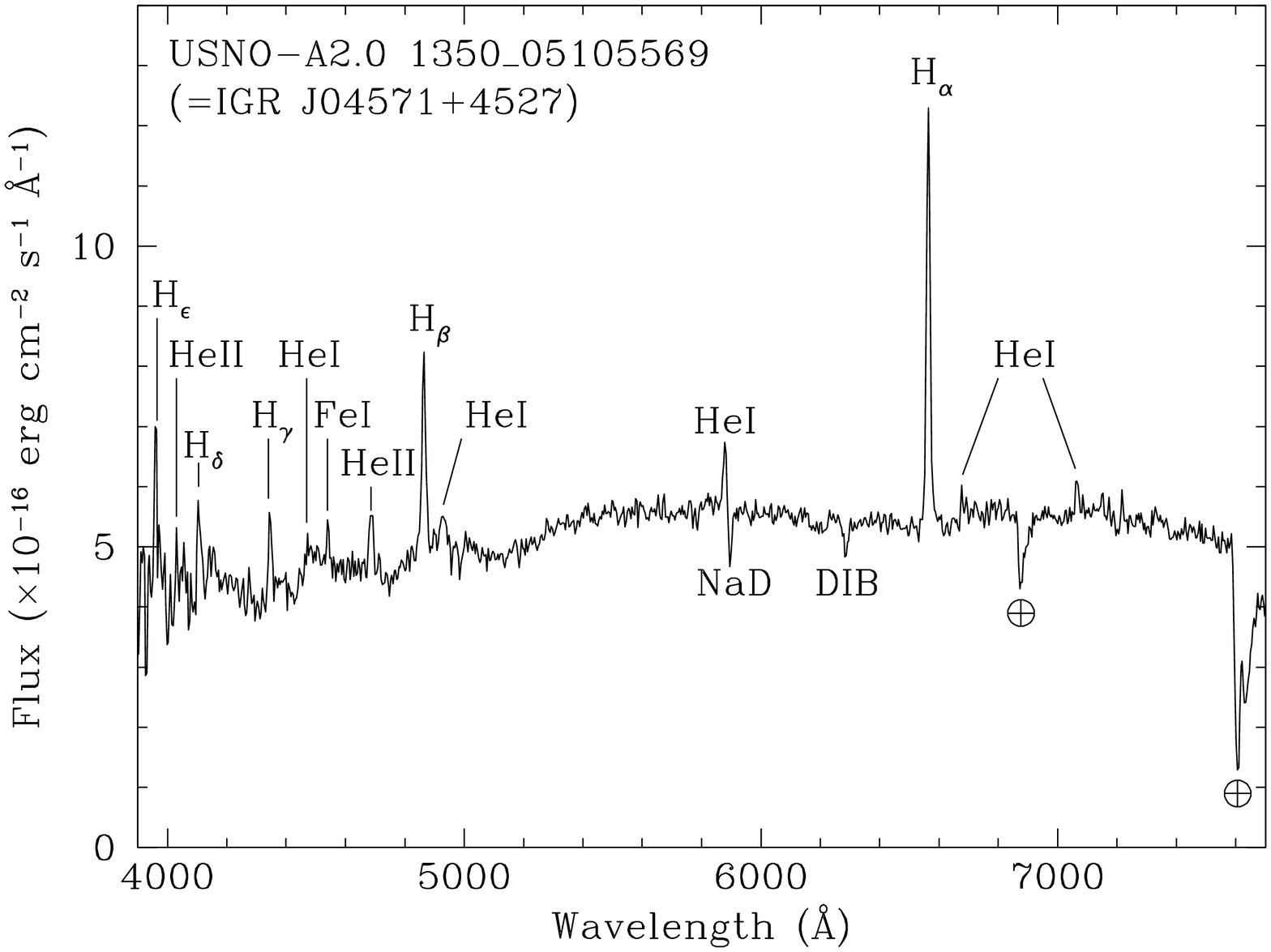,width=9cm,angle=270}}
\mbox{\psfig{file=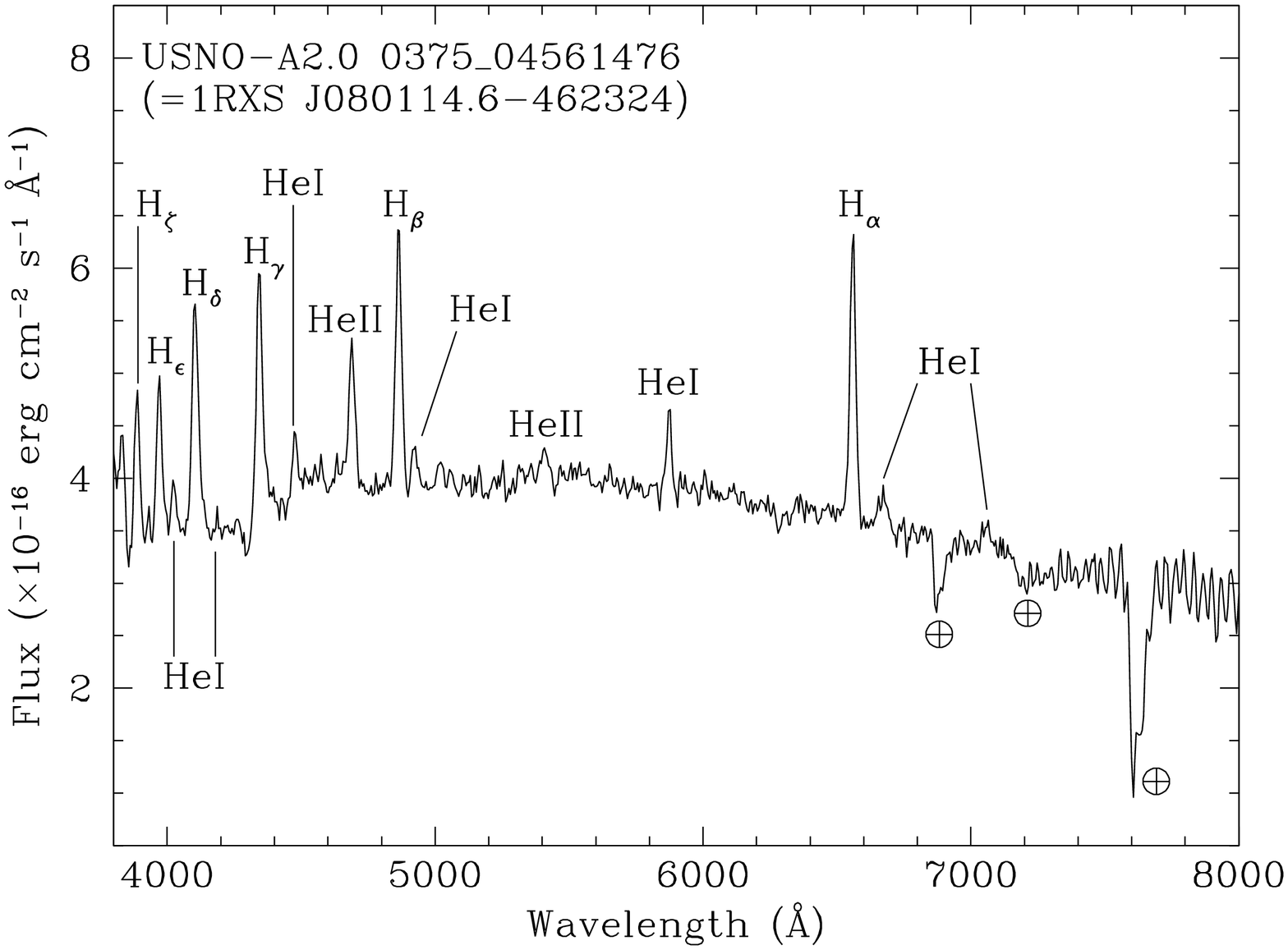,width=9cm,angle=270}}

\vspace{-.9cm}
\mbox{\psfig{file=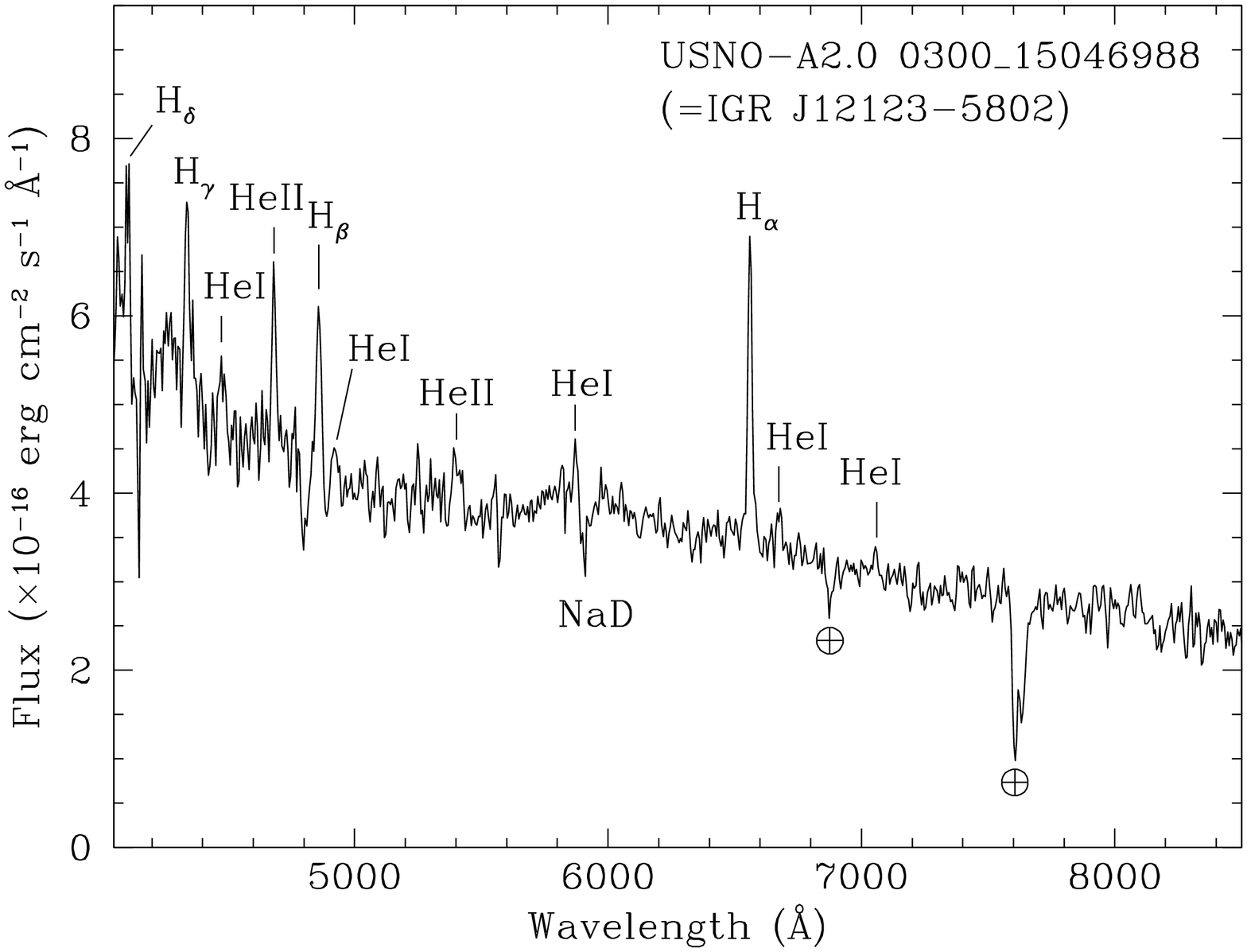,width=9cm,angle=270}}
\mbox{\psfig{file=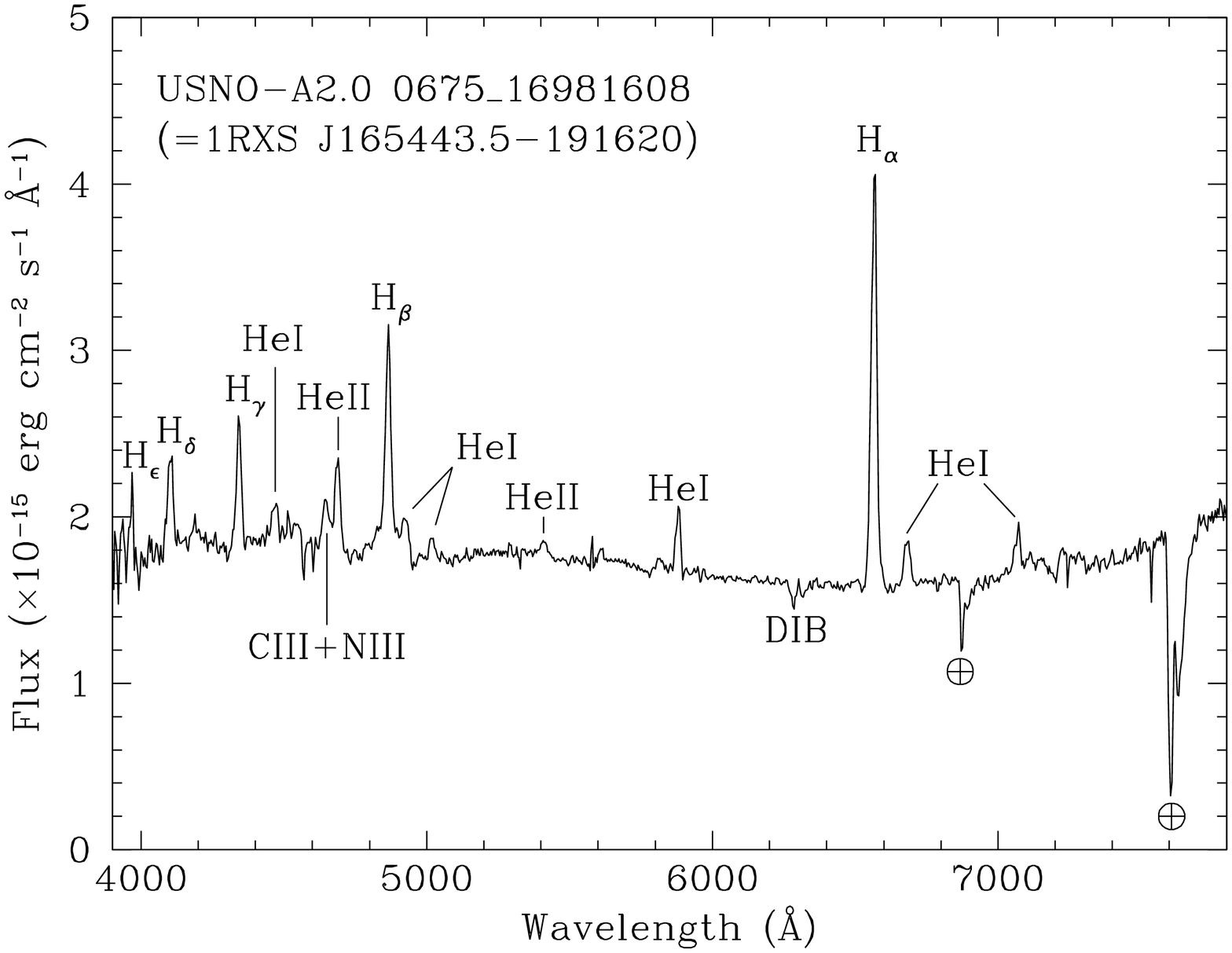,width=9cm,angle=270}}

\vspace{-.9cm}
\mbox{\psfig{file=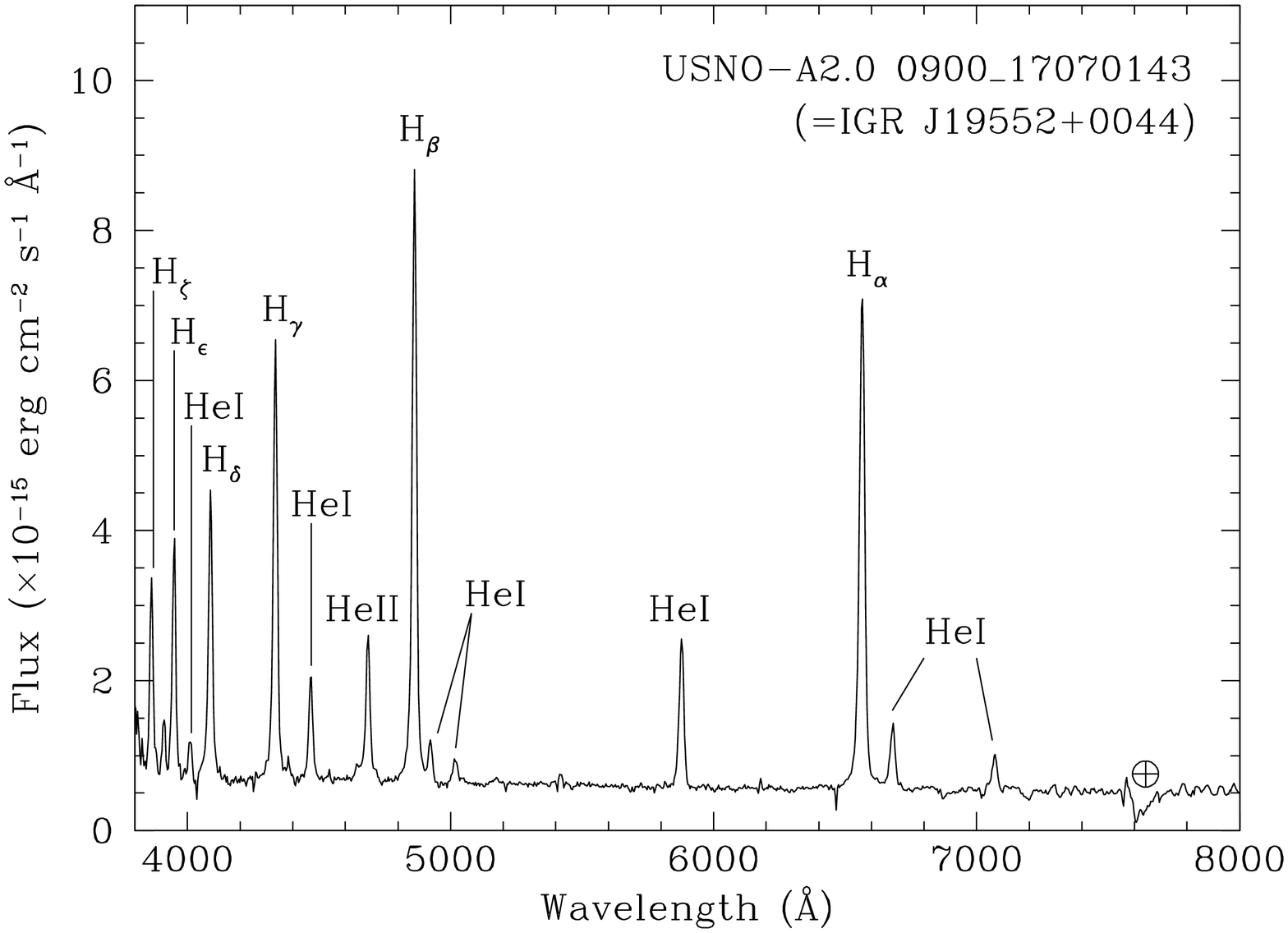,width=9cm,angle=270}}
\mbox{\psfig{file=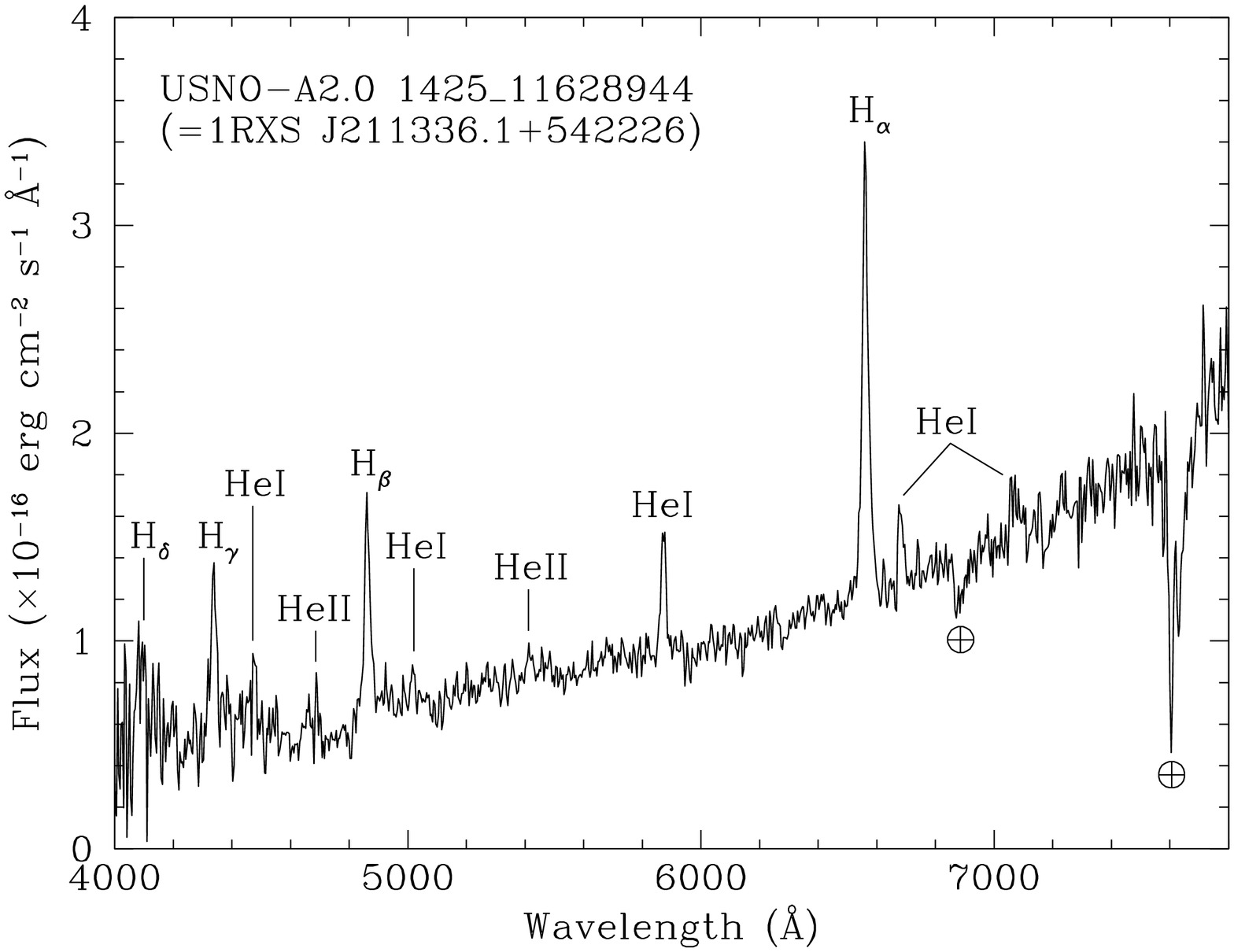,width=9cm,angle=270}}

\caption{Spectra (not corrected for the intervening Galactic absorption) 
of the optical counterparts of the 6 CVs belonging to the sample of {\it 
INTEGRAL} sources presented in this paper. For each spectrum, the main 
spectral features are labeled. The symbol $\oplus$ indicates atmospheric 
telluric absorption bands.}

\end{figure*}

We identify 6 objects of our sample (IGR J04571+4527, 1RXS 
J080114.6$-$462324, IGR J12123$-$5802, 1RXS J165443.5$-$191620, 
IGR J19552+0044, and 1RXS J211336.1+542226) as dwarf nova CVs because of 
the characteristics of their optical spectra (Fig. 6). All of them show 
Balmer emission lines up to at least H$_\delta$, as well as helium lines 
in emission. All of the detected lines are consistent with a redshift $z$ 
= 0, indicating that these sources are within our Galaxy. 

The main spectral diagnostic lines of these objects, as well as the main 
astrophysical parameters which can be inferred from the available optical 
and X--ray observational data, are reported in Table 2. The X--ray 
luminosities listed in this table for the various objects were computed 
using the fluxes reported in Voges et al. (1999, 2000), Bird et al. (2010), 
Cusumano et al. (2010), Krivonos et al. (2010), and Landi et al. (2010).

In the spectra of some of these sources (namely 1RXS
J080114.6$-$462324, IGR J12123$-$5802, and 1RXS J165443.5$-$191620), the 
He {\sc ii} $\lambda$4686 / H$_\beta$ equivalent width (EW) ratio is 
$\ga$0.5 and the EWs of these emission lines are around (or larger 
than) 10 \AA. This indicates that these sources are likely magnetic 
CVs belonging to the intermediate polar (IP) subclass (see Warner 1995 and 
references therein). To a lesser extent, a tentative IP classification
can also be made for IGR J19552+0044 and 1RXS J211336.1+542226
given the strength of their He {\sc ii} emission, although their 
He {\sc ii}/H$_\beta$ EW ratios are $<$0.5. The spectral characteristics 
of IGR J04571+4527 (namely the He {\sc ii}/H$_\beta$ EW ratio and 
the He {\sc ii} EW) instead imply that it is a non-magnetic CV.

These IP classifications however need confirmation by the measurement of 
both the orbital period and the spin period of the white dwarf (WD) 
harbored in these systems, as optical spectroscopy is sometimes 
insufficient to firmly establish the magnetic nature of CVs (see e.g.,
Pretorius 2009 and de Martino et al. 2010).

\begin{table*}[th!]
\caption[]{Synoptic table containing the main results concerning the 6 CVs
(see Fig. 6) identified in the present sample of {\it INTEGRAL} sources.}
\scriptsize
\vspace{-.3cm}
\begin{center}
\begin{tabular}{lcccccccccr}
\noalign{\smallskip}
\hline
\hline
\noalign{\smallskip}
\multicolumn{1}{c}{Object} & \multicolumn{2}{c}{H$_\alpha$} & 
\multicolumn{2}{c}{H$_\beta$} & \multicolumn{2}{c}{He {\sc ii} $\lambda$4686} & 
$R$ & $A_V$ & $d$ & \multicolumn{1}{c}{$L_{\rm X}$} \\
\cline{2-7}
\noalign{\smallskip} 
 & EW & Flux & EW & Flux & EW & Flux & mag & (mag) & (pc) & \\

\noalign{\smallskip}
\hline
\noalign{\smallskip}

IGR J04571+4527 & 19.4$\pm$0.6 & 10.3$\pm$0.3 & 9.5$\pm$0.5 & 4.8$\pm$0.2 & 3.6$\pm$0.4 & 1.67$\pm$0.17 & 
 17.5 & $\sim$0 & $\sim$500 & 0.10 (0.1--2.4) \\
 & & & & & & & & & & 53 (17--60) \\
 & & & & & & & & & & 67 (14--150) \\

& & & & & & & & & & \\ 

1RXS J080114.6$-$462324 & 19.9$\pm$0.6 & 7.2$\pm$0.2 & 16.7$\pm$0.8 & 6.5$\pm$0.3 & 9.7$\pm$0.7 & 3.8$\pm$0.3 & 
 14.8 & $\sim$0 & $\sim$150 & 0.064 (0.1--2.4) \\
 & & & & & & & & & & 1.1 (0.2--12) \\
 & & & & & & & & & & 0.43--1.4 (2--10) \\
 & & & & & & & & & & 1.6 (20--100) \\

& & & & & & & & & & \\ 

IGR J12123$-$5802 & 20.3$\pm$1.0 & 7.1$\pm$0.4 & 12.7$\pm$1.3 & 5.1$\pm$0.5 & 7.4$\pm$0.7 & 3.5$\pm$0.3 & 
 15.4 & $\sim$0 & $\sim$190 & 1.9 (2--10) \\
 & & & & & & & & & & 1.6 (20--40) \\ 
 & & & & & & & & & & $<$1.6 (40--100) \\ 

& & & & & & & & & \\ 

1RXS J165443.5$-$191620 & 38.7$\pm$1.2 & 60.9$\pm$1.8 & 18.5$\pm$0.9 & 33.2$\pm$1.7 & 7.5$\pm$0.5 & 13.4$\pm$0.9 & 
 15.6 & $\sim$0 & $\sim$210 & 0.17 (0.1--2.4) \\
 & & & & & & & & & & 7.7 (20--100) \\ 

& & & & & & & & & \\ 

IGR J19552+0044 & 271$\pm$8 & 164$\pm$5 & 264$\pm$8 & 168$\pm$5 & 39$\pm$2 & 60$\pm$3 & 
 16.0 & $\sim$0 & $\sim$250 & 0.66 (0.1--2.4) \\
 & & & & & & & & & & 4.2 (0.2--12) \\ 
 & & & & & & & & & & 6.2 (20--40) \\ 
 & & & & & & & & & & $<$4.9 (40--100) \\ 

& & & & & & & & & \\ 

1RXS J211336.1+542226
& 46$\pm$2 & 5.6$\pm$0.3 & 43$\pm$3 & 2.61$\pm$0.18 & 10$\pm$2 & 5.0$\pm$1.0 & 
18.8 & $\sim$0 & $\sim$910 & 1.3 (0.1--2.4) \\
 & & & & & & & & & & 99 (20--100) \\ 

\noalign{\smallskip} 
\hline
\noalign{\smallskip} 
\multicolumn{11}{l}{Note: EWs are expressed in \AA, line fluxes are
in units of 10$^{-15}$ erg cm$^{-2}$ s$^{-1}$, whereas X--ray luminosities
are in units of 10$^{31}$ erg s$^{-1}$ and the} \\
\multicolumn{11}{l}{reference band (between round brackets) is expressed 
in keV.} \\
\noalign{\smallskip} 
\hline
\hline
\noalign{\smallskip} 
\end{tabular} 
\end{center}
\end{table*}

\subsection{X--ray binaries}

\begin{figure*}[th!]
\mbox{\psfig{file=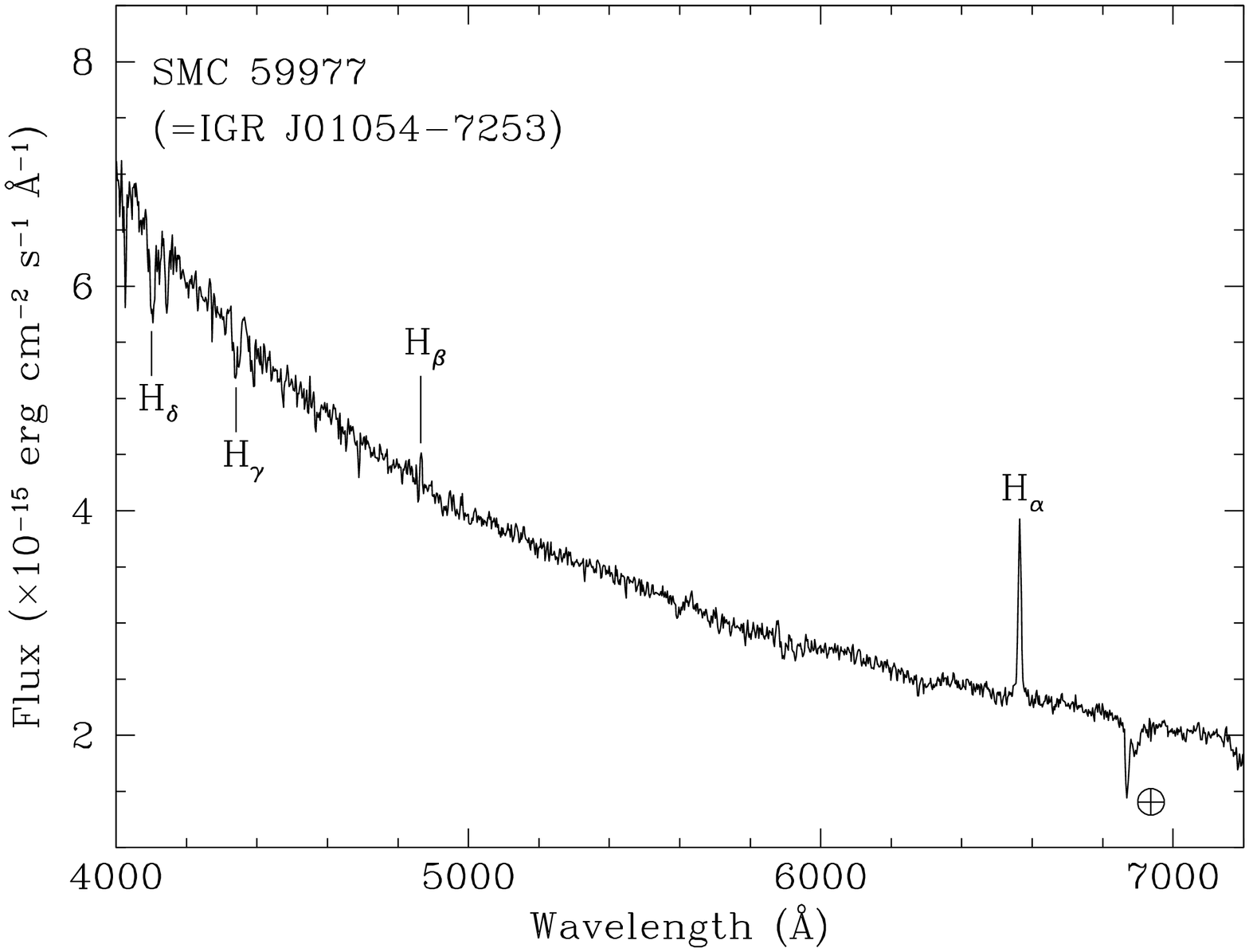,width=9cm,angle=270}}
\mbox{\psfig{file=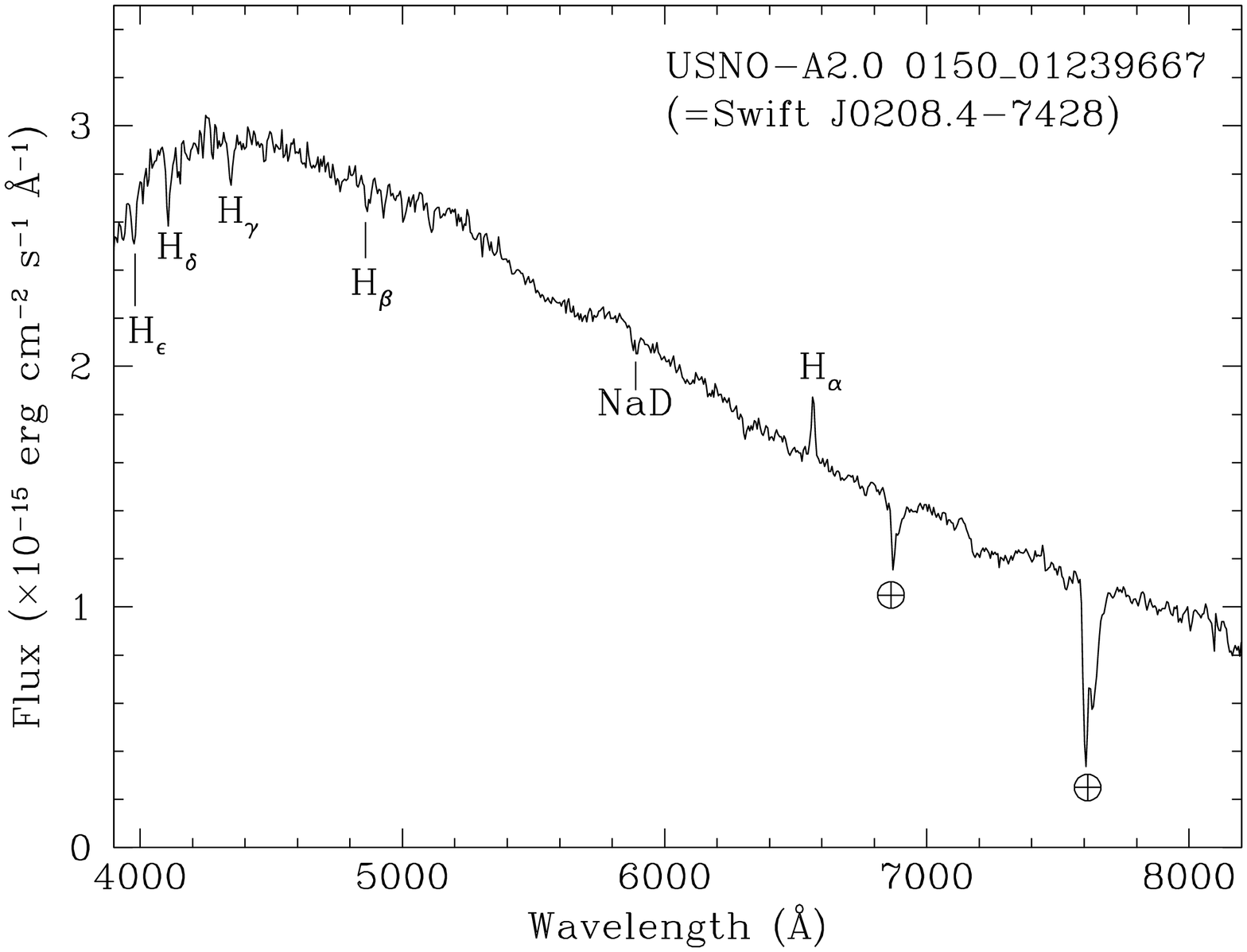,width=9cm,angle=270}}

\vspace{-.9cm}
\mbox{\psfig{file=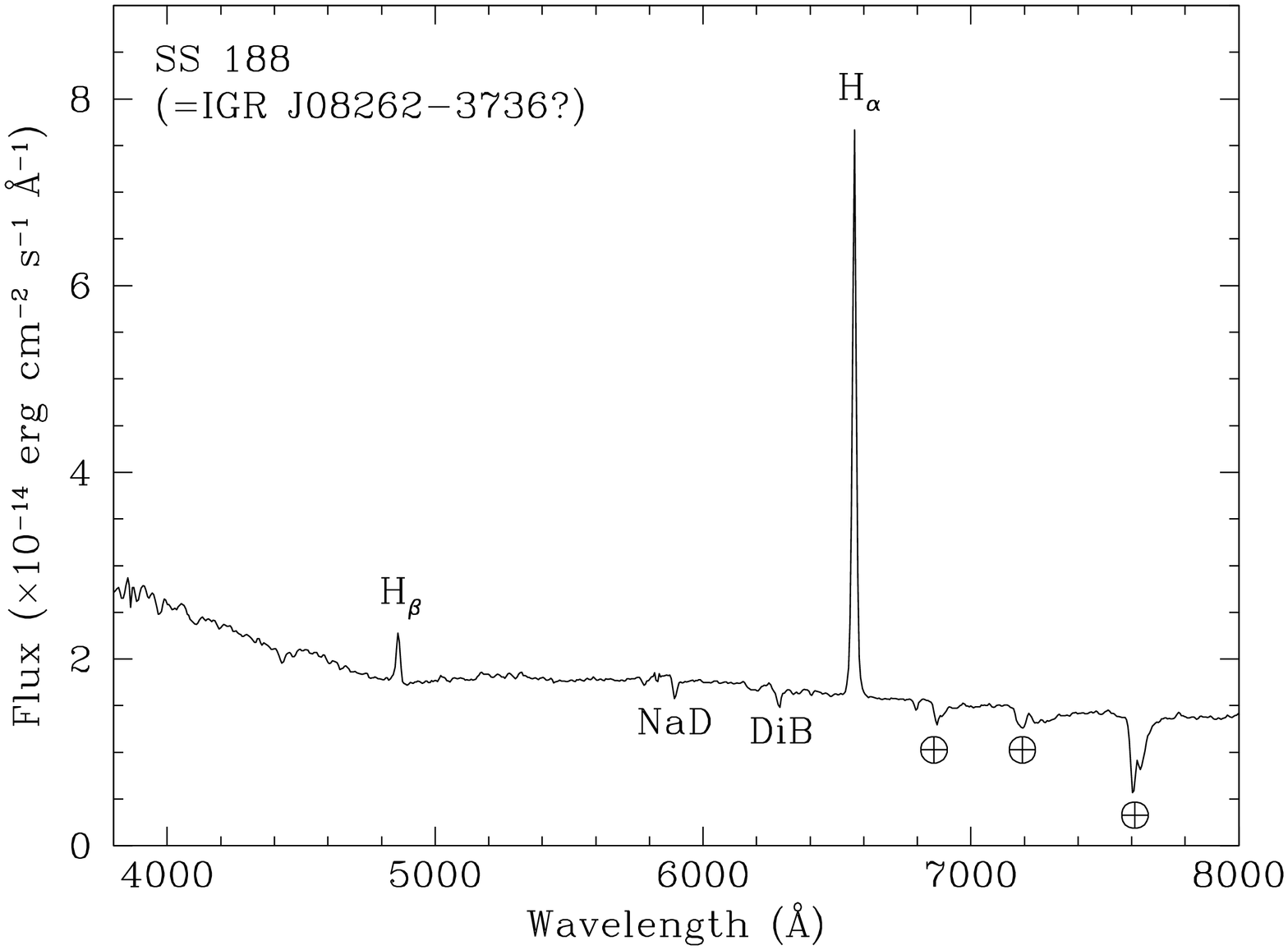,width=9cm,angle=270}}
\mbox{\psfig{file=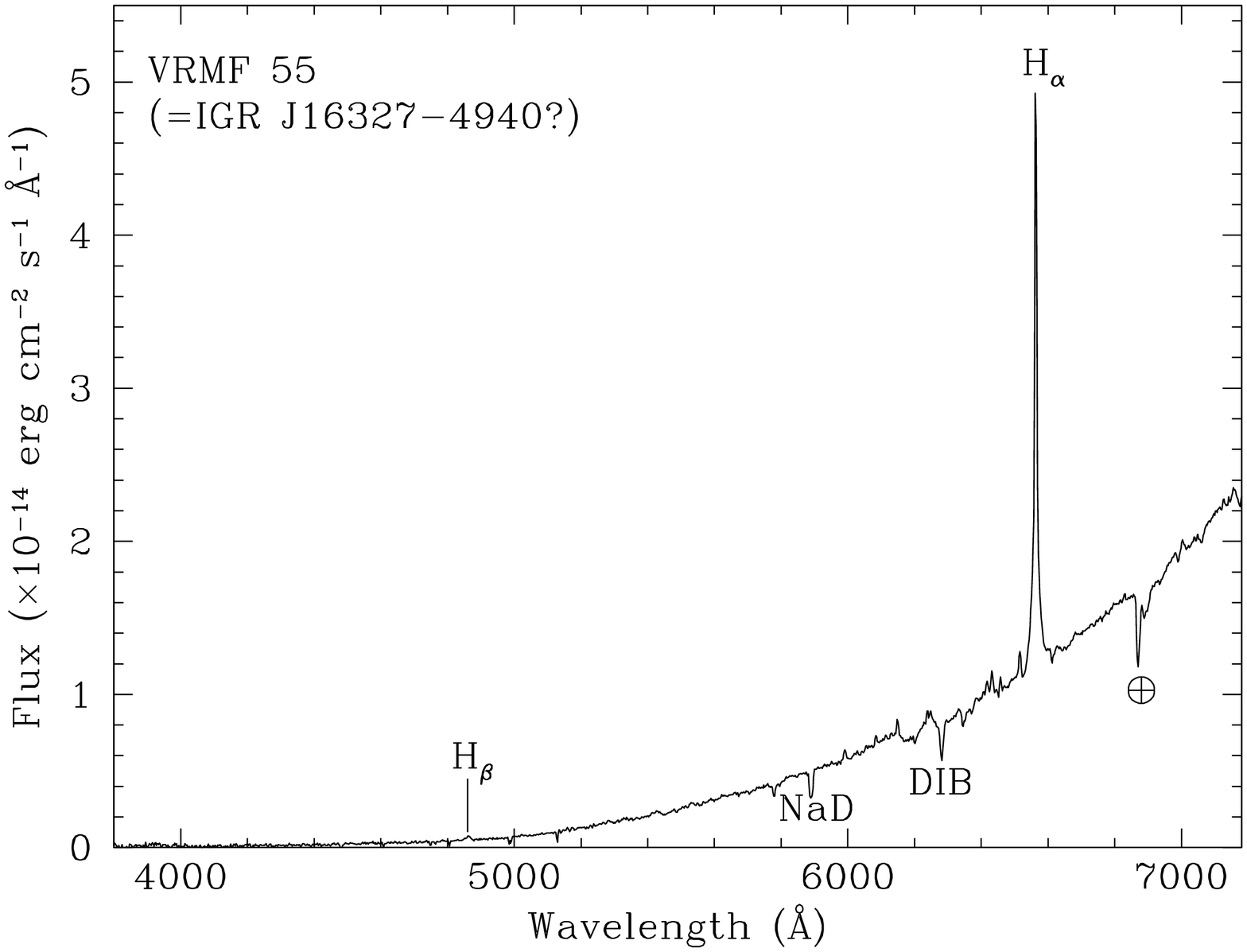,width=9cm,angle=270}}

\vspace{-.9cm}
\mbox{\psfig{file=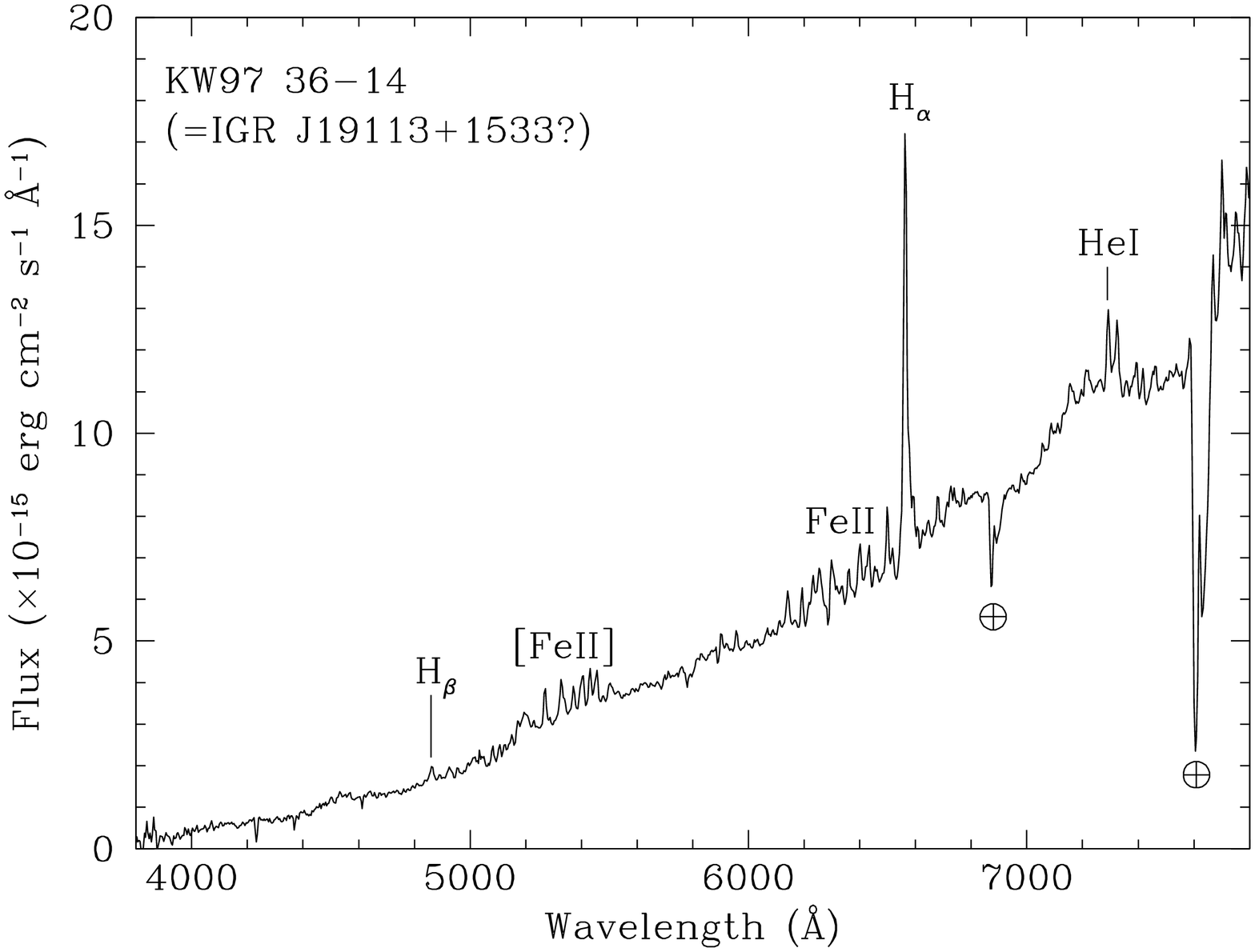,width=9cm,angle=270}}
\mbox{\psfig{file=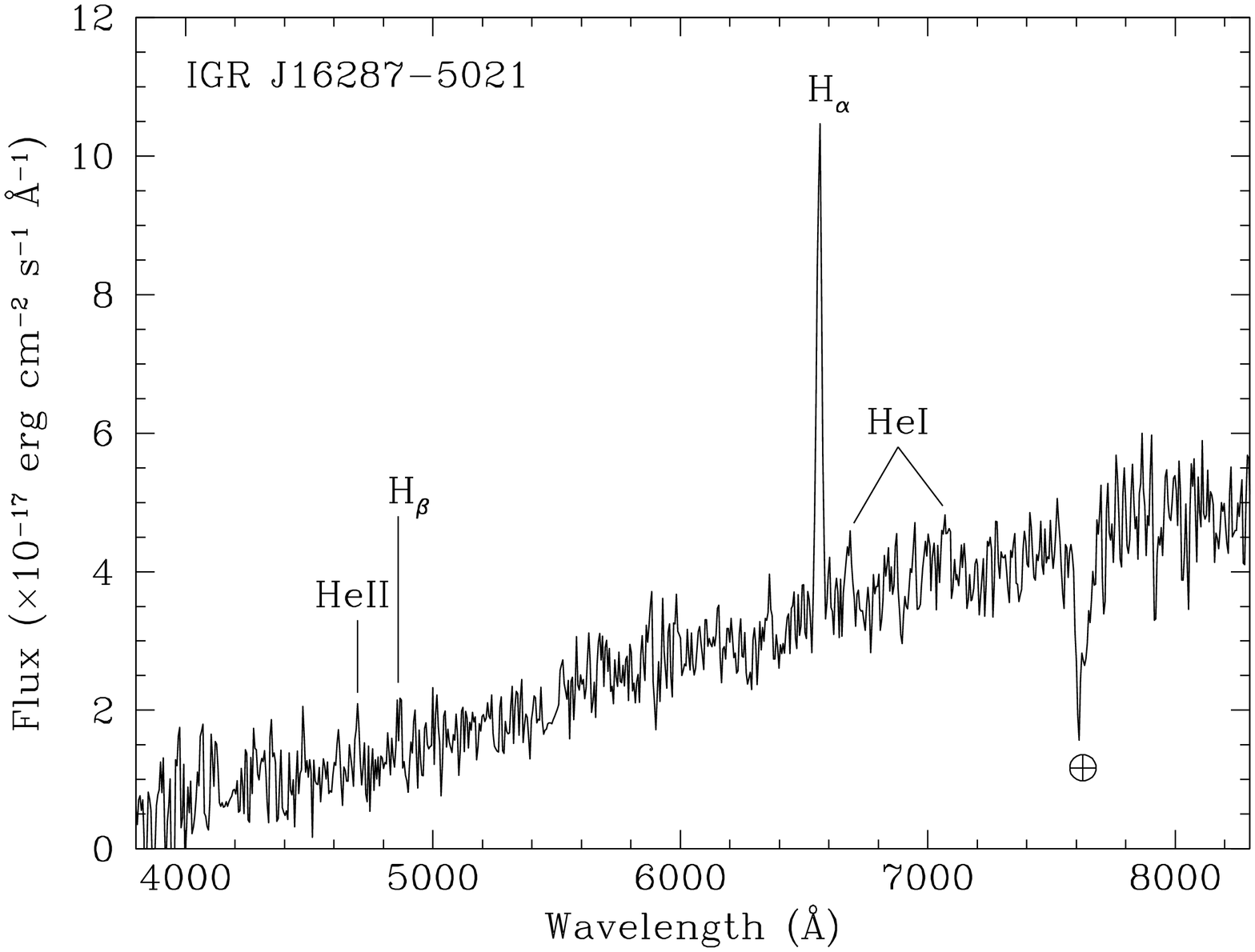,width=9cm,angle=270}}
\vspace{-.5cm}
\caption{Spectra (not corrected for the intervening Galactic absorption) 
of the optical counterparts of the 6 X--ray binaries belonging to
the sample of {\it INTEGRAL} sources presented in this paper.
For each spectrum, the main spectral features are labeled. The 
symbol $\oplus$ indicates atmospheric telluric absorption bands.}
\end{figure*}

\begin{table*}[th!]
\caption[]{Synoptic table containing the main results concerning the 6
X--ray binaries (see Fig. 7) identified in the present sample of {\it 
INTEGRAL} sources.}
\hspace{-1.2cm}
\scriptsize
\vspace{-.5cm}
\begin{tabular}{lccccccccccr}
\noalign{\smallskip}
\hline
\hline
\noalign{\smallskip}
\multicolumn{1}{c}{Object} & \multicolumn{2}{c}{H$_\alpha$} & 
\multicolumn{2}{c}{H$_\beta$} &
\multicolumn{2}{c}{He {\sc ii} $\lambda$4686} &
Optical & $A_V$ & $d$ & Spectral & \multicolumn{1}{c}{$L_{\rm X}$} \\
\cline{2-7}
\noalign{\smallskip} 
 & EW & Flux & EW & Flux & EW & Flux & mag. & (mag) & (kpc) & type & \\

\noalign{\smallskip}
\hline
\noalign{\smallskip}

IGR J01054$-$7253 & 7.15$\pm$0.15 & 16.8$\pm$0.3 & 0.47$\pm$0.12 & 2.0$\pm$0.5 & in abs. & in abs. &
 14.81$^{\rm a}$ ($V$) & $\sim$0.7 & 60$^{\rm b}$ & B0\,III & 0.13 (0.1--2.4) \\
 & & & & & & & & & & & 700$^{\rm c}$ (3--10) \\
 & & & & & & & & & & & 77 (20--100) \\

& & & & & & & & & & & \\ 

Swift J0208.4$-$7428 & 2.9$\pm$0.3 & 4.5$\pm$0.5 & in abs. & in abs. & $<$0.5 & $<$0.2 &
 14.75$^{\rm d}$ ($V$) & 0.18$^{\rm e}$ & 60$^{\rm b}$ & B1\,III & 19 (0.2--12) \\
 & & & & & & & & & & & 6.5 (3--10) \\
 & & & & & & & & & & & 108 (15--35) \\

& & & & & & & & & & & \\ 

IGR J08262$-$3736 & 65$\pm$2 & 1050$\pm$30 & 5.5$\pm$0.4 & 96$\pm$7 & in abs.? & in abs.? &
 12.9$^{\rm f}$ ($V$) & $\sim$3.3$^{\rm e}$ & $\sim$6.1 & OB V & 0.43 (20--100) \\

& & & & & & & & & & & \\ 

IGR J16327$-$4940 & 31.3$\pm$0.9 & 370$\pm$11 & 7.6$\pm$0.7 & 3.8$\pm$0.4 & in abs. & in abs. &
 15.5 ($R$) & 11.2 & $\approx$2 & OB giant & $<$0.0072 (20--40) \\
 & & & & & & & & & & & 0.018 (40--100) \\

& & & & & & & & & & & \\ 

IGR J19113+1533 & 22.5$\pm$1.1 & 159$\pm$8 & 3.6$\pm$0.7 & 5.9$\pm$1.2 & $<$1.2 & $<$1.5 &
 13.3 ($R$) & 7.0 & $\sim$9.1 & sgB[e] & $<$0.15 (20--40) \\
 & & & & & & & & & & & $<$0.28 (40--100) \\

\noalign{\smallskip} 
\hline
\noalign{\smallskip}

IGR J16287$-$5021 & 52$\pm$3 & 1.75$\pm$0.09 & 19$\pm$6 & 0.23$\pm$0.07 & 15$\pm$5 & 0.16$\pm$0.05 &
 19.5$^{\rm g}$ ($R$) & 3.1 & $<$19 & --- & $<$1.3 (0.2--12) \\
 & & & & & & & & & & & $<$2.9 (0.3--10) \\
 & & & & & & & & & & & $<$2.8 (2--10) \\
 & & & & & & & & & & & $<$3.8 (17--60) \\
 & & & & & & & & & & & $<$2.3 (20--40) \\
 & & & & & & & & & & & $<$1.2 (40--100) \\

\noalign{\smallskip} 
\hline
\noalign{\smallskip}
\multicolumn{12}{l}{Note: EWs are expressed in \AA, line fluxes are
in units of 10$^{-15}$ erg cm$^{-2}$ s$^{-1}$, whereas X--ray luminosities
are in units of 10$^{35}$ erg s$^{-1}$ and the reference} \\
\multicolumn{12}{l}{band (between round brackets) is expressed in keV.} \\
\multicolumn{12}{l}{$^{\rm a}$: from Massey (2002); $^{\rm b}$: from Harries et al. (2003);
$^{\rm c}$: from Coe et al. (2010); $^{\rm d}$: from Demers \& Irwin (1991);} \\
\multicolumn{12}{l}{$^{\rm e}$: only the Galactic reddening along the line of sight was 
assumed; $^{\rm f}$: from Pettersson (1987); $^{\rm g}$: this work.} \\
\noalign{\smallskip}
\hline
\hline
\end{tabular} 
\end{table*}

Six objects of our sample can be classified as X--ray binaries, given
their optical spectral shape and characteristics (see Fig. 7), that is, 
Balmer emission lines at redshift 0 superimposed on an intrinsically blue 
continuum (in some cases modified by interstellar reddening).

Table 3 collects the relevant optical spectral information about these 6
sources, along with their main parameters inferred from the available 
X--ray and optical data. X--ray luminosities in Table 3 were calculated 
using the fluxes in Saxton et al. (2008), Bozzo et al. (2009), 
Rodriguez et al. (2009), Tomsick et al. (2009), Bird et al. (2010), 
Krivonos et al. (2010), and McBride et al. (2010). In this table, we also 
report the results of our $R$-band photometry of the counterpart of IGR 
J16287$-$5021, described in Sect. 3, which infers a magnitude 
$R$ = 19.5$\pm$0.1.

Five of the sources (IGR J01054$-$7253, Swift J0208.4$-$7428, IGR 
J08262$-$3736, IGR J16327$-$4940, and IGR J19113+1533) can be classified 
as HMXBs due to their overall early-type star spectral appearance, which 
is typical of this class of objects (see e.g. Papers II-VII).

The spectra of the last three sources appear substantially reddened, 
which is indicative of interstellar dust along their lines of 
sight. This is quite common in X--ray binaries detected with {\it 
INTEGRAL} (e.g., Papers III-VII) and indicates that these objects 
are relatively far from Earth. The confirmation of reddening towards these 
sources comes from either their H$_\alpha$/H$_\beta$ line ratio (see Table 
3) or their optical colors. For IGR J08262$-$3736 and IGR J16327$-$4940, 
we find a reddening compatible with the Galactic one along the line of 
sight of the object (Schlegel et al. 1998).

The Balmer line ratio of KW97 36$-$14, the putative optical counterpart of 
IGR J19113+1533, instead indicates an absorption that is larger than the 
Galactic one, implying that there is additional absorbing material local 
to the source. The spectral appearance of this optical object resembles 
that of IGR J16318$-$4848 (Filliatre \& Chaty 2004), a heavily absorbed 
X--ray binary hosting a supergiant B[e] star, albeit the source of our 
sample is much less affected by reddening. Assuming these two sources to 
be similar, and the distance to IGR J16318$-$4848 as reported 
in Rahoui et al. (2008), the relevant parameters in Table 3 for IGR 
J19113+1533 were computed. We note that, if pointed soft X--ray 
observations confirm the association of IGR J19113+1533 with KW97 36$-$14, 
this would be the second supergiant B[e]/X--ray binary detected by {\it 
INTEGRAL} at hard X--rays.

For the first 4 HMXBs of Table 3, we obtained the constraints for 
distance, reddening, spectral type, and X--ray luminosity shown in the 
table by considering the absolute magnitudes of early-type stars and by 
applying the method described in Papers III and IV for the classification 
of this type of X--ray sources.
In particular, the optical counterparts of IGR J01054$-$7253 and Swift 
J0208.4$-$7428 are consistent with being blue giant stars at the distance 
of the SMC (60 kpc: see e.g. Harries et al. 2003) in terms of line 
redshift, magnitude, and optical colors. This result for IGR J01054$-$7253 
was independently confirmed by the X--ray timing analysis of Townsend et 
al. (2009): these authors indeed classify this source as a Be/X--ray 
binary on the basis of its orbital and neutron star spin periodicities.
For Swift J0208.4$-$7428, the use of our spectroscopy with the available 
optical photometry (Demers \& Irwin 1991) allowed us to 
state that this is a B-type star and thus improve the preliminary 
classification given to the putative optical counterpart of this source by 
McBride et al. (2010).
The two remaining cases (IGR J08262$-$3736 and IGR J16327$-$4940)
are instead broadly classified as OB stars due to the strength of their 
H$_\alpha$ emission lines, which is much larger than the typical 
values seen in blue supergiants (Leitherer 1988); the Galactic Arm
model of Leitch \& Vasisht (1998) is then used to infer a likely
distance of these stars and hence their luminosity class.
The lack of further detailed photometric optical information and
of higher-resolution spectroscopy does not allow us to refine 
our classification for the putative counterparts of these two
{\it INTEGRAL} sources.

The HMXB nature of IGR J16287$-$5021 can instead be excluded because its 
spectral shape differs from those of the HMXBs in the present 
sample (see Fig. 7, lower right panel); it resembles instead those 
of reddened LMXBs detected with {\it INTEGRAL} (see Papers V-VII).
The presence of Balmer and helium emission lines and no apparent 
absorption lines on top of a reddened (but intrinsically blue) continuum 
are indicative of an accretion disk dominating the total optical 
emission; the relative faintness of the optical/NIR counterpart also
indicates that the size of the system (and thus of the donor star) is much 
smaller than that of a HMXB. In addition, the reddening itself suggests 
that this source lies far from Earth.

This is confirmed by the X--ray spectrum of this object (Tomsick et al. 
2009), which also reveals absorption toward the source direction. A CV 
identification would instead make the relatively large optical and X--ray 
absorption mentioned above incompatible with the relatively small distance 
of the source ($\sim$500 pc) derived from its $R$-band magnitude. We thus 
conclude that this hard X--ray source is a LMXB. Despite this 
classification, we are unable to provide meaningful constraints on the 
distance to this source: the assumptions of Sect. 3 only allow us to 
determine a distance of 19 kpc (which can be considered as an upper limit 
given its large value; see Table 3).

We finally note that none of these systems are associated with a radio 
source. This implies that they are X--ray binaries that do not display
collimated (jet-like) outflows, that is, that no system is a microquasar.

\subsection{AGNs}

\begin{table*}
\caption[]{Synoptic table containing the main results concerning the 18
broad emission-line AGNs (Figs. 8-10) identified or observed in the present 
sample of {\it INTEGRAL} sources.}
\scriptsize
\begin{center}
\begin{tabular}{lcccccrcr}
\noalign{\smallskip}
\hline
\hline
\noalign{\smallskip}
\multicolumn{1}{c}{Object} & $F_{\rm H_\alpha}$ & $F_{\rm H_\beta}$ &
$F_{\rm [OIII]}$ & Class & $z$ &
\multicolumn{1}{c}{$D_L$ (Mpc)} & $E(B-V)_{\rm Gal}$ & \multicolumn{1}{c}{$L_{\rm X}$} \\
\noalign{\smallskip}
\hline
\noalign{\smallskip}

IGR J00158+5605 & * & 27.7$\pm$1.3 & 30.0$\pm$0.9 & Sy1.5 & 0.169 & 875.5 & 0.427 & 1.8 (0.1--2.4) \\
 & * & [88$\pm$4] & [94$\pm$3] & & & & & 6.1 (0.2--2) \\
 & & & & & & & & 35 (20--40) \\
 & & & & & & & & $<$26 (40--100) \\

 & & & & & & & & \\

IGR J02086$-$1742 & * & 120$\pm$12 & 38.5$\pm$1.9 & Sy1.2 & 0.129 & 651.8 & 0.028 & 1.1 (0.1--2.4) \\
 & * & [126$\pm$13] & [42$\pm$2] & & & & & 33 (2--10) \\
 & & & & & & & & 140 (20--100) \\

 & & & & & & & & \\

IGR J06058$-$2755 & * & 58$\pm$6 & 66$\pm$3 & Sy1.5 & 0.090 & 443.1 & 0.030 & 5.0 (0.1--2.4) \\
 & * & [64$\pm$6] & [72$\pm$4] & & & & & 30 (0.2--12) \\
 & & & & & & & & 36 (17--60) \\

 & & & & & & & & \\

Swift J0845.0$-$3531 & * & 26.3$\pm$1.3 & 5.4$\pm$0.3 & Sy1.2 & 0.137 & 695.7 & 0.531 & 5.9 (0.1--2.4) \\
 & * & [112$\pm$6] & [22.9$\pm$1.6] & & & & & 78 (14--150) \\
 & & & & & & & & 50 (17--60) \\

 & & & & & & & & \\

IGR J09094+2735 & * & 4.7$\pm$0.3 & 1.36$\pm$0.04 & NLSy1 & 0.2844 & 1572.4 & 0.030 & 6.7 (0.1--2.4) \\
 & * & [4.8$\pm$0.3] & [1.51$\pm$0.05] & & & & & 920 (17--60) \\

 & & & & & & & & \\

PKS 1143$-$696 & * & 48$\pm$2 & 1.0$\pm$0.1 & Sy1.2 & 0.244 & 1320.2 & 0.385 & 23 (0.1--2.4) \\
 & * & [134$\pm$7] & [2.4$\pm$0.2] & & & & & 120 (2--10) \\
 & & & & & & & & 230 (17--60) \\
 & & & & & & & & 280 (20--100) \\

 & & & & & & & & \\

IGR J12107+3822 & * & 109$\pm$7 & 64.2$\pm$1.9 & Sy1.5 & 0.0229 & 107.5 & 0.021 & 0.31 (0.2--12) \\
 & * & [115$\pm$7] & [68$\pm$2] & & & & & 1.8 (14--150) \\
 & & & & & & & & 1.6 (17--60) \\

 & & & & & & & & \\

IGR J1248.2$-$5828 & * & 1.1$\pm$0.3 & 7.8$\pm$0.5 & Sy1.9 & 0.028 & 131.9 & 0.624 & 0.85 (2--10) \\
 & * & [5.8$\pm$1.6] & [54$\pm$4] & & & & & 1.6 (17--60) \\
 & & & & & & & & 2.1 (20--100) \\

 & & & & & & & & \\

IGR J13168$-$7157 & * & 67$\pm$10 & 47.1$\pm$1.4 & Sy1.5 & 0.070 & 339.9 & 0.255 & 0.60 (0.1--2.4) \\
 & * & [160$\pm$20] & [103$\pm$3] & & & & & 11 (0.2--12) \\
 & & & & & & & & 13 (17--60) \\

 & & & & & & & & \\

IGR J13187+0322 & --- & 17.2$\pm$1.7 & 1.5$\pm$0.3 & Type 1 & 0.6058 & 3846.2 & 0.031 & $<$540 (20--40) \\
 & --- & [17.2$\pm$1.7] & [1.5$\pm$0.3] & QSO & & & & $<$1300 (40--100) \\

 & & & & & & & & \\

Swift J1513.8$-$8125 & * & 103$\pm$5 & 37.3$\pm$1.1 & Sy1.2 & 0.069 & 334.8 & 0.272 & 20 (14--150) \\
 & * & [223$\pm$11] & [83$\pm$2] & & & & & 34 (14--195) \\
 & & & & & & & & 15 (17--60) \\

 & & & & & & & & \\

IGR J15311$-$3737 & --- & 48$\pm$2 & 6.9$\pm$0.3 & Sy1 & 0.127 & 640.8 & 0.320 & 2.3 (0.1--2.4) \\
 & --- & [120$\pm$4] & [16.6$\pm$0.5] & & & & & 12--16 (0.2--12) \\
 & & & & & & & & 44 (20--100) \\

 & & & & & & & & \\

IGR J18078+1123 & * & 64$\pm$4 & 12.6$\pm$0.6 & Sy1/1.2 & 0.078 & 380.9 & 0.131 & 1.2 (0.1--2.4) \\
 & * & [90$\pm$6] & [18.6$\pm$0.9] & & & & & 51 (17--60) \\

 & & & & & & & & \\

IGR J19077$-$3925 & * & 4.8$\pm$0.7 & 20.1$\pm$1.2 & Sy1.9 & 0.0760 & 370.6 & 0.107 & 1.3 (0.1--2.4) \\
 & * & [6.2$\pm$0.9] & [28.0$\pm$1.4] & & & & & 24 (20--100) \\

 & & & & & & & & \\

1RXS J191928.5$-$295808 & * & 42$\pm$8 & 122$\pm$6 & Sy1.5/1.8 & 0.1669 & 863.5 & 0.139 & 12 (0.1--2.4) \\
 & * & [56$\pm$11] & [183$\pm$9] & & & & & 120 (17--60) \\
 & & & & & & & & 67 (20--40) \\
 & & & & & & & & $<$42 (40--100) \\

 & & & & & & & & \\

1RXS J211928.4+333259 & * & 10$\pm$2 & 15.2$\pm$0.8 & Sy1.5 & 0.051 & 244.4 & 0.217 & 0.44 (0.1--2.4) \\
 & * & [18$\pm$3] & [29.6$\pm$1.5] & & & & & 6.8 (0.2--12) \\
 & & & & & & & & 5.6 (14--150) \\
 & & & & & & & & 11 (14--195) \\
 & & & & & & & & 9.1 (17--60) \\
 & & & & & & & & 11 (20--100) \\

 & & & & & & & & \\

1RXS J213944.3+595016 & * & 3.0$\pm$0.2 & 2.01$\pm$0.10 & Sy1.5 & 0.114 & 570.4 & 1.271 & 1.7 (0.1--2.4) \\
 & * & [112$\pm$8] & [72$\pm$4] & & & & & 23 (2--10) \\
 & & & & & & & & 14 (14--150) \\
 & & & & & & & & 40 (20--100) \\

 & & & & & & & & \\

IGR J22292+6647 & * & 2.25$\pm$0.11 & 1.24$\pm$0.06 & Sy1.5 & 0.112 & 559.6 & 1.089 & 0.63 (0.1--2.4) \\
 & * & [52$\pm$3] & [26.3$\pm$1.3] & & & & & 12 (0.2--12) \\
 & & & & & & & & 25 (2--10) \\
 & & & & & & & & 55 (14--150) \\
 & & & & & & & & 100 (14--195) \\
 & & & & & & & & 42 (17--60) \\
 & & & & & & & & 42 (20--100) \\

\noalign{\smallskip} 
\hline
\noalign{\smallskip} 
\multicolumn{9}{l}{Note: emission-line fluxes are reported both as 
observed and (between square brackets) corrected for the intervening Galactic} \\ 
\multicolumn{9}{l}{absorption $E(B-V)_{\rm Gal}$ along the object line of sight 
(from Schlegel et al. 1998). Line fluxes are in units of 10$^{-15}$ erg cm$^{-2}$ s$^{-1}$,} \\
\multicolumn{9}{l}{whereas X--ray luminosities are in units of 10$^{43}$ erg s$^{-1}$ 
and the reference band (between round brackets) is expressed in keV.} \\ 
\multicolumn{9}{l}{The typical error in the redshift measurement is $\pm$0.001 
but for the SDSS and 6dFGS spectra, for which an uncertainty} \\
\multicolumn{9}{l}{of $\pm$0.0003 can be assumed.} \\
\multicolumn{9}{l}{$^*$: heavily blended with [N {\sc ii}] lines} \\
\noalign{\smallskip} 
\hline
\hline
\end{tabular} 
\end{center} 
\end{table*}

\begin{table*}[t!]
\caption[]{Synoptic table containing the main results concerning the 13
narrow emission-line AGNs and of the single XBONG (see Figs. 11-13) identified 
or observed in the present sample of {\it INTEGRAL} sources.}
\scriptsize
\begin{center}
\begin{tabular}{lcccccrccr}
\noalign{\smallskip}
\hline
\hline
\noalign{\smallskip}
\multicolumn{1}{c}{Object} & $F_{\rm H_\alpha}$ & $F_{\rm H_\beta}$ & $F_{\rm [OIII]}$ & Class & $z$ &
\multicolumn{1}{c}{$D_L$} & \multicolumn{2}{c}{$E(B-V)$} &
\multicolumn{1}{c}{$L_{\rm X}$} \\
\cline{8-9}
\noalign{\smallskip}
 & & & & & & (Mpc) & Gal. & AGN & \\
\noalign{\smallskip}
\hline
\noalign{\smallskip}

IGR J00465$-$4005 & 1.8$\pm$0.4 & 0.17$\pm$0.05 & 3.63$\pm$0.18 & Sy2 & 0.201 & 1061.3 & 0.011 & 1.26 & 16 (2--10) \\
 & [2.1$\pm$0.6] & [0.21$\pm$0.07] & [3.71$\pm$0.19] & & & & & & 480 (20--100) \\

 & & & & & & & & & \\

IGR J01545+6437 & 27$\pm$3 & 2.6$\pm$0.3 & 17.0$\pm$0.9 & Sy2 & 0.034 & 160.9 & 0.866 & 0.38 & 1.4 (20--40) \\
 & [170$\pm$20] & [41$\pm$5] & [240$\pm$12] & & & & & & $<$1.2 (40--100) \\

 & & & & & & & & & \\

IGR J03344+1506 & 42$\pm$2 & 5.1$\pm$0.5 & 20.7$\pm$1.0 & Sy2 & 0.021 & 98.4 & 0.316 & 0.81 & 0.0095 (0.1--2.4) \\
 & [87$\pm$4] & [14.0$\pm$1.4] & [54$\pm$3] & & & & & & $<$0.79 (20--40) \\
 & & & & & & & & & 2.3 (40--100) \\

 & & & & & & & & & \\

IGR J04451$-$0445 & 5.2$\pm$0.5 & $<$0.5 & $<$0.7 & likely Sy2 & 0.076 & 370.6 & 0.048 & $>$1.05 & 47 (20--100) \\
 & [5.6$\pm$0.3] & [$<$0.7] & [$<$0.9] & & & & & & \\

 & & & & & & & & & \\

IGR J05253+6447 & 4.5$\pm$0.5 & in abs. & 6.0$\pm$0.3 & likely Sy2 & 0.071 & 345.0 & 0.190 & --- & 7.7 (0.2--2) \\
 & [6.7$\pm$0.7] & $''$ & [11.1$\pm$0.6] & & & & & & $<$8.6 (20--40) \\
 & & & & & & & & & 19 (40--100) \\

 & & & & & & & & & \\

IGR J06293$-$1359 & 34.8$\pm$1.7 & 3.3$\pm$0.2 & 19.0$\pm$0.6 & Sy2 & 0.033 & 156.0 & 0.323 & 1.28 & 4.8 (20--40) \\
 & [73$\pm$4] & [7.4$\pm$0.5] & [51.7$\pm$1.6] & & & & & & $<$4.1 (40--100) \\

 & & & & & & & & & \\

IGR J08557+6420 & 6.2$\pm$0.4 & in abs. & 8.8$\pm$0.4 & likely Sy2 & 0.037 & 175.5 & 0.103 & --- & 7.4 (17--60) \\
 & [7.9$\pm$0.6] & $''$ & [12.3$\pm$0.6] & & & & & & \\

 & & & & & & & & & \\

MCG +04$-$26$-$006 & 21$\pm$2 & 8.2$\pm$0.6 & 15.4$\pm$0.8 & LINER & 0.020 & 93.7 & 0.030 & 0 & 0.23 (2--10) \\
 & [22$\pm$2] & [8.6$\pm$0.6] & [17.0$\pm$0.9] & & & & & & 1.4 (14--150) \\
 & & & & & & & & & 3.3 (20--100) \\

 & & & & & & & & & \\

IGR J14301$-$4158 & 22.5$\pm$1.8 & 5.7$\pm$0.9 & 39.6$\pm$1.2 & Sy2 & 0.0387 & 183.8 & 0.117 & 0.29 & 3.7 (20--100) \\
 & [29$\pm$2] & [7.6$\pm$1.2] & [56.1$\pm$1.7] & & & & & & \\

 & & & & & & & & & \\

IGR J15549$-$3740 & 67.7$\pm$1.8 & 7.93$\pm$0.12 & 46$\pm$2 & Sy2 & 0.0194 & 90.8 & 0.475 & 0.67 & 1.7 (20--100) \\
 & [200$\pm$6] & [36$\pm$5] & [195$\pm$10] & & & & & & \\

 & & & & & & & & & \\

IGR J17009+3559 & --- & in abs. & $<$0.33 & XBONG & 0.113 & 565.0 & 0.018 & --- & 8.0 (1--10) \\
 & --- & $''$ & [$<$0.34] & & & & & & 52 (17--60) \\

 & & & & & & & & & \\

IGR J18308+0928 & 13.1$\pm$1.3 & $<$2 & 16.0$\pm$1.6 & Sy2 & 0.019 & 88.9 & 0.243 & $>$0.83 & 1.5 (14--150) \\
 & [26$\pm$3] & $<$4 & [33$\pm$3] & & & & & & 3.7 (14--195) \\
 & & & & & & & & & 0.081 (17--60) \\

 & & & & & & & & & \\

IGR J18311$-$3337 & 40.2$\pm$1.8 & 6.0$\pm$0.6 & 82$\pm$2 & Sy2 & 0.0687 & 333.3 & 0.185 & 0.94 & 22 (20--100) \\
 & [60$\pm$3] & [8.4$\pm$0.8] & [141$\pm$4] & & & & & & \\

 & & & & & & & & & \\

IGR J19118$-$1707 & 132$\pm$4 & 32$\pm$2 & 83$\pm$4 & LINER & 0.0234 & 109.9 & 0.137 & 0.25 & 0.87 (20--40) \\
 & [179$\pm$5] & 49$\pm$3 & [126$\pm$6] & & & & & & $<$0.82 (40--100) \\

\noalign{\smallskip} 
\hline
\noalign{\smallskip} 
\multicolumn{10}{l}{Note: emission-line fluxes are reported both as 
observed and (between square brackets) corrected for the intervening Galactic} \\ 
\multicolumn{10}{l}{absorption $E(B-V)_{\rm Gal}$ along the object line of sight 
(from Schlegel et al. 1998). Line fluxes are in units of 10$^{-15}$ erg cm$^{-2}$ s$^{-1}$,} \\
\multicolumn{10}{l}{whereas X--ray luminosities are in units of 10$^{43}$ erg s$^{-1}$ 
and the reference band (between round brackets) is expressed in keV.} \\ 
\multicolumn{10}{l}{The typical error in the redshift measurement is $\pm$0.001 
but for the SDSS and 6dFGS spectra, for which an uncertainty} \\
\multicolumn{10}{l}{of $\pm$0.0003 can be assumed.} \\
\noalign{\smallskip} 
\hline
\hline
\end{tabular}
\end{center}
\end{table*}

\begin{figure*}[th!]
\mbox{\psfig{file=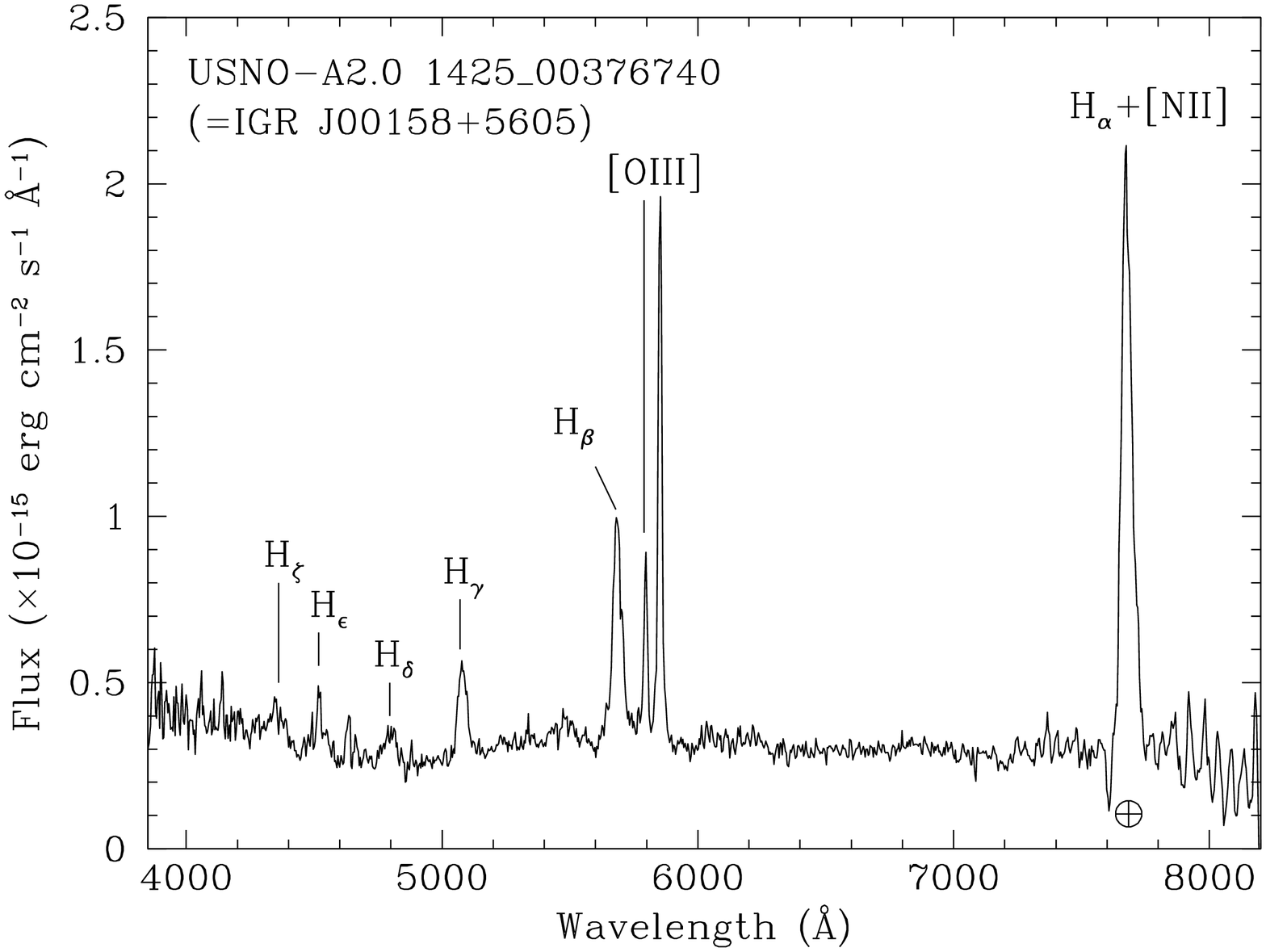,width=9cm,angle=270}}
\mbox{\psfig{file=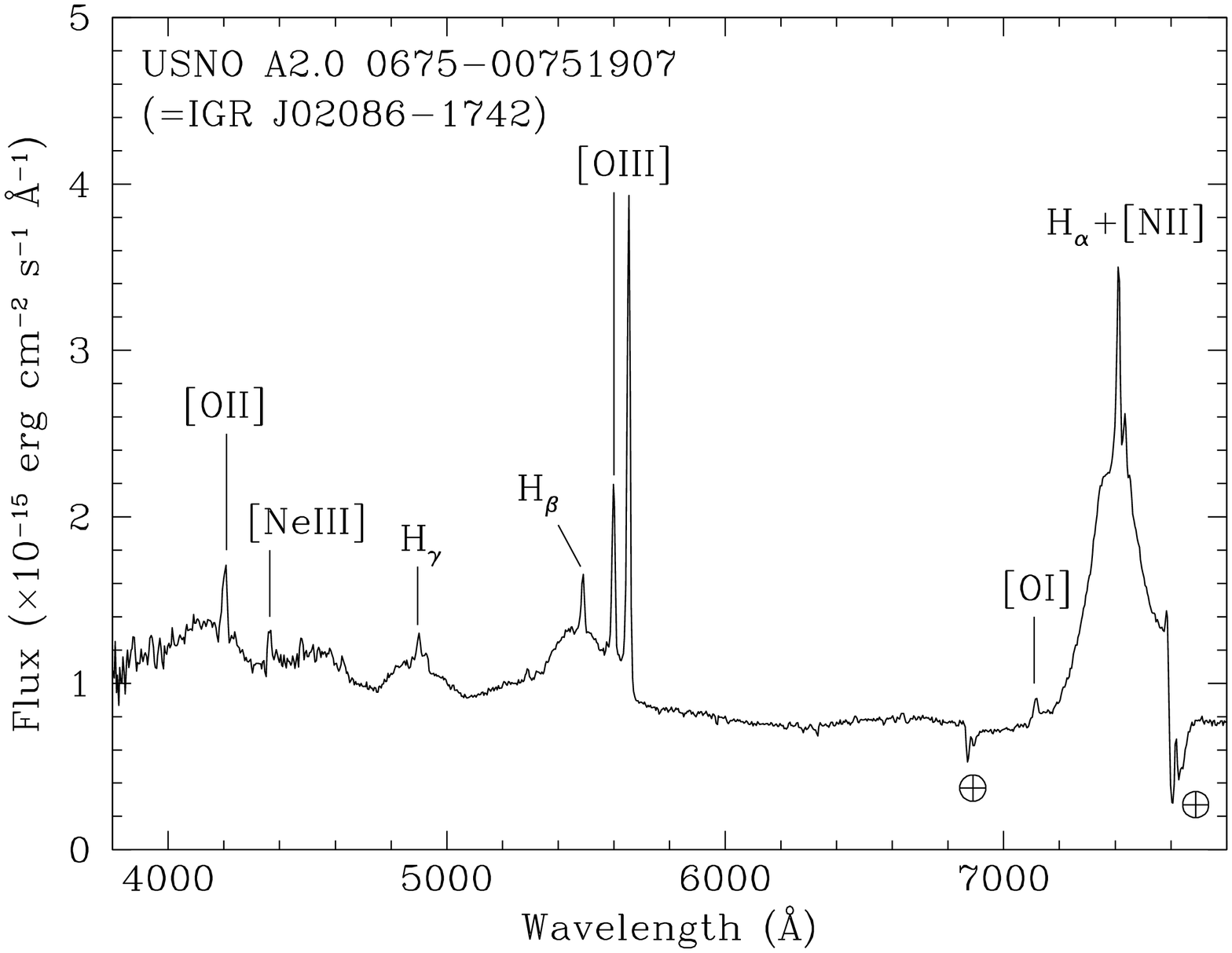,width=9cm,angle=270}}

\vspace{-.9cm}
\mbox{\psfig{file=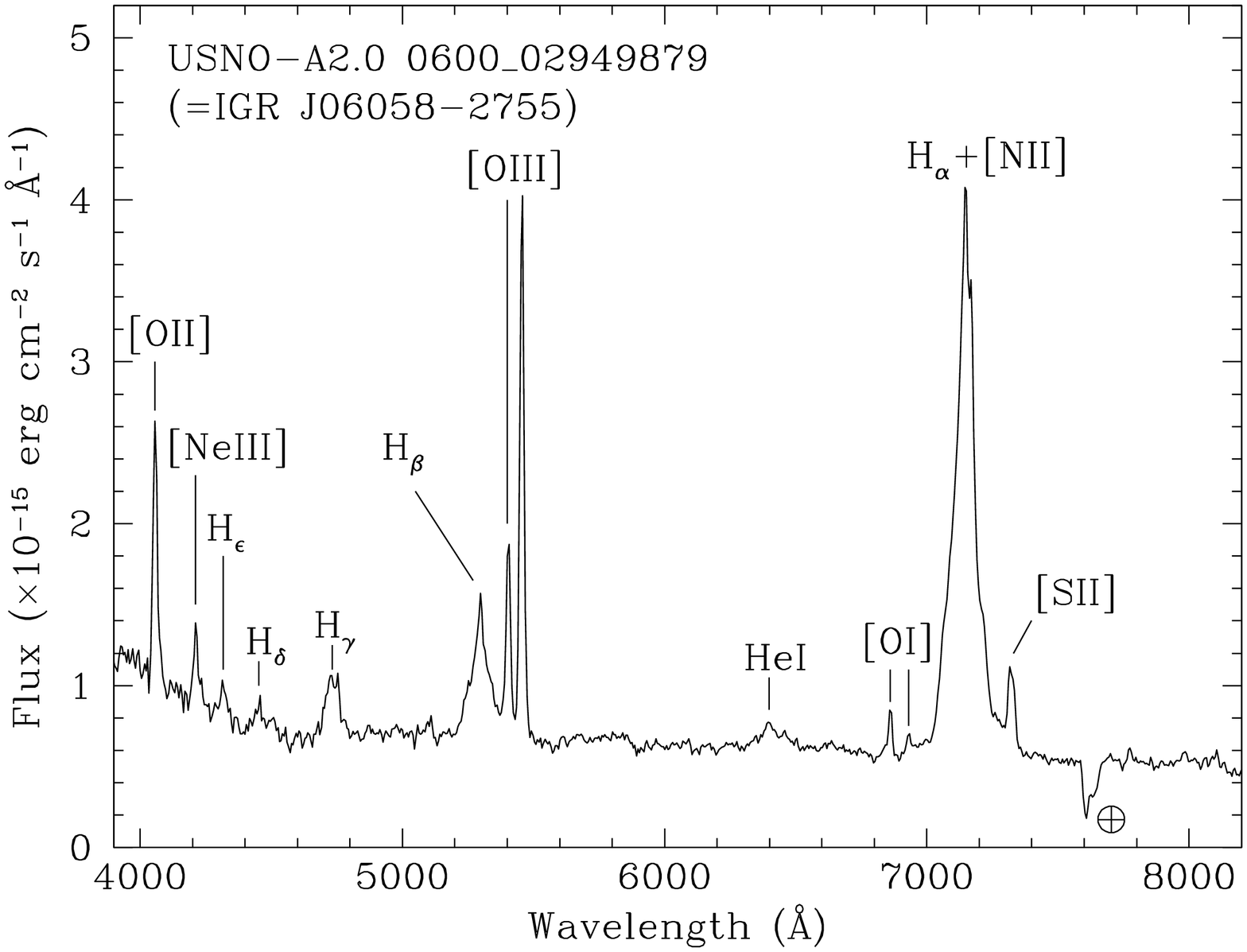,width=9cm,angle=270}}
\mbox{\psfig{file=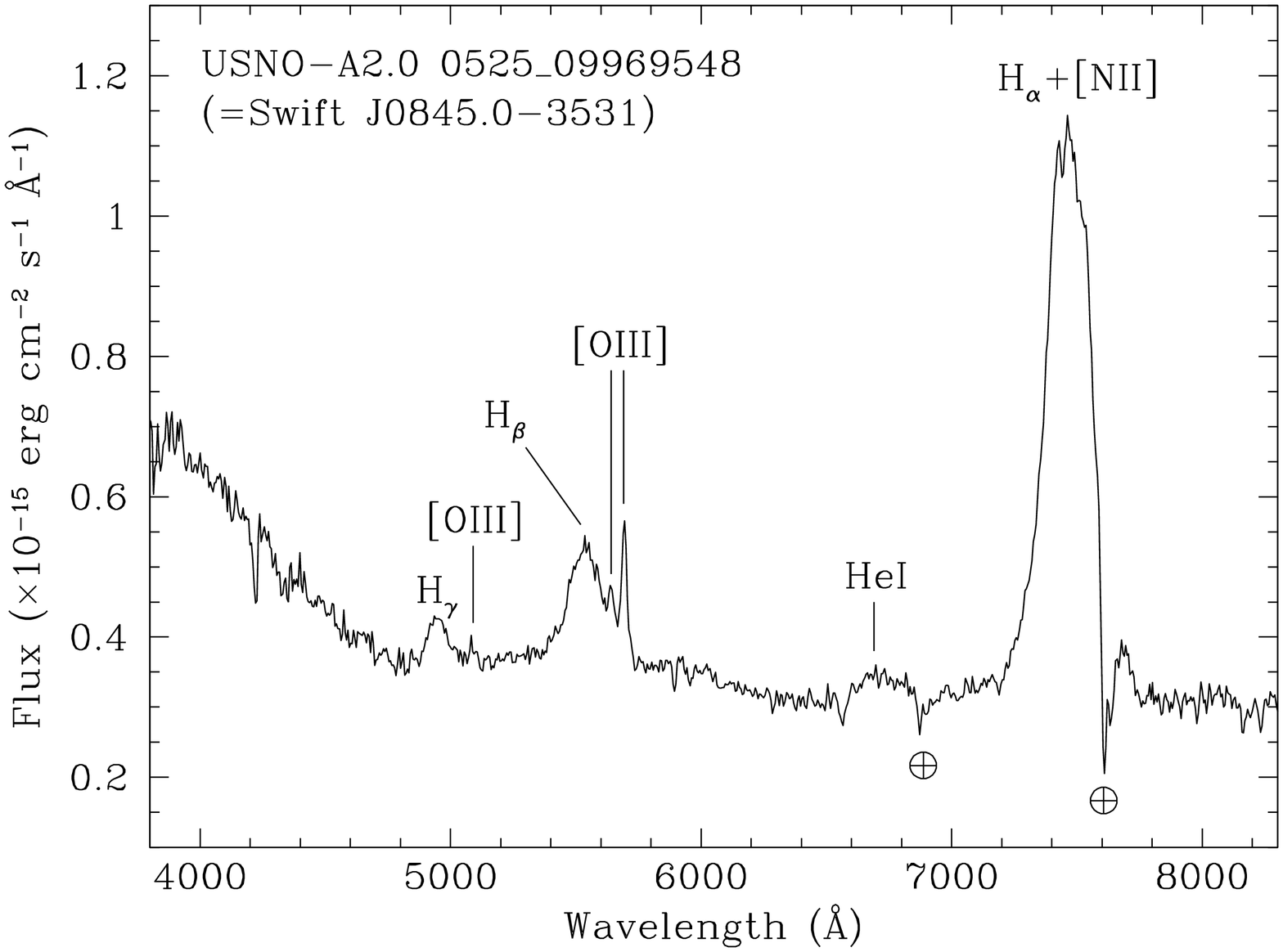,width=9cm,angle=270}}

\vspace{-.9cm}
\mbox{\psfig{file=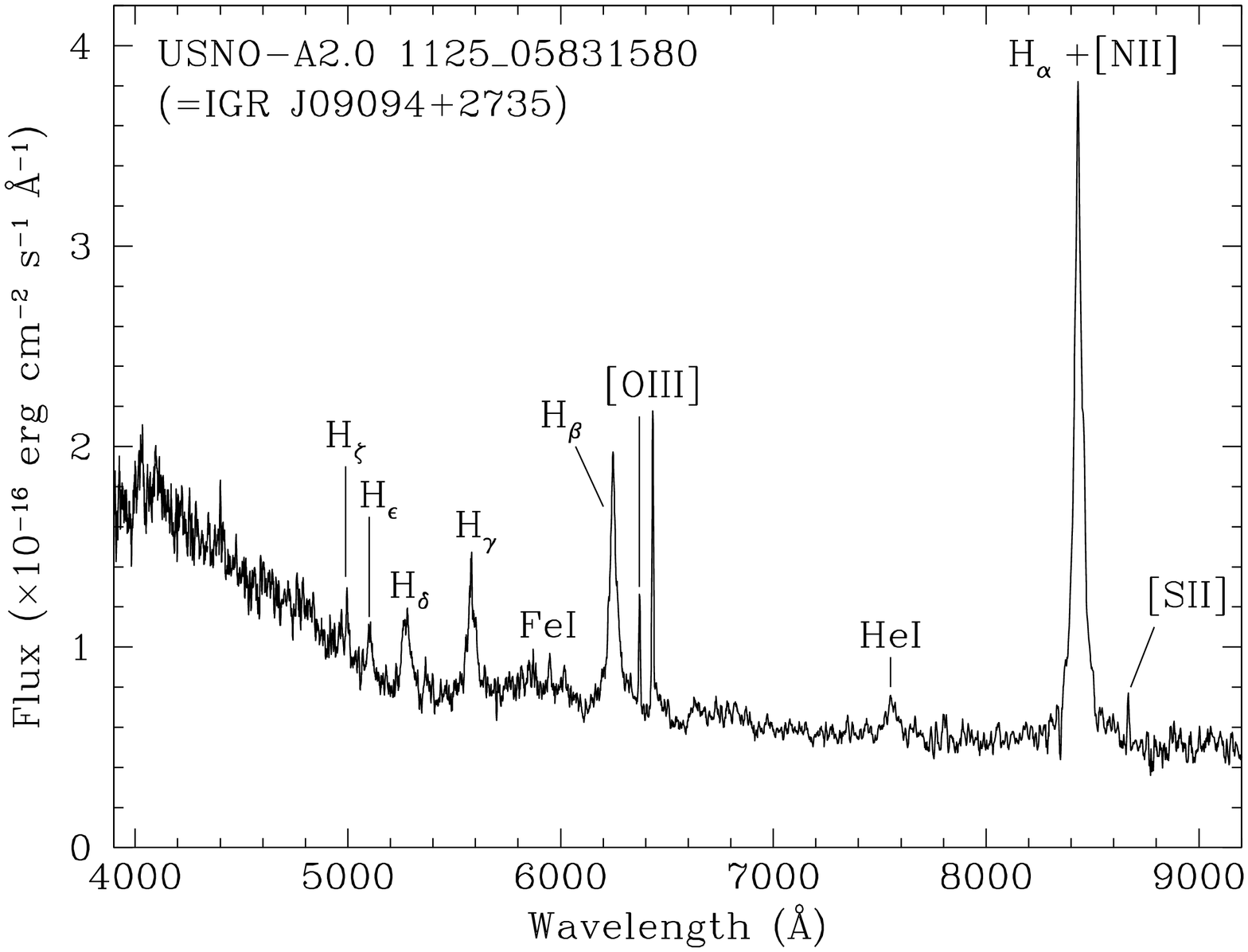,width=9cm,angle=270}}
\mbox{\psfig{file=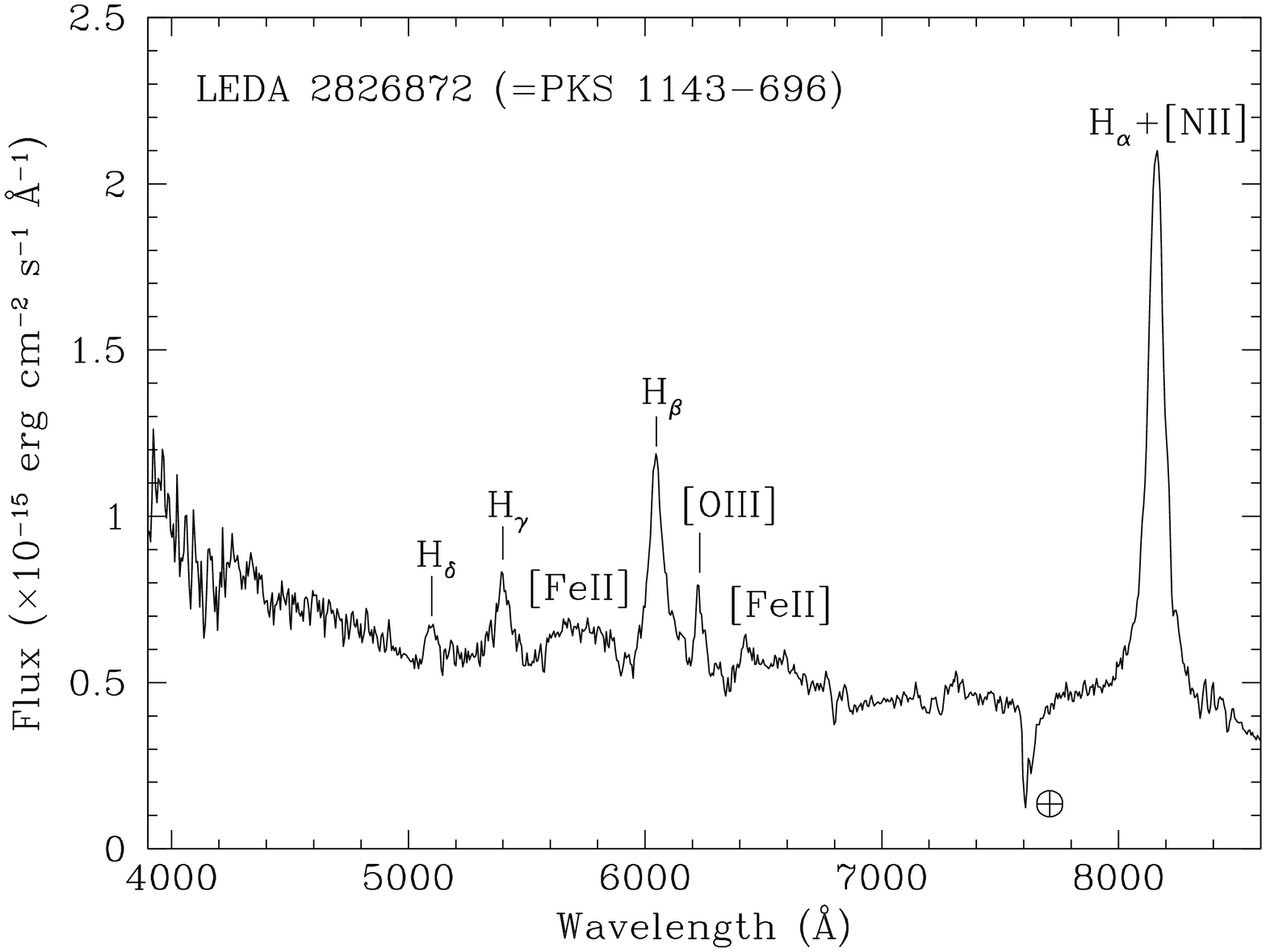,width=9cm,angle=270}}

\caption{Spectra (not corrected for the intervening Galactic absorption) 
of the optical counterparts of 6 broad emission-line AGNs belonging to 
the sample of {\it INTEGRAL} sources presented in this paper.
For each spectrum, the main spectral features are labeled. The 
symbol $\oplus$ indicates atmospheric telluric absorption bands.
The SDSS spectrum has been smoothed using a Gaussian filter with
$\sigma$ = 5 \AA.
}
\end{figure*}

\begin{figure*}[th!]
\mbox{\psfig{file=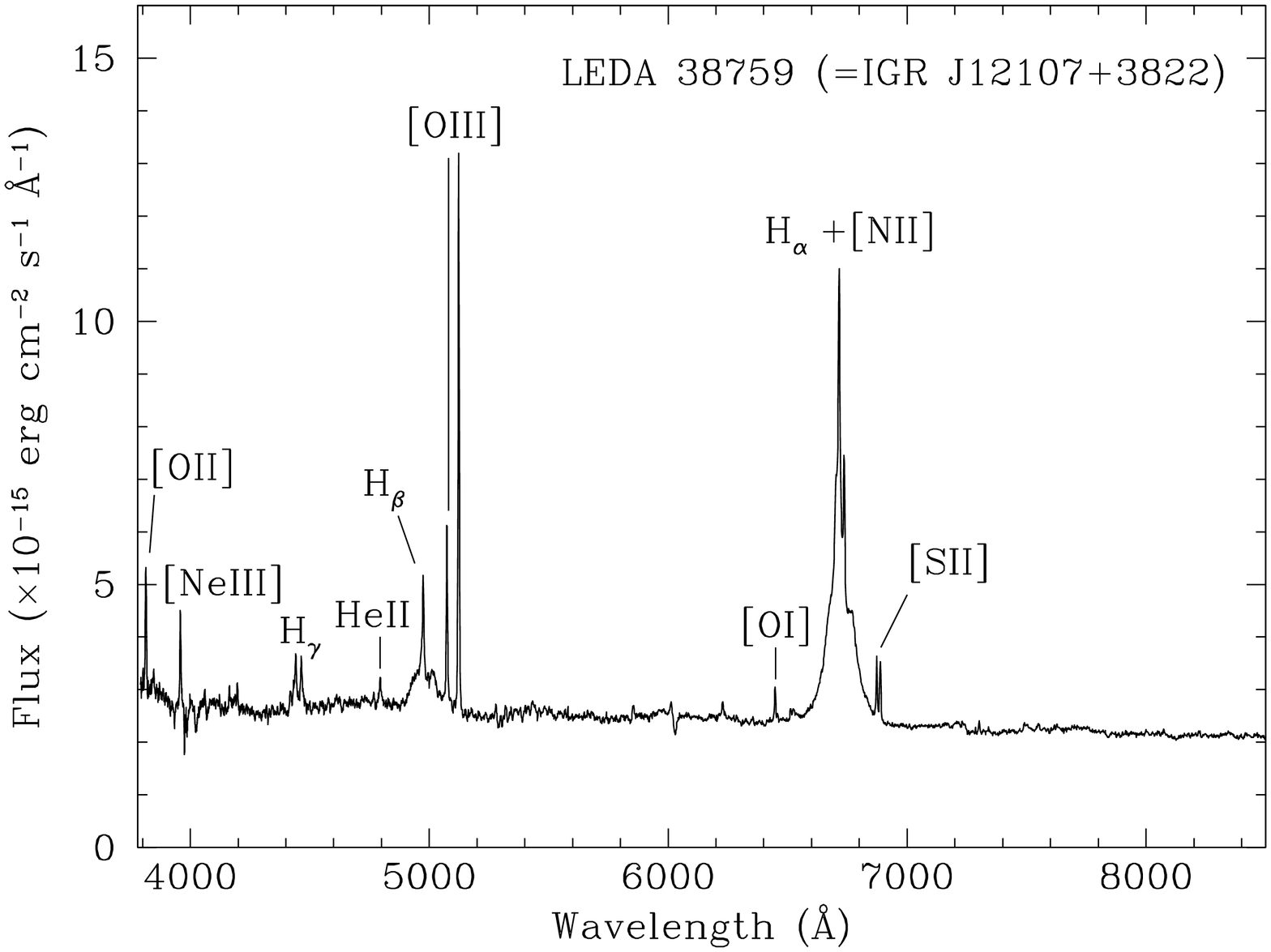,width=9cm,angle=270}}
\mbox{\psfig{file=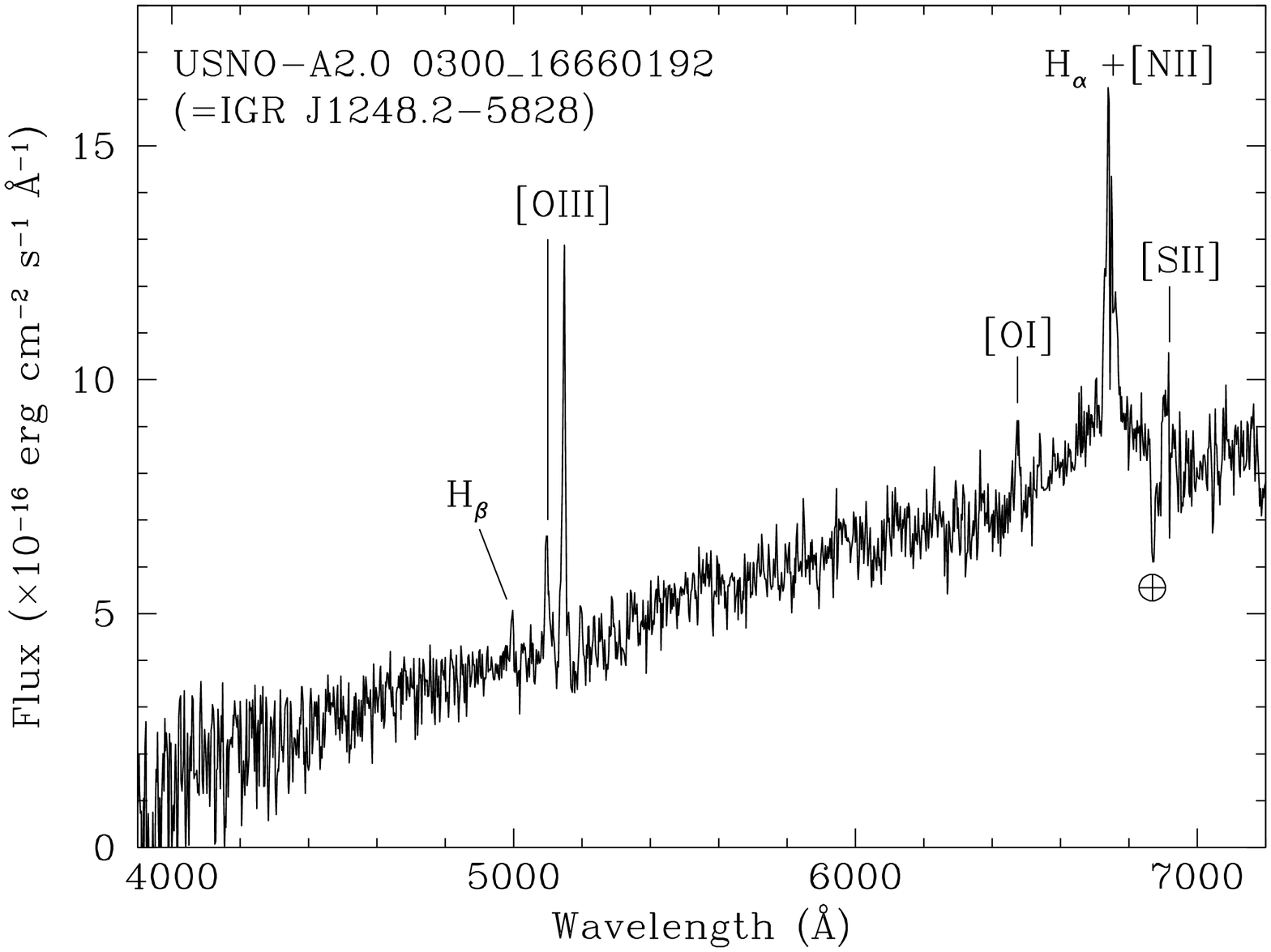,width=9cm,angle=270}}

\vspace{-.9cm}
\mbox{\psfig{file=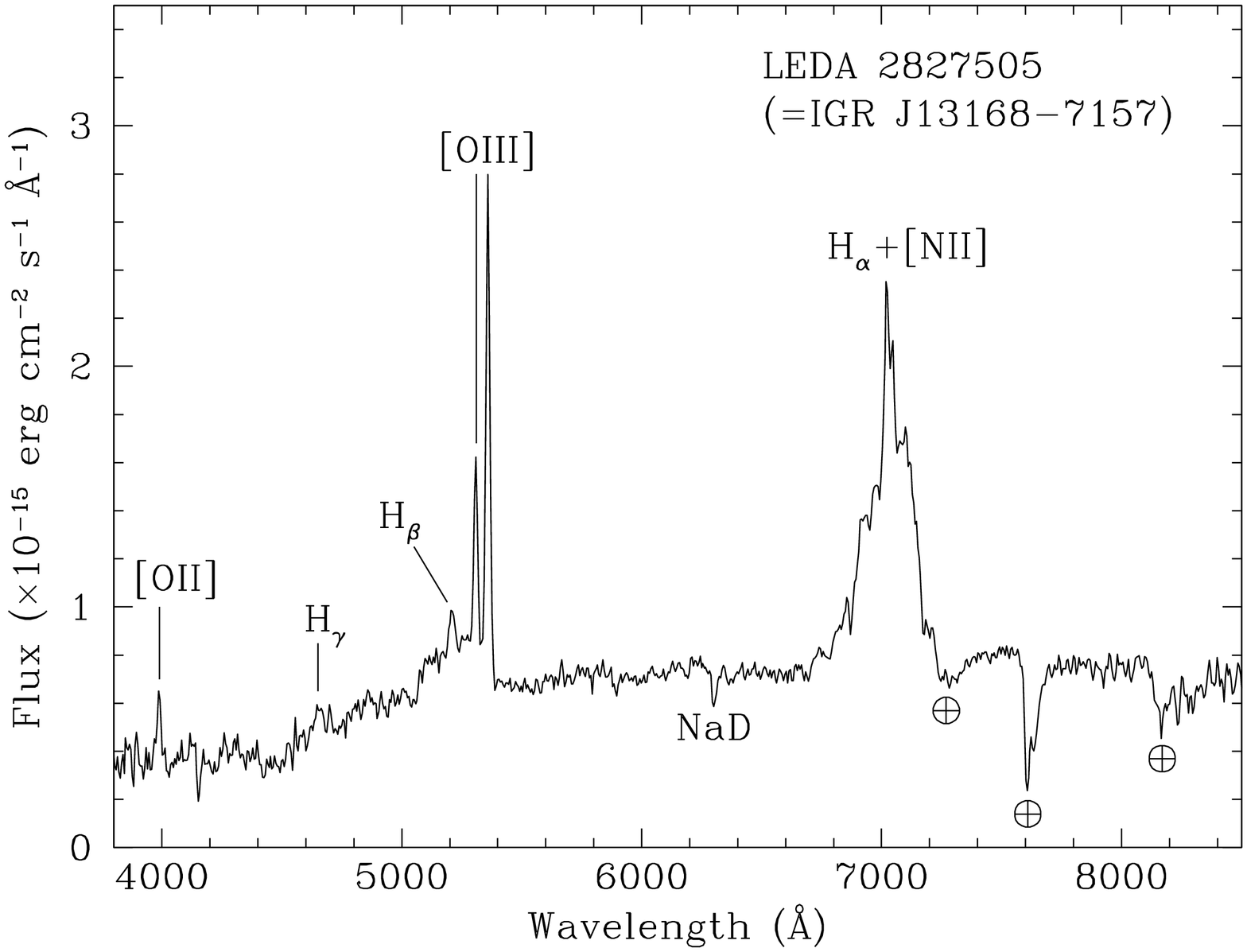,width=9cm,angle=270}}
\mbox{\psfig{file=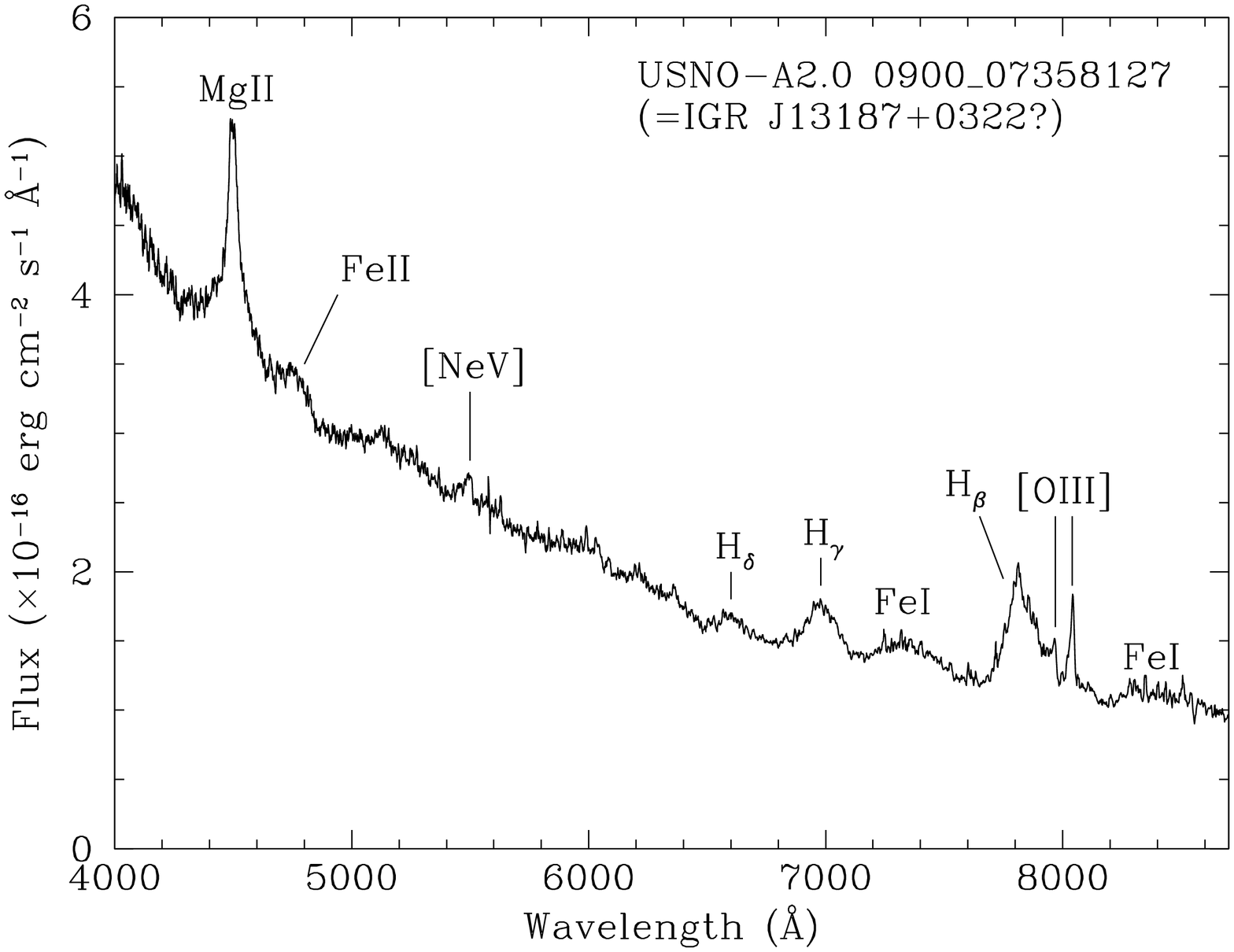,width=9cm,angle=270}}

\vspace{-.9cm}
\mbox{\psfig{file=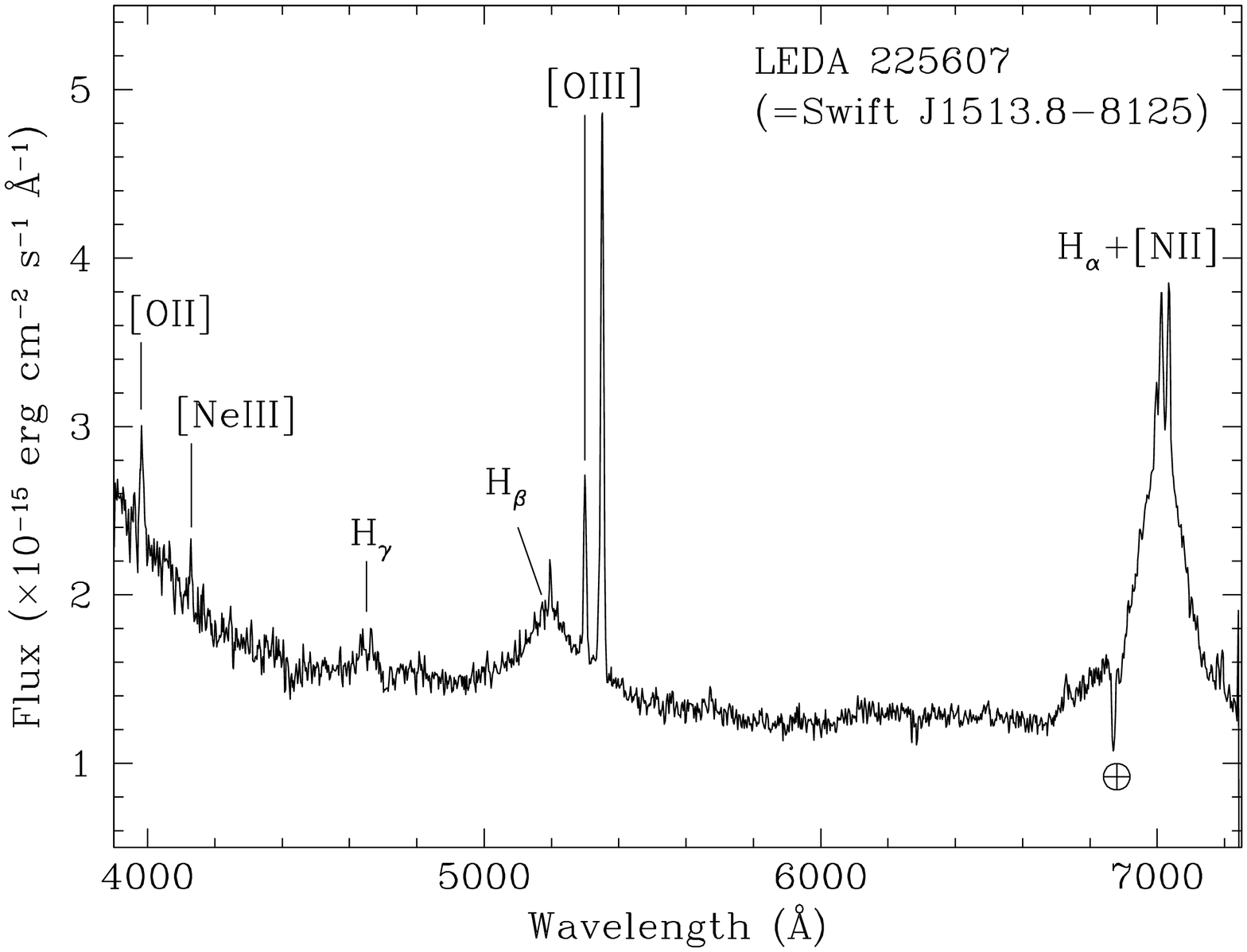,width=9cm,angle=270}}
\mbox{\psfig{file=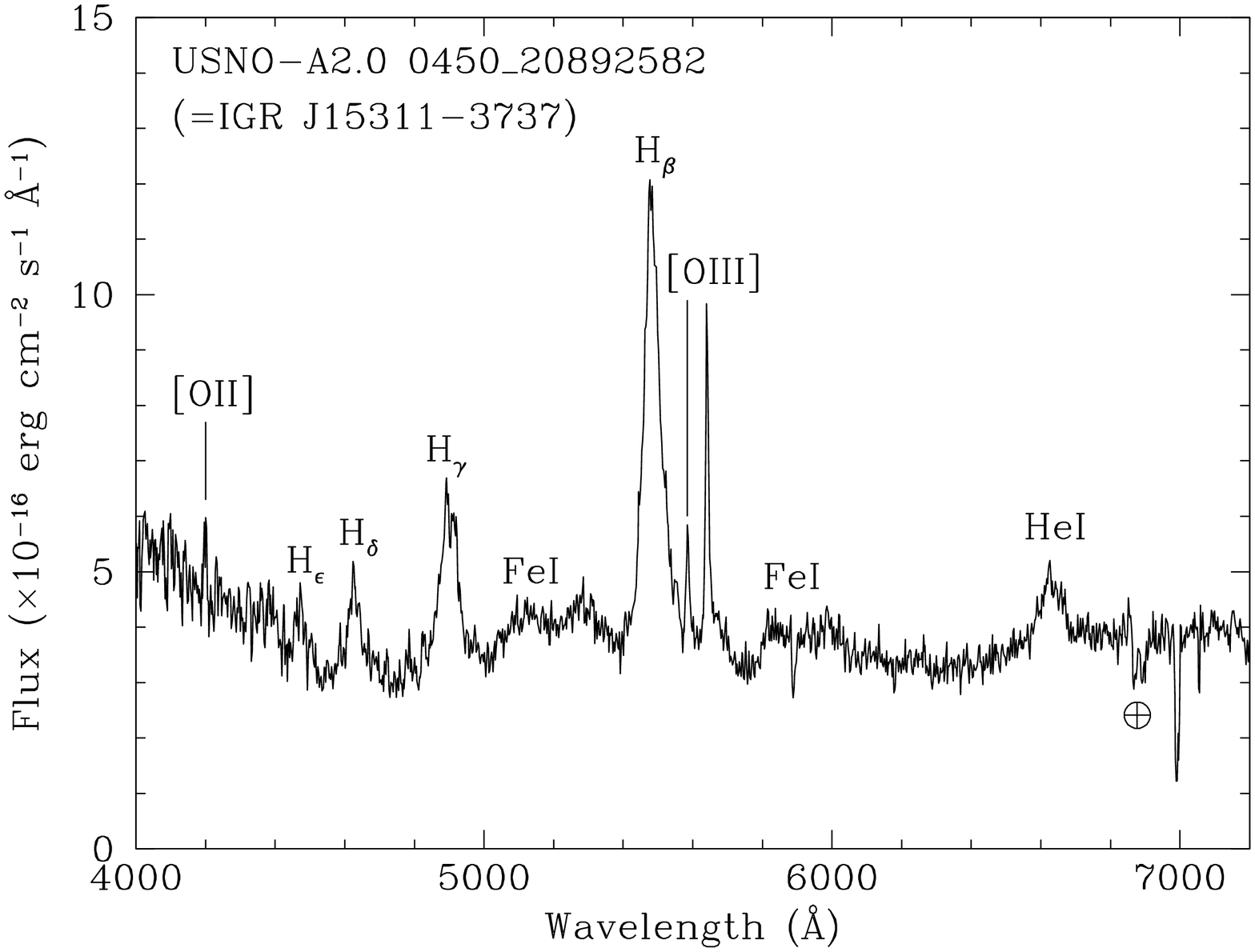,width=9cm,angle=270}}

\caption{Spectra (not corrected for the intervening Galactic absorption) 
of the optical counterparts of 6 additional broad emission-line AGNs 
belonging to the sample of {\it INTEGRAL} sources presented in this paper.
For each spectrum, the main spectral features are labeled. The 
symbol $\oplus$ indicates atmospheric telluric absorption bands.
The SDSS spectrum has been smoothed using a Gaussian filter with
$\sigma$ = 5 \AA.
}
\end{figure*}

\begin{figure*}[th!]
\mbox{\psfig{file=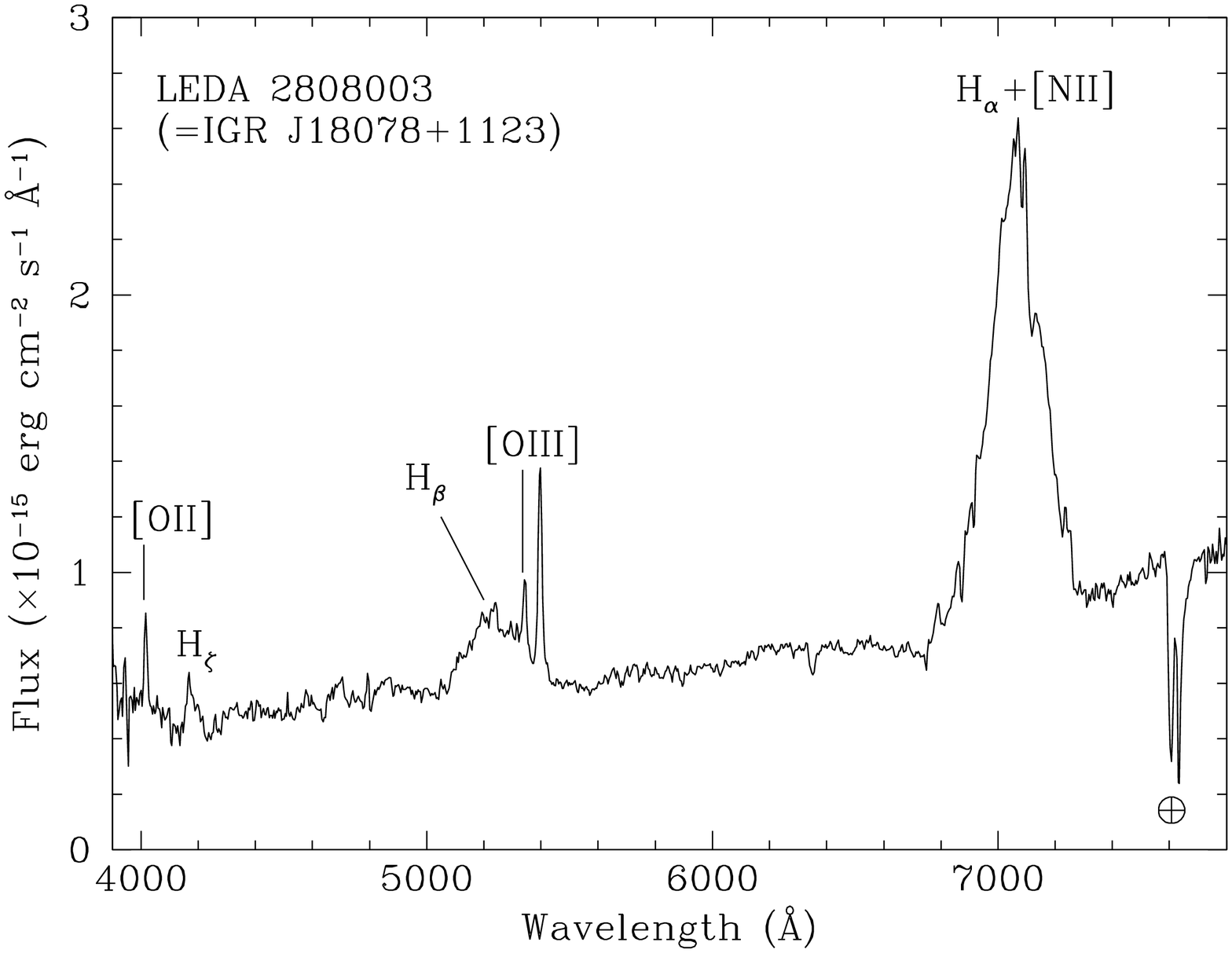,width=9cm,angle=270}}
\mbox{\psfig{file=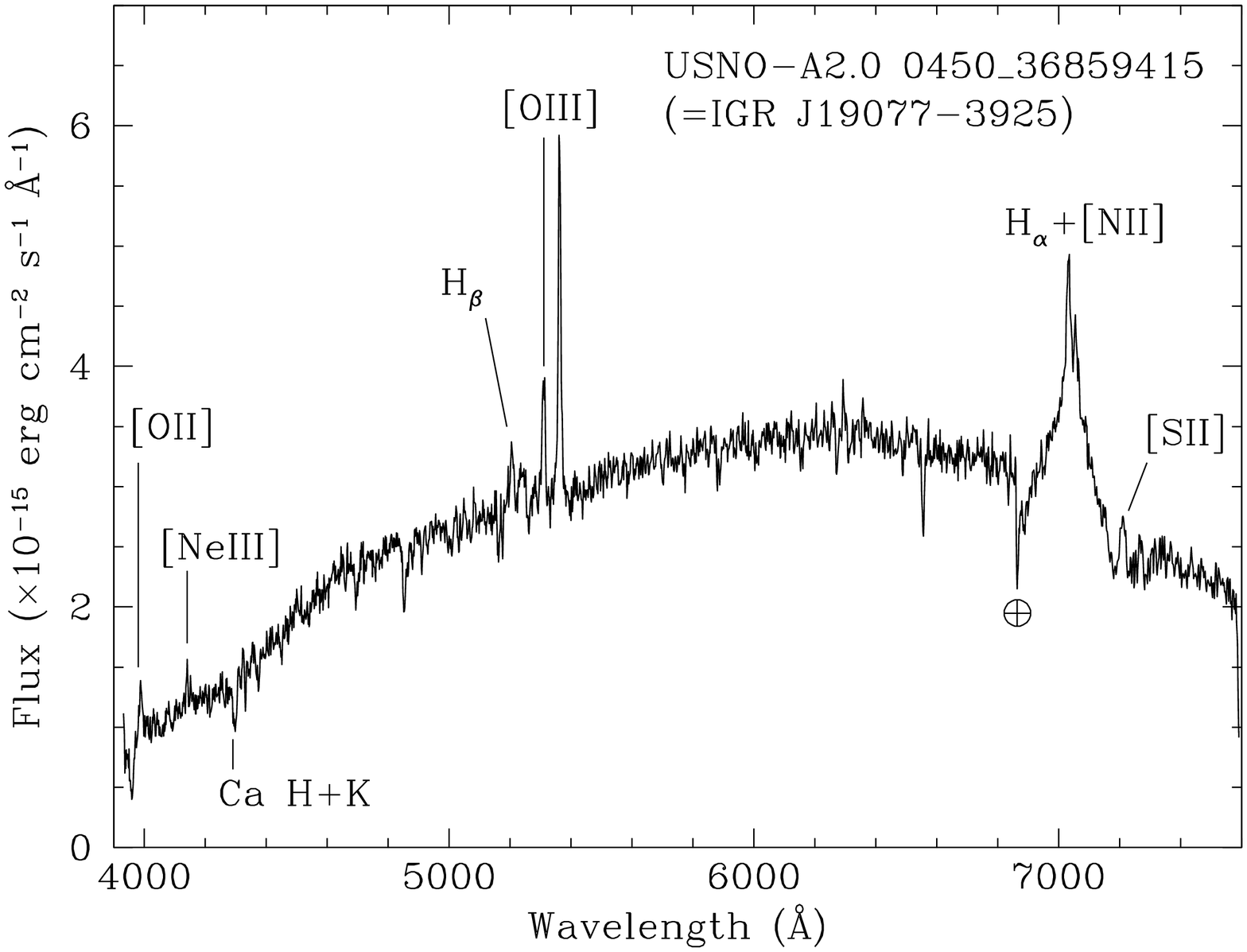,width=9cm,angle=270}}

\vspace{-.9cm}
\mbox{\psfig{file=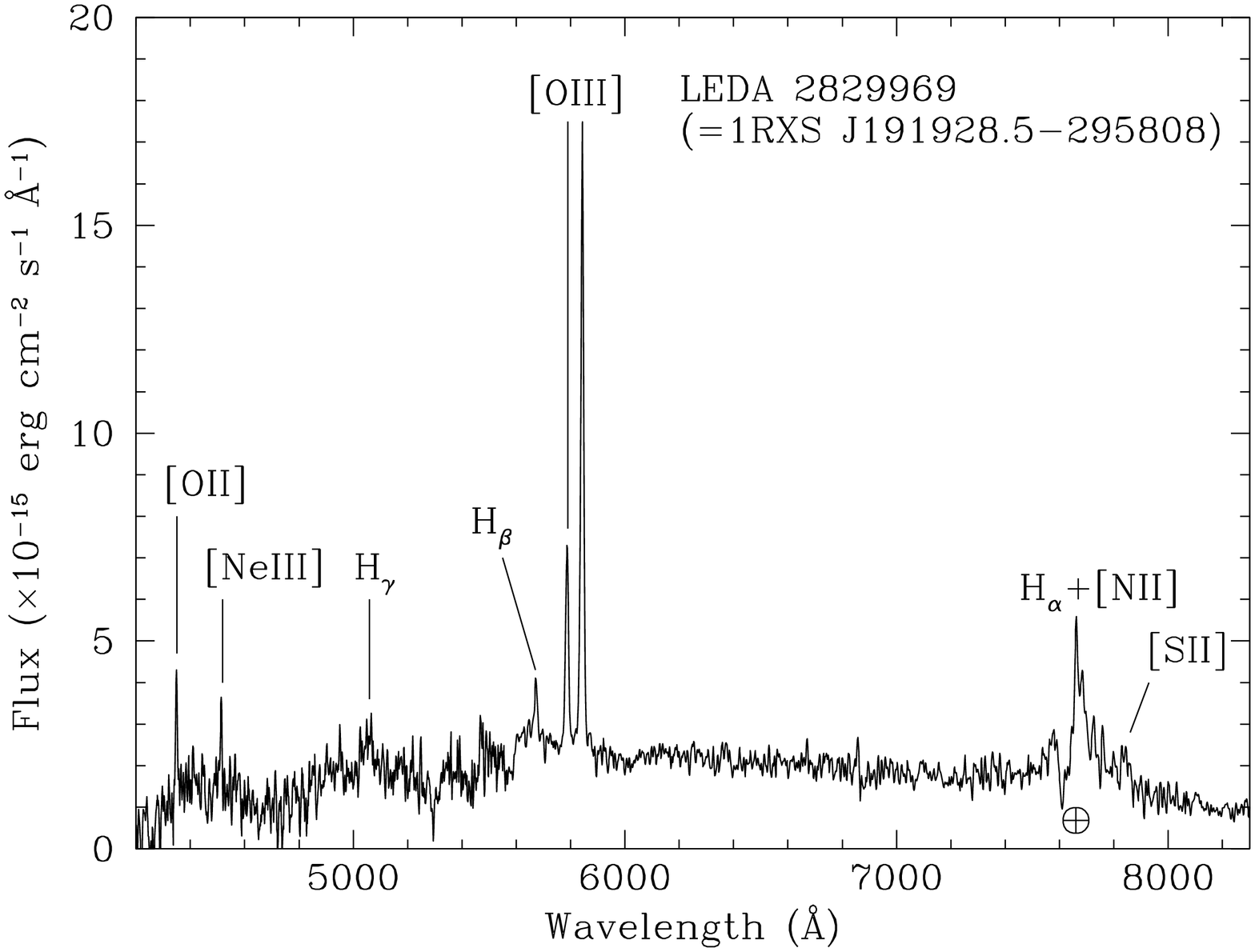,width=9cm,angle=270}}
\mbox{\psfig{file=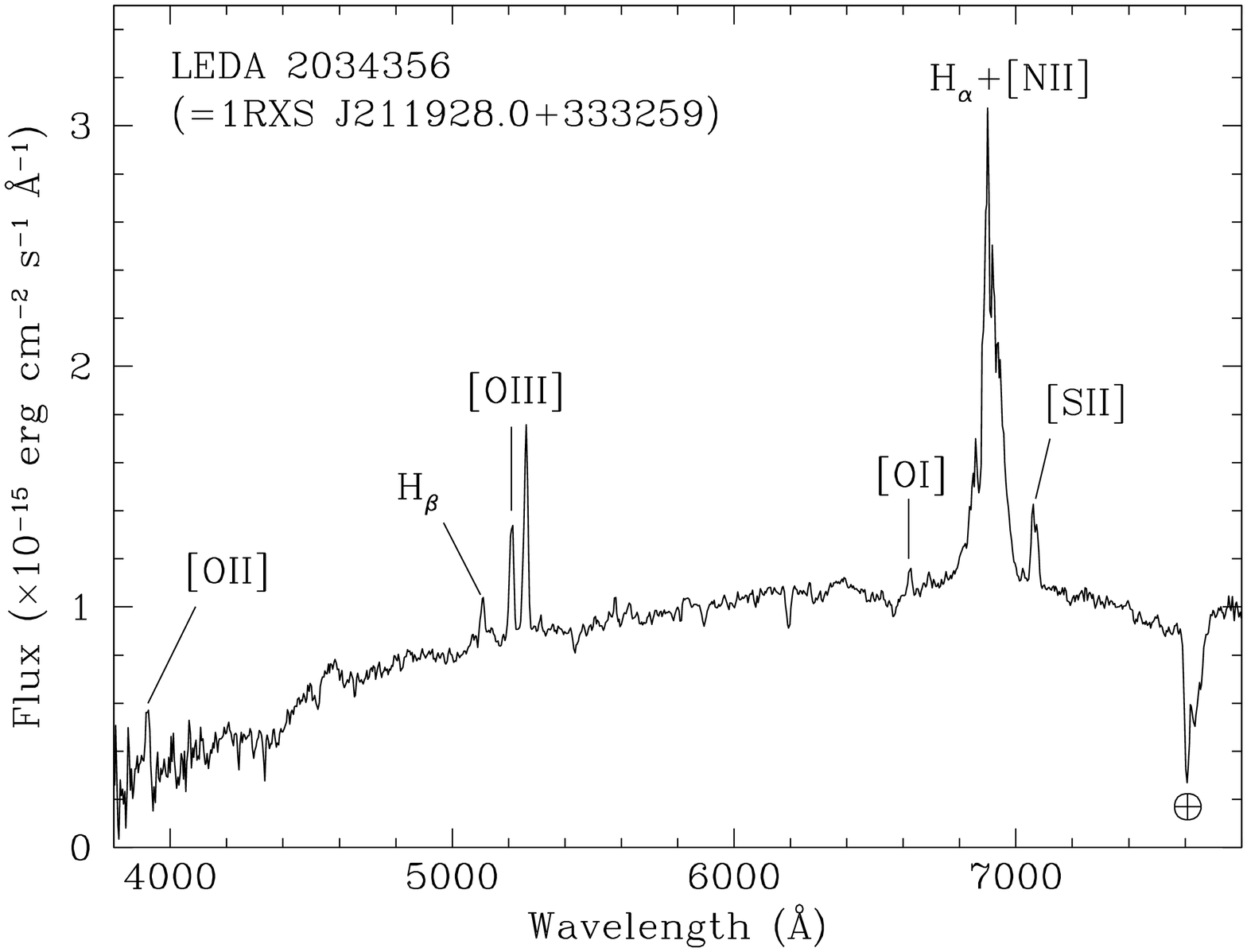,width=9cm,angle=270}}

\vspace{-.9cm}
\mbox{\psfig{file=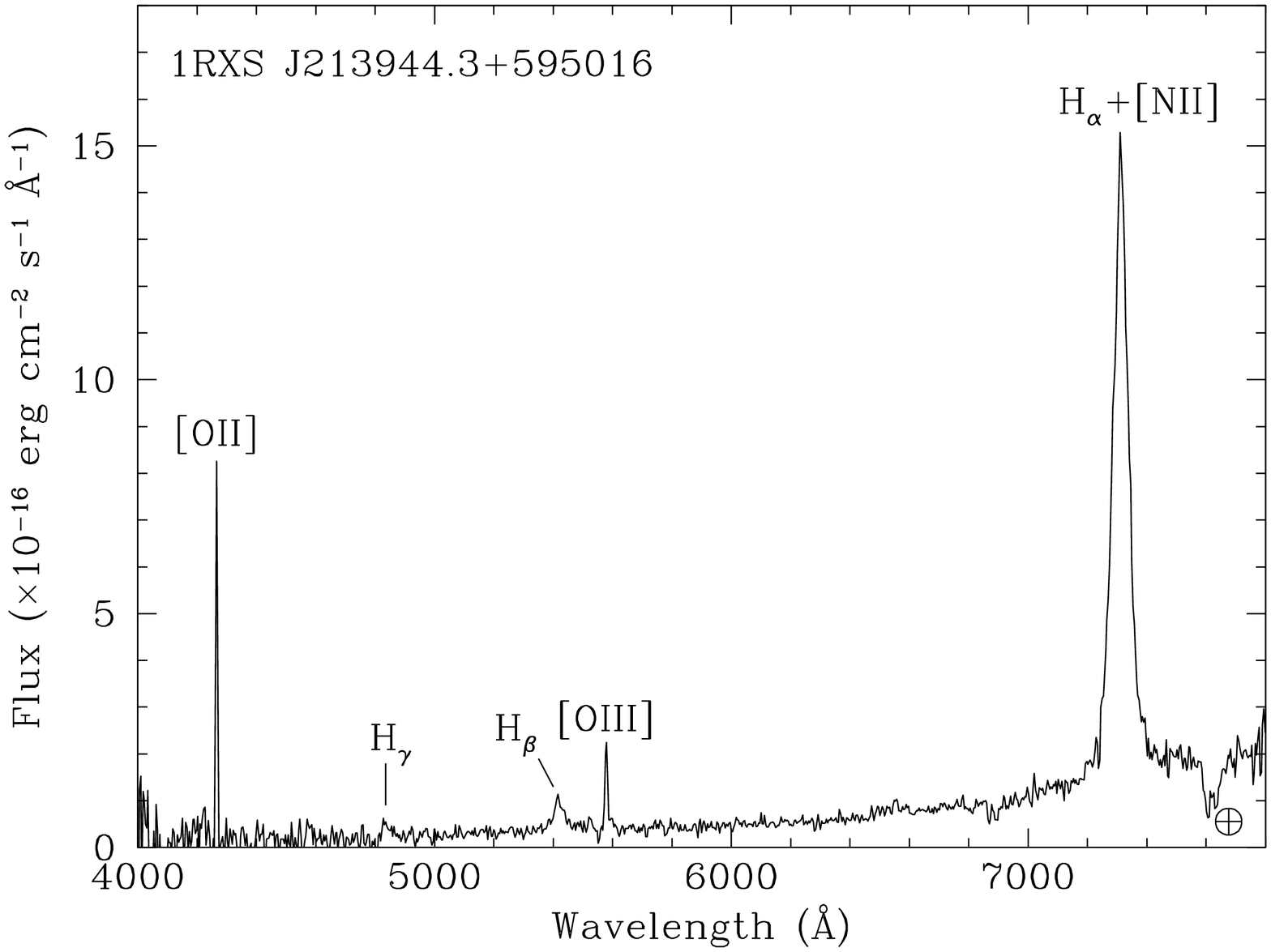,width=9cm,angle=270}}
\mbox{\psfig{file=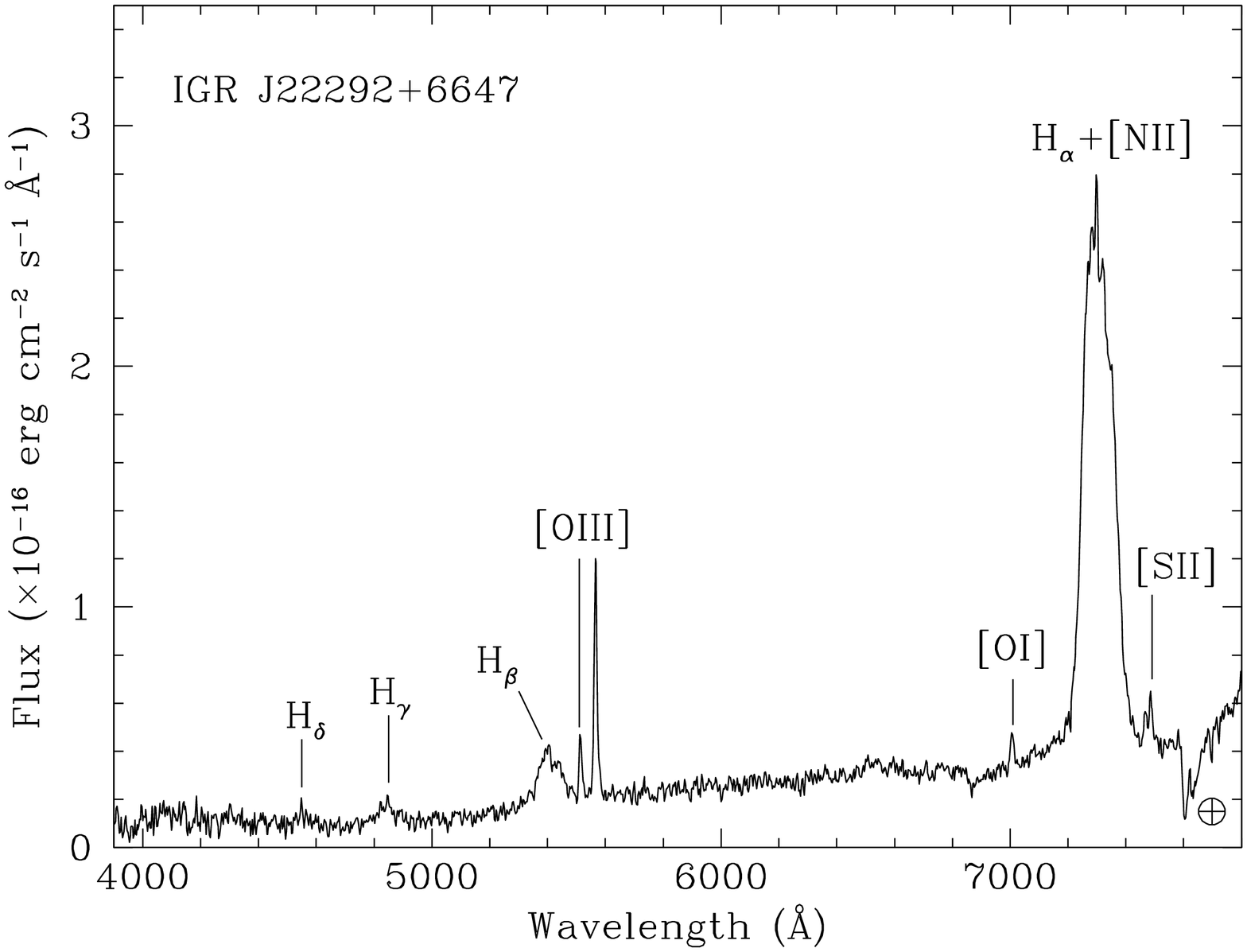,width=9cm,angle=270}}

\caption{Spectra (not corrected for the intervening Galactic absorption) 
of the optical counterparts of the remaining 6 broad emission-line AGNs 
belonging to the sample of {\it INTEGRAL} sources presented in this paper.
For each spectrum, the main spectral features are labeled. The 
symbol $\oplus$ indicates atmospheric telluric absorption bands.
The 6dFGS spectrum of 1RXS J191928.5$-$295808 has been smoothed using a 
Gaussian filter with $\sigma$ = 3 \AA.
}
\end{figure*}

\begin{figure*}[th!]
\mbox{\psfig{file=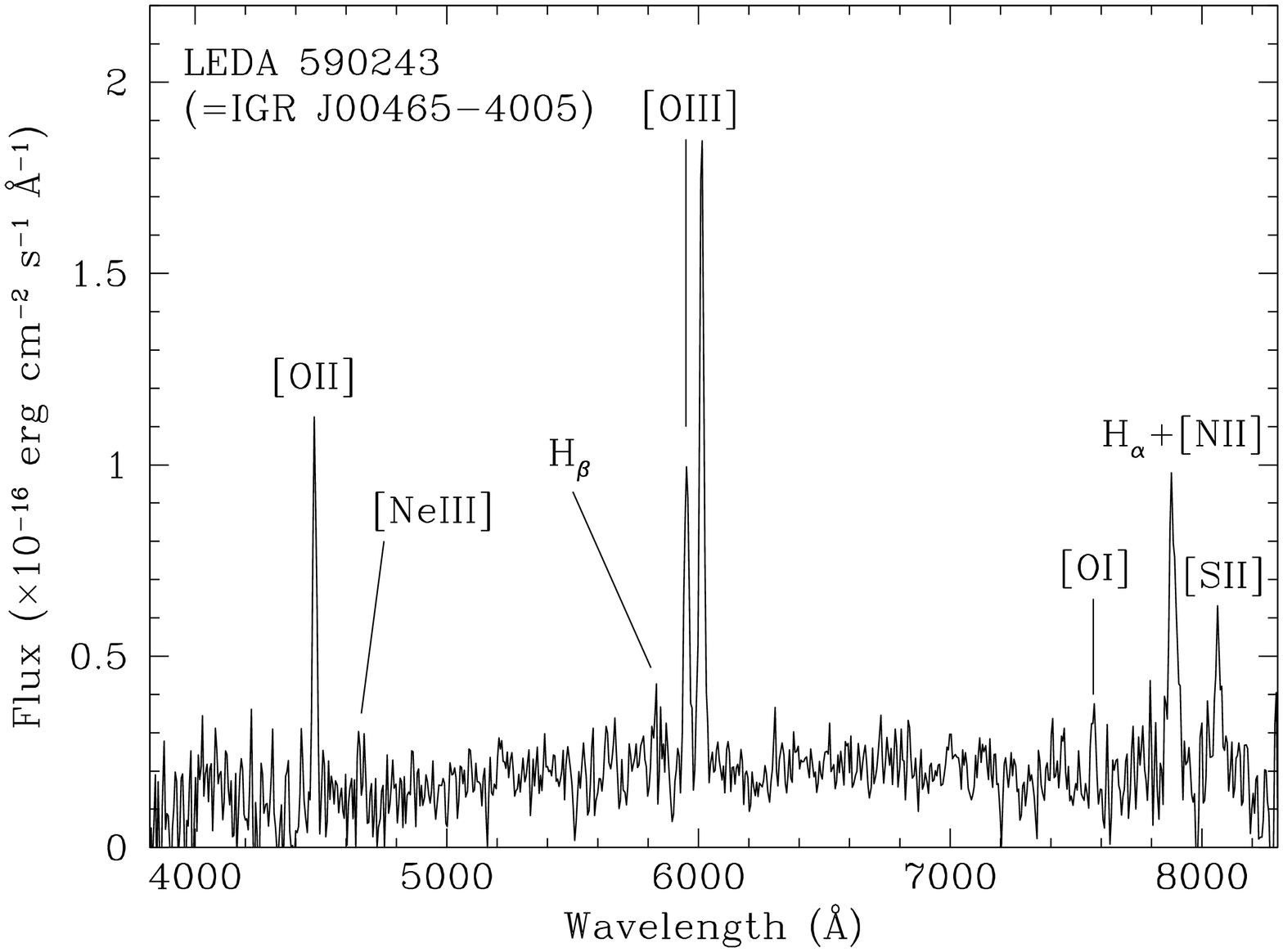,width=9cm,angle=270}}
\mbox{\psfig{file=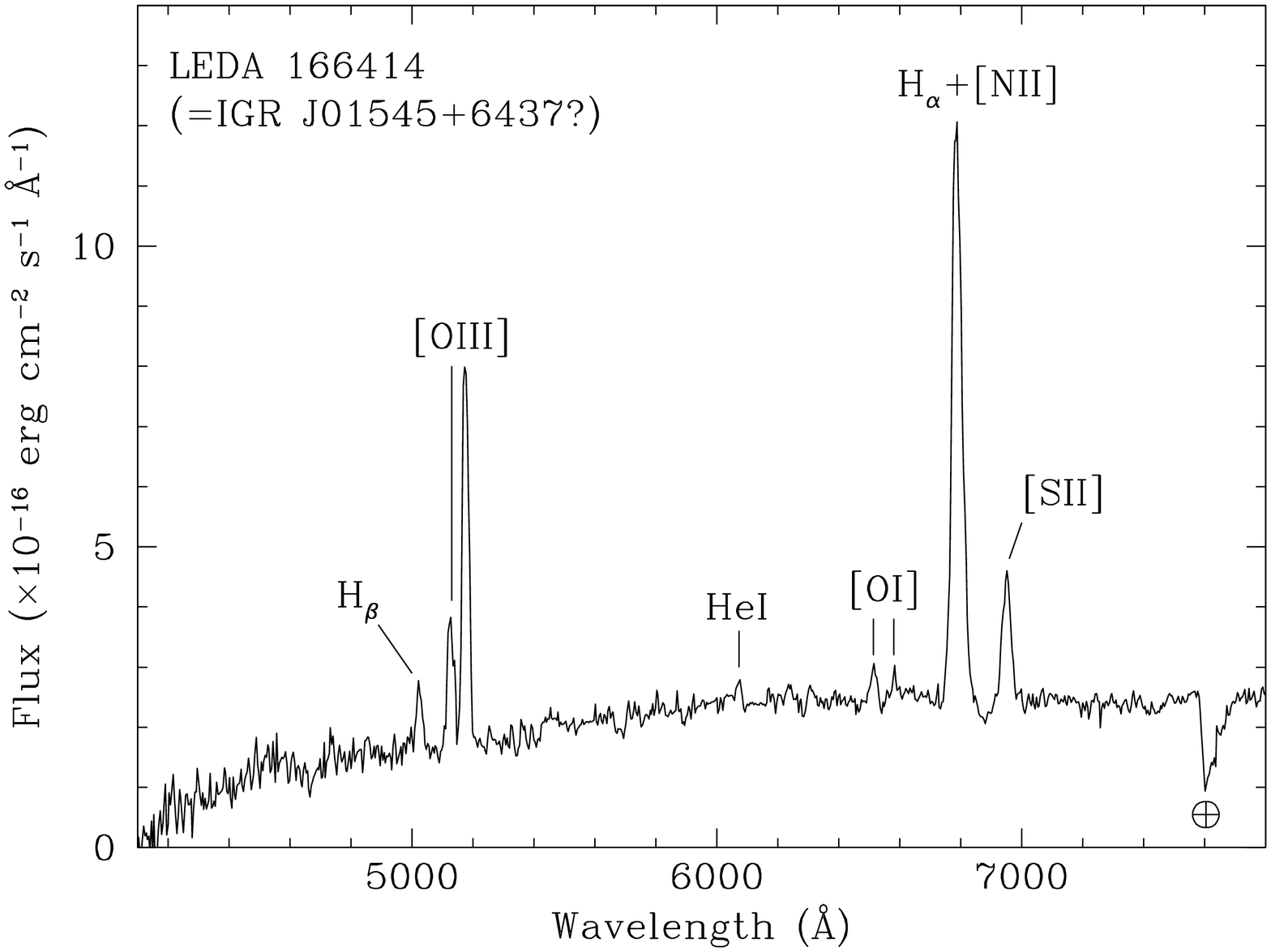,width=9cm,angle=270}}

\vspace{-.9cm}
\mbox{\psfig{file=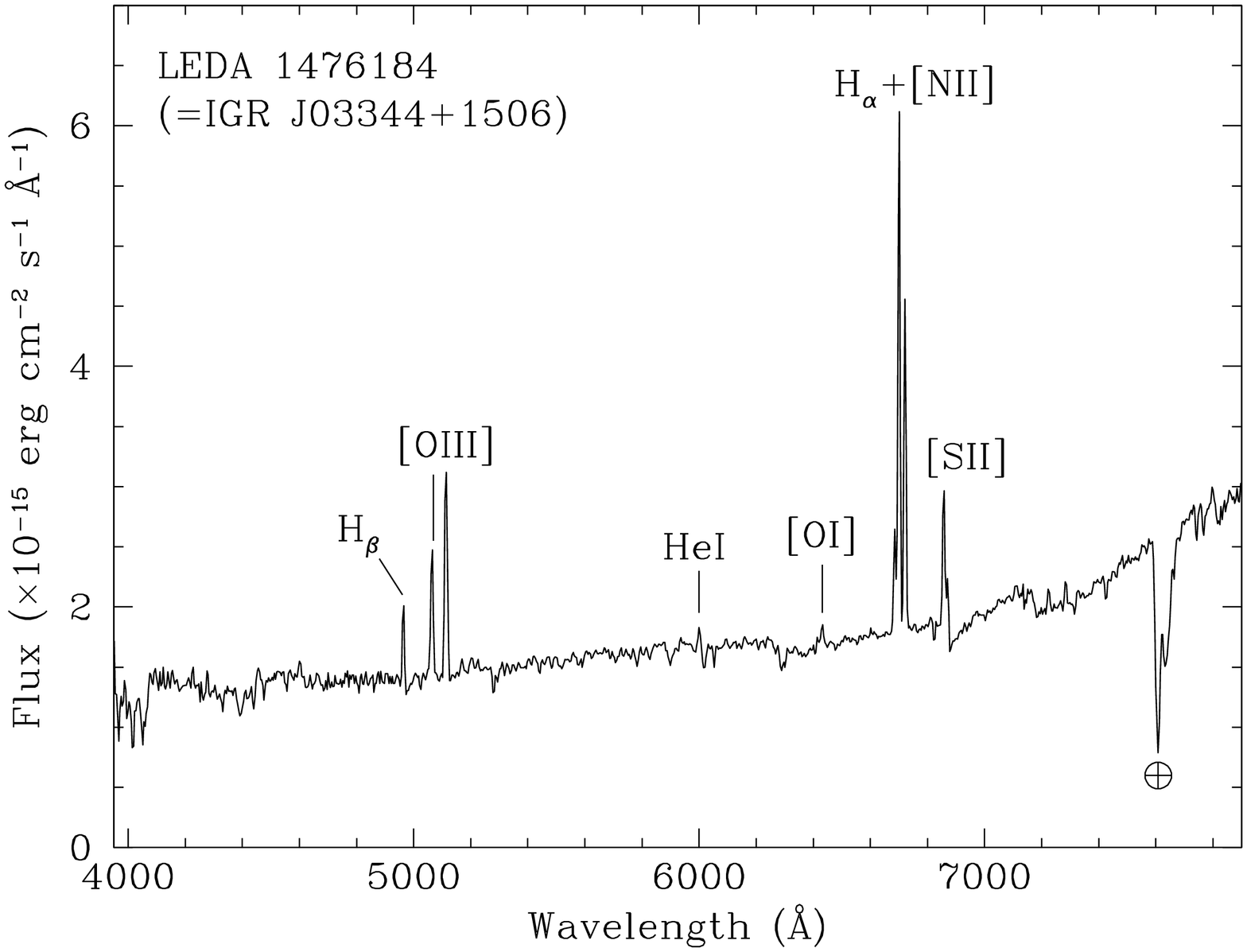,width=9cm,angle=270}}
\mbox{\psfig{file=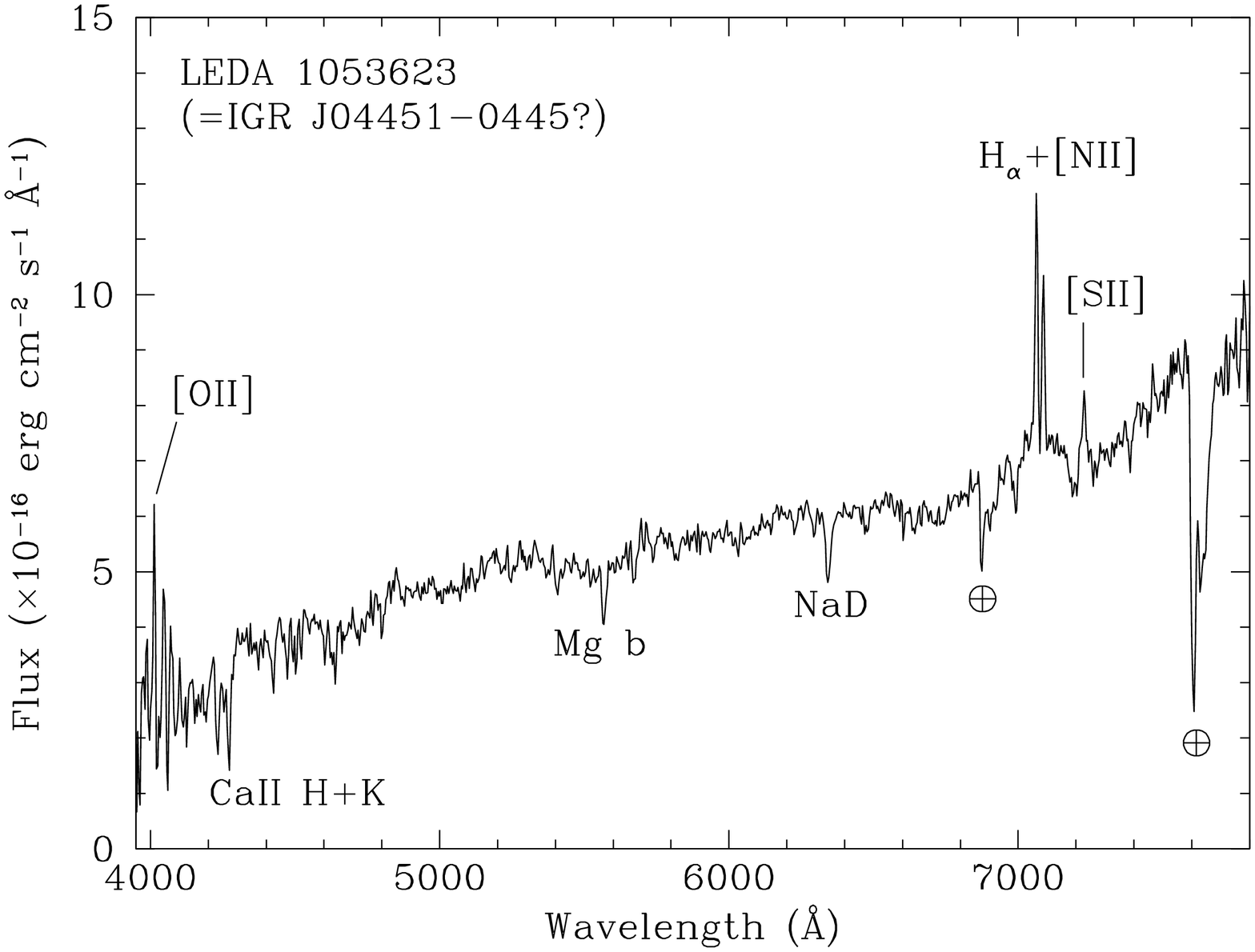,width=9cm,angle=270}}

\vspace{-.9cm}
\mbox{\psfig{file=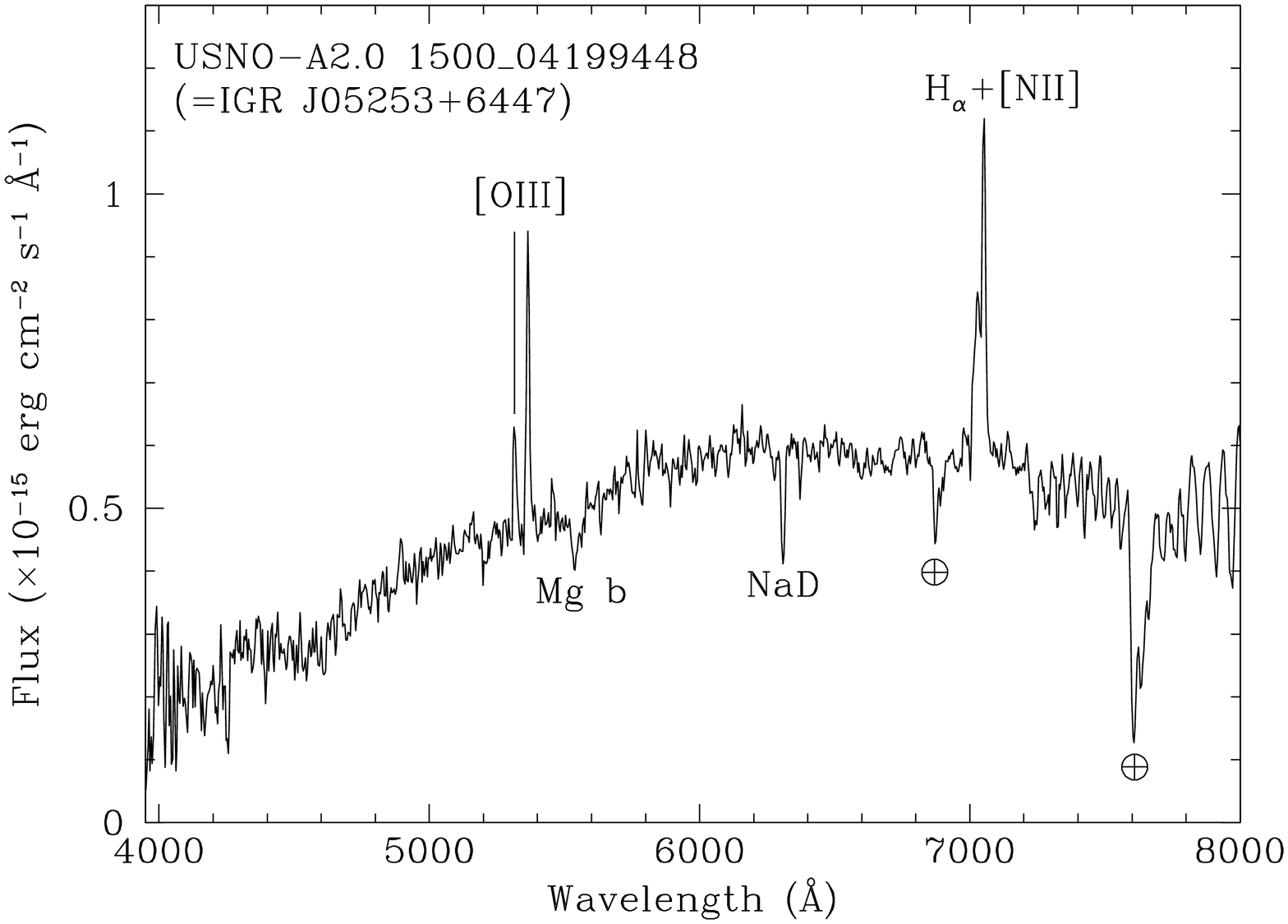,width=9cm,angle=270}}
\mbox{\psfig{file=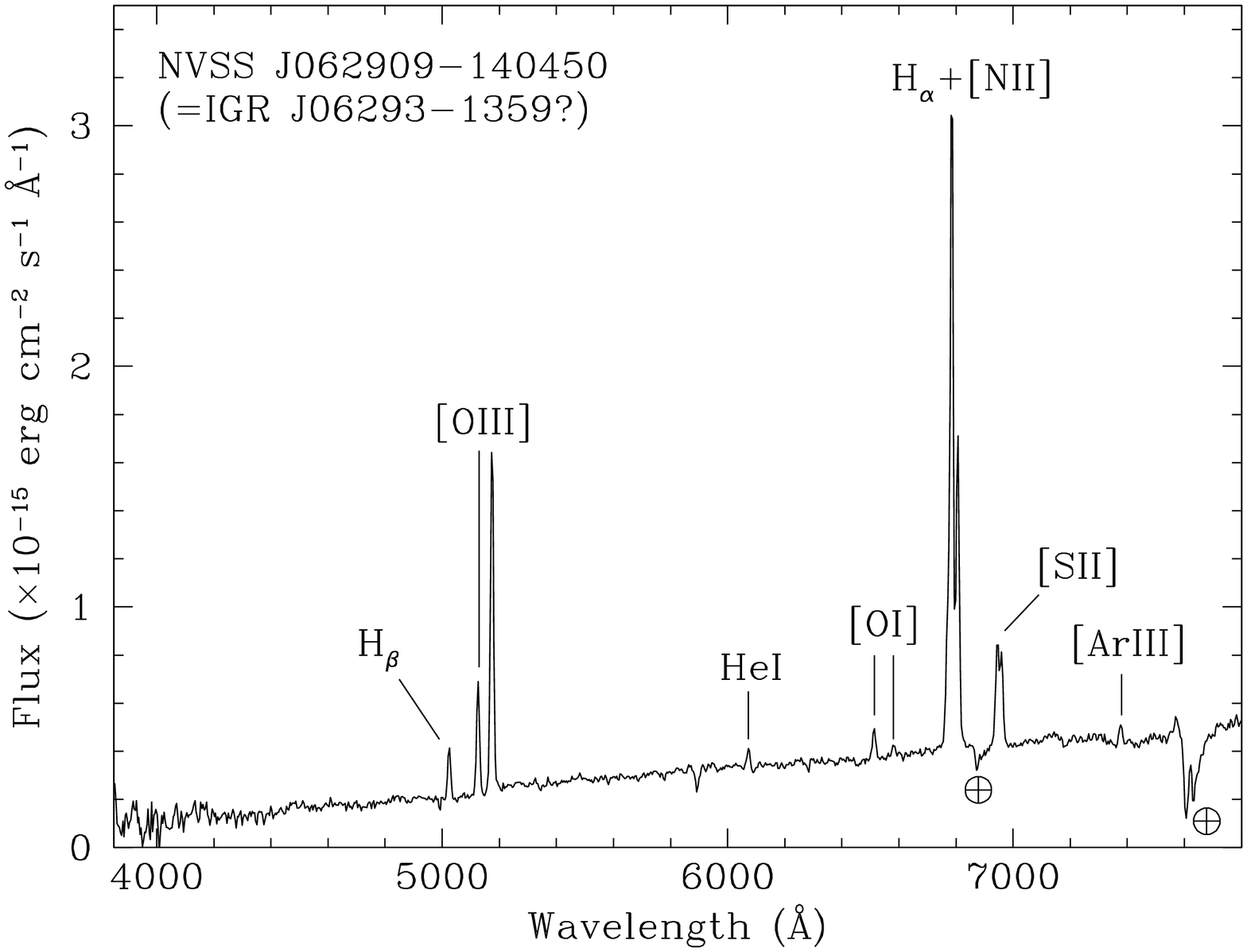,width=9cm,angle=270}}

\caption{Spectra (not corrected for the intervening Galactic absorption) 
of the optical counterparts of 6 Sy2 AGNs belonging to the sample of {\it 
INTEGRAL} sources presented in this paper.
For each spectrum, the main spectral features are labeled. The 
symbol $\oplus$ indicates atmospheric telluric absorption bands.
}
\end{figure*}

\begin{figure*}[th!]
\mbox{\psfig{file=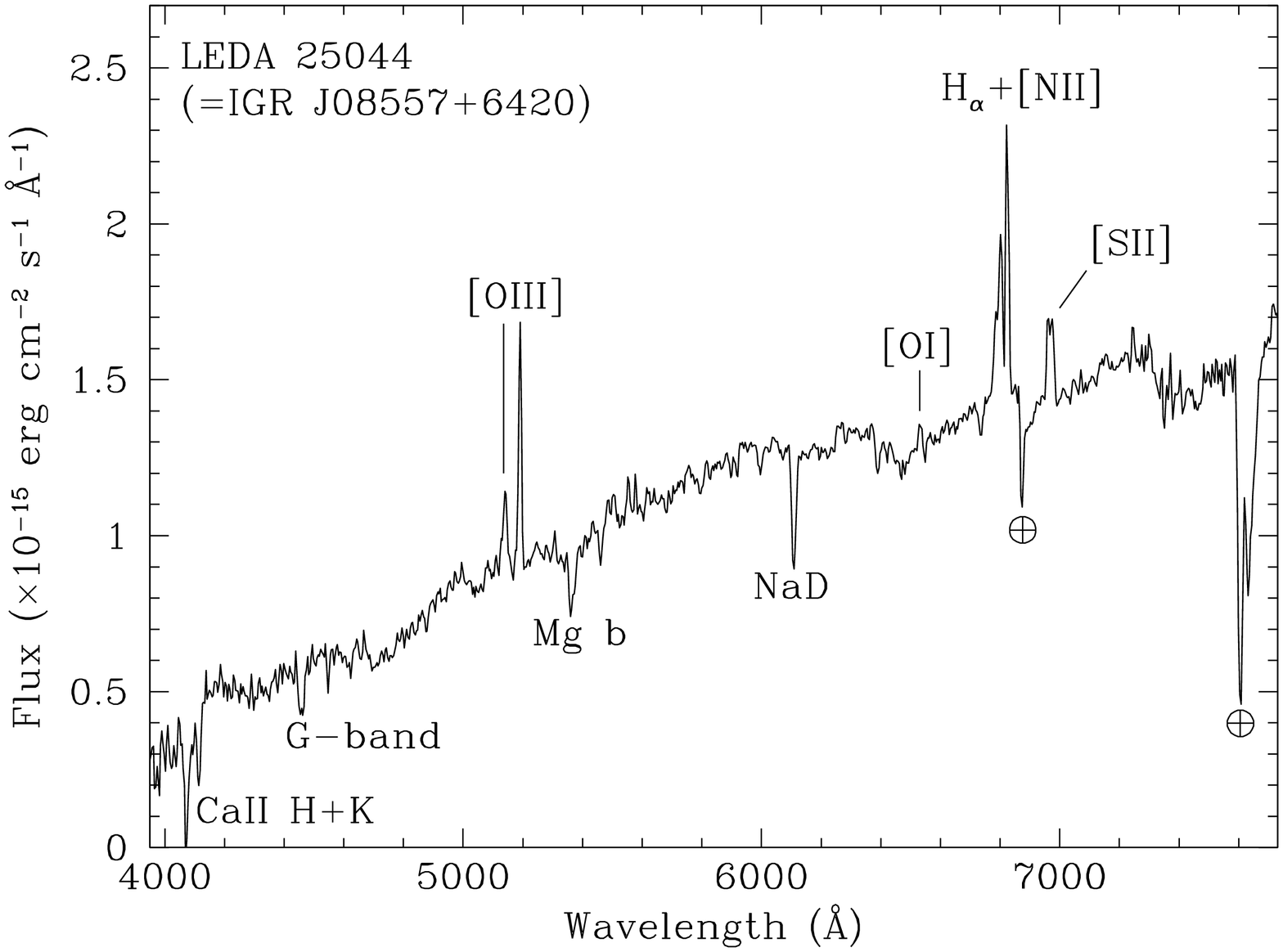,width=9cm,angle=270}}
\mbox{\psfig{file=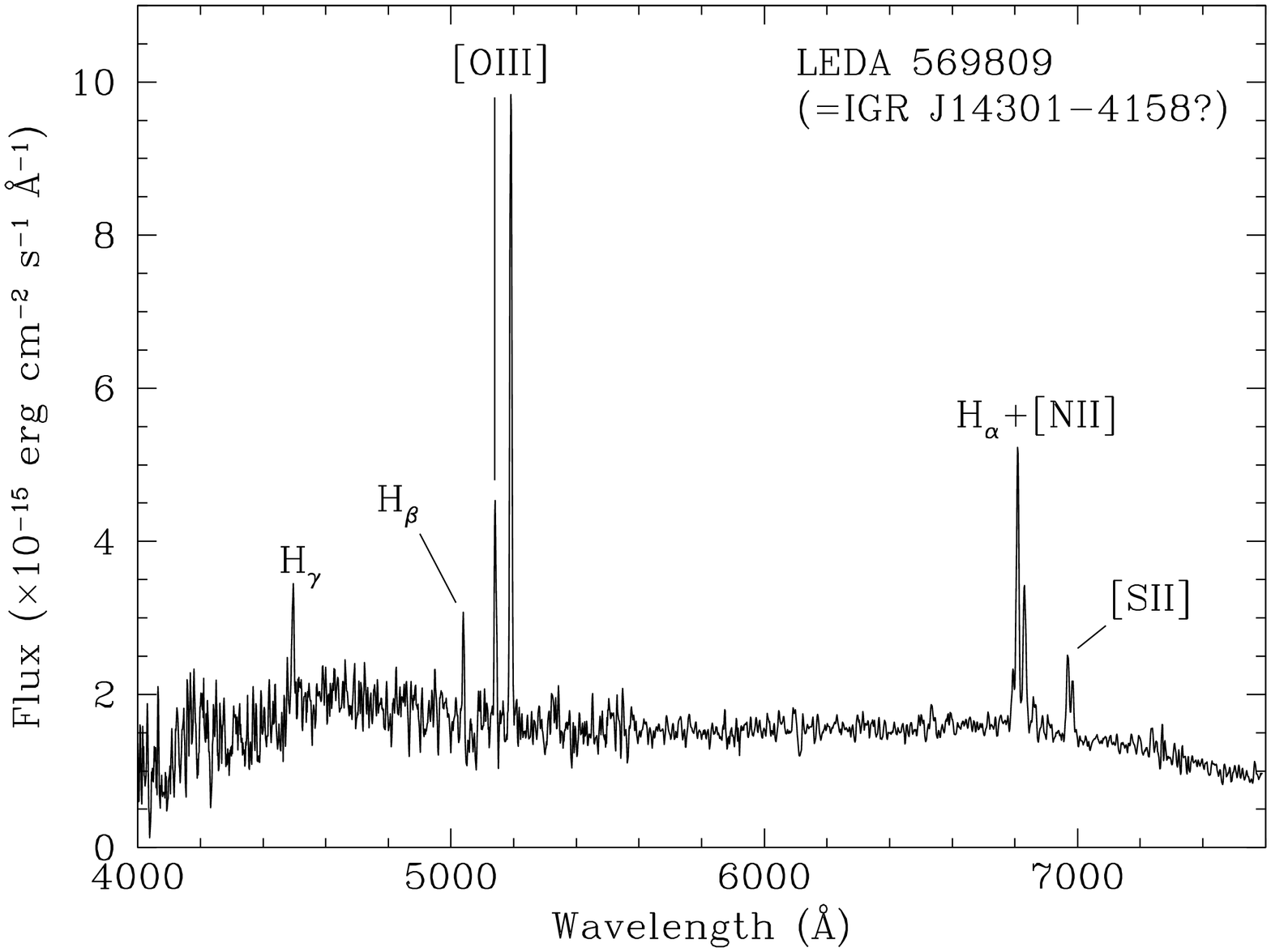,width=9cm,angle=270}}

\vspace{-.9cm}
\mbox{\psfig{file=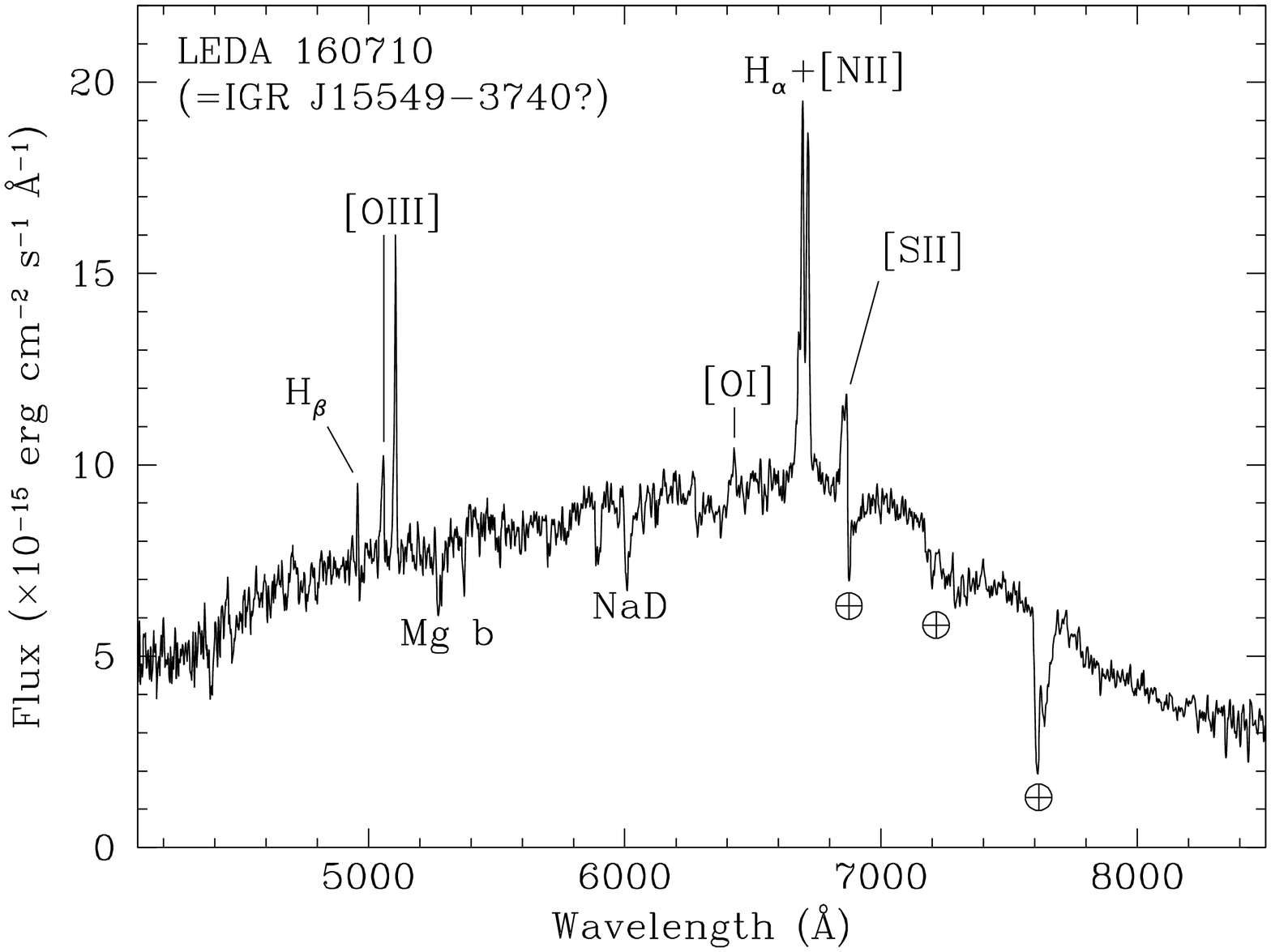,width=9cm,angle=270}}
\mbox{\psfig{file=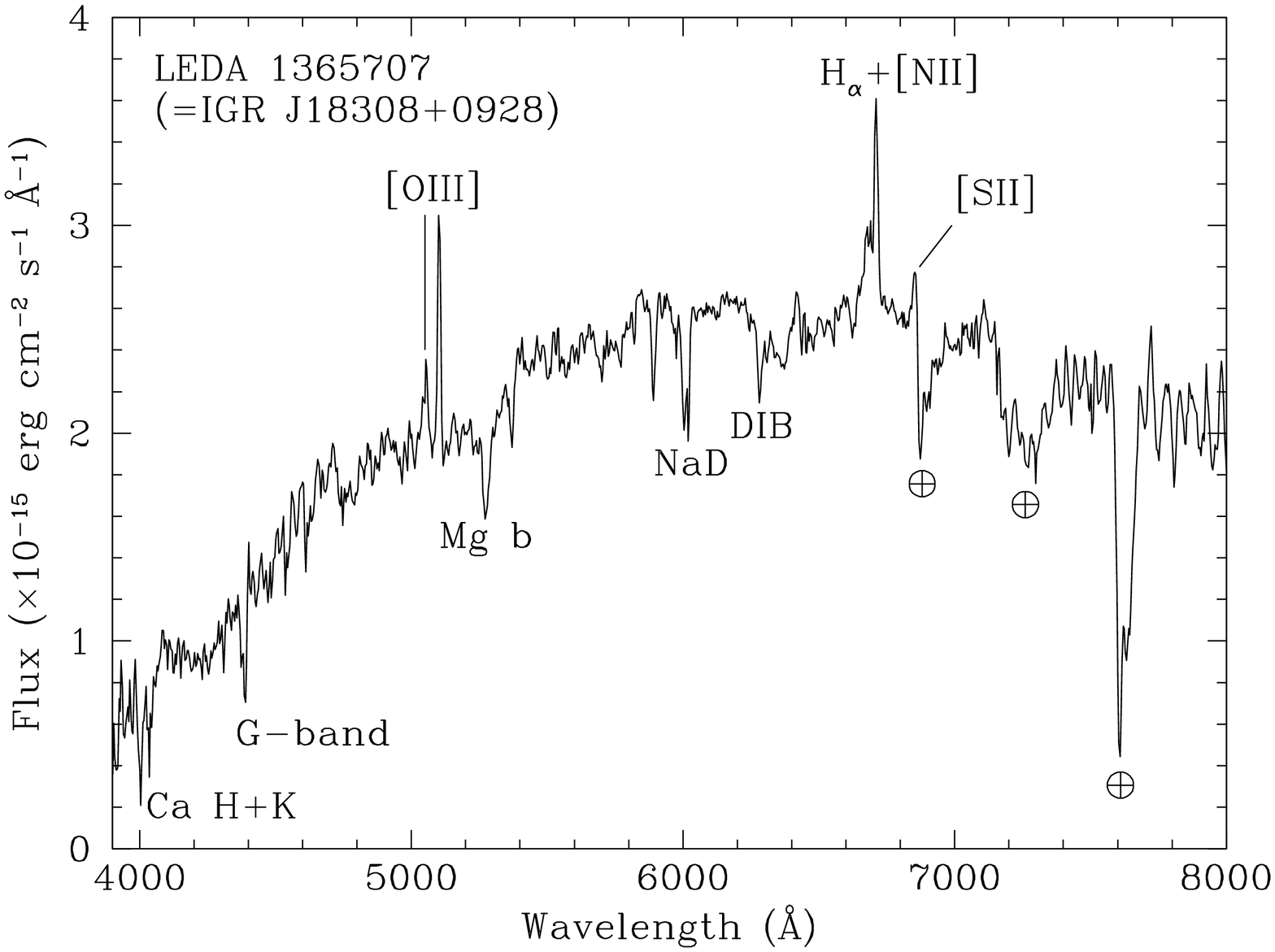,width=9cm,angle=270}}

\vspace{-.9cm}
\mbox{\psfig{file=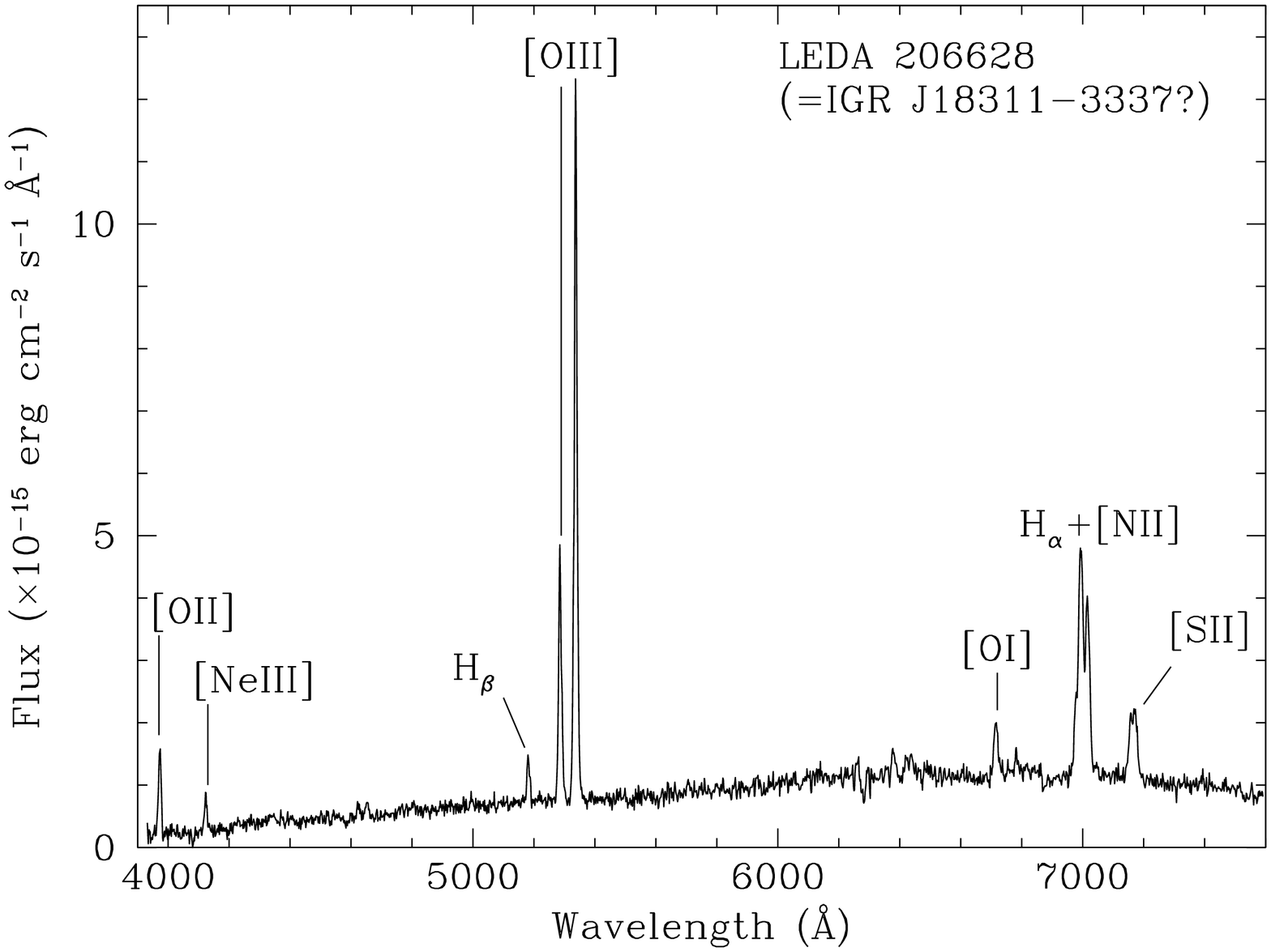,width=9cm,angle=270}}
\mbox{\psfig{file=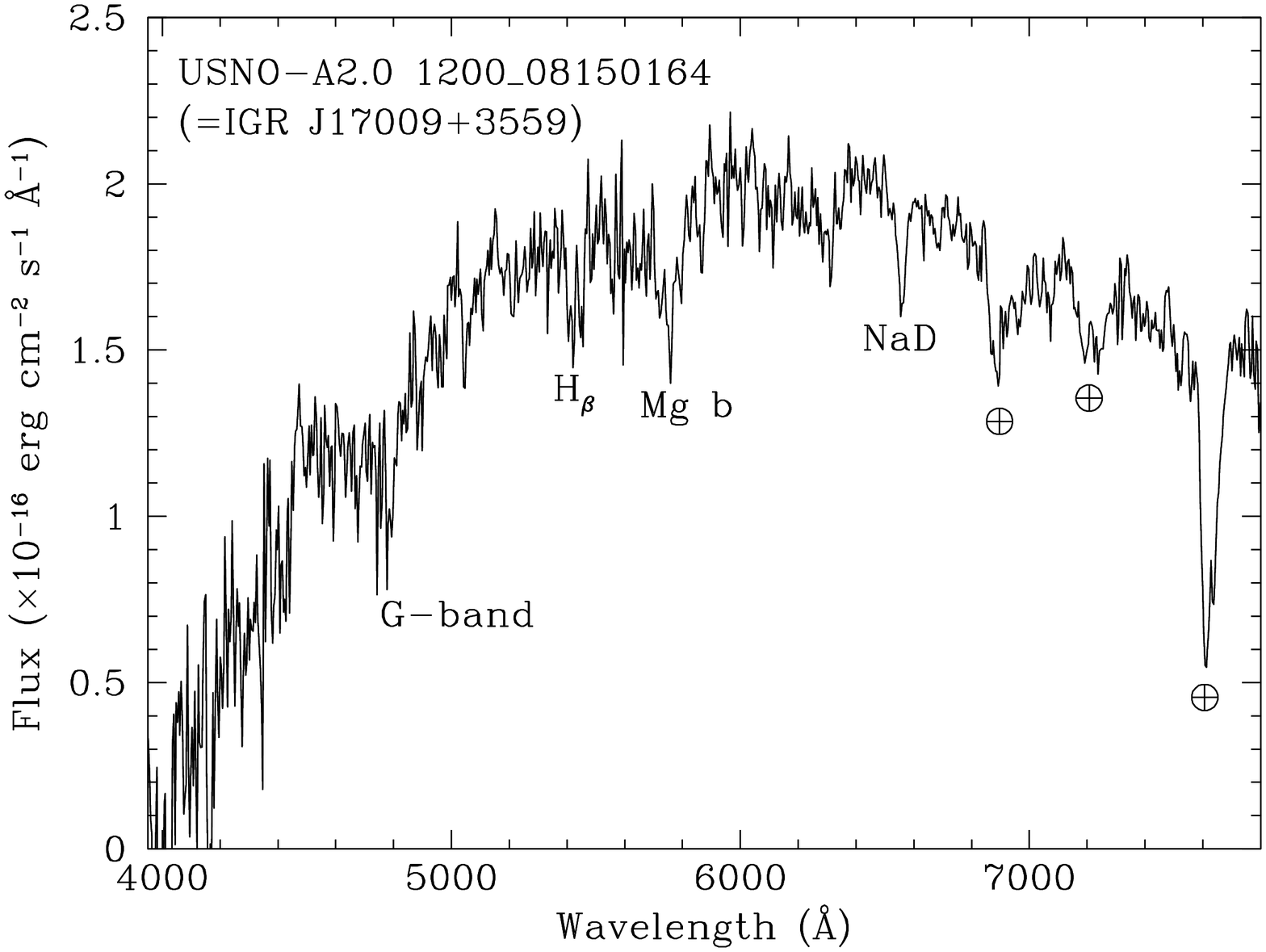,width=9cm,angle=270}}

\caption{Spectra (not corrected for the intervening Galactic absorption) 
of the optical counterparts of the remaining 5 Sy2 AGNs and of the 
XBONG (in the lower right panel) belonging to 
the sample of {\it INTEGRAL} sources presented in this paper.
For each spectrum, the main spectral features are labeled. The 
symbol $\oplus$ indicates atmospheric telluric absorption bands.
The 6dFGS spectra of IGR J14301$-$4158 and IGR J15549$-$3740 have been 
smoothed using a Gaussian filter with $\sigma$ = 3 \AA.
}
\end{figure*}

\begin{figure*}[th!]
\mbox{\psfig{file=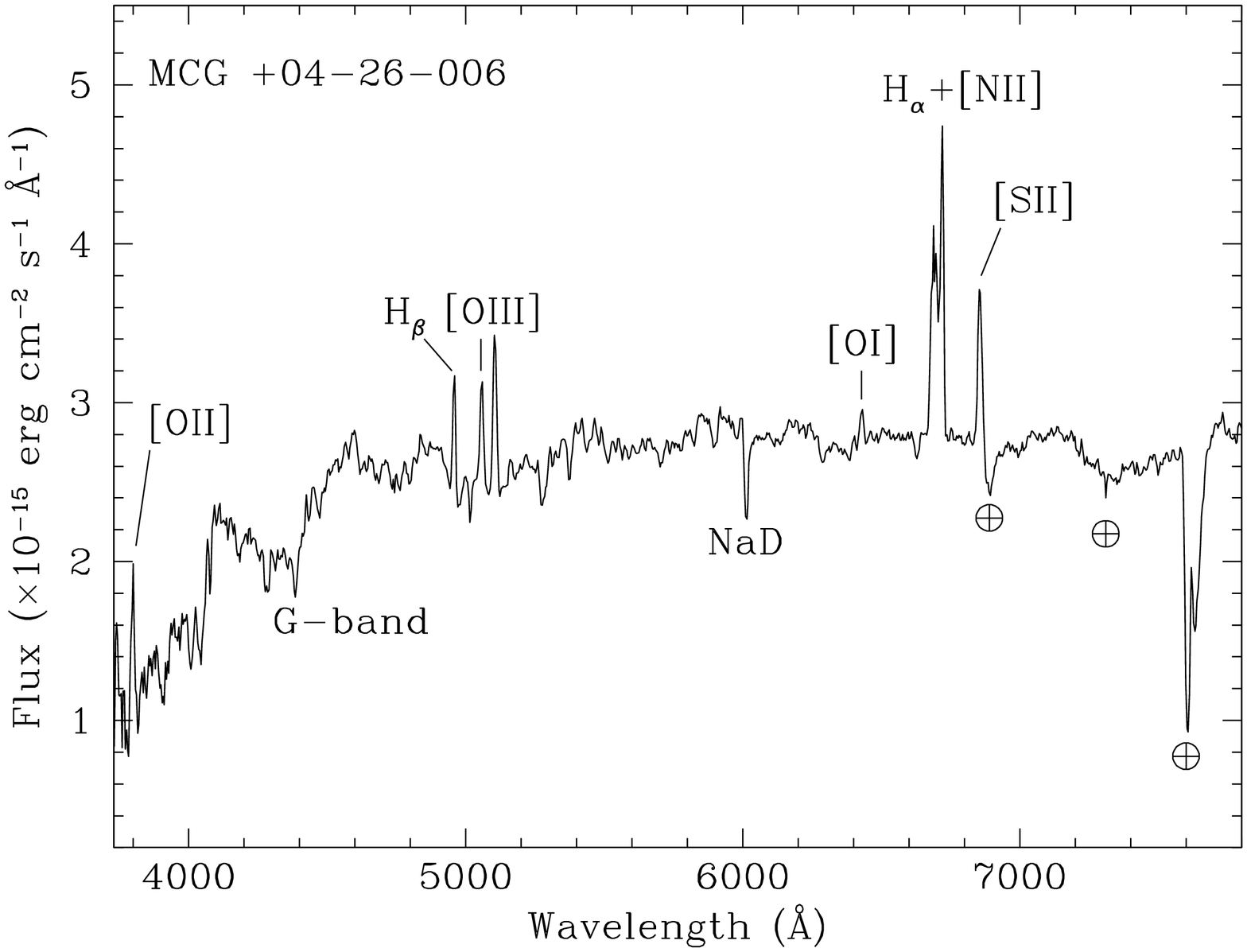,width=9cm,angle=270}}
\mbox{\psfig{file=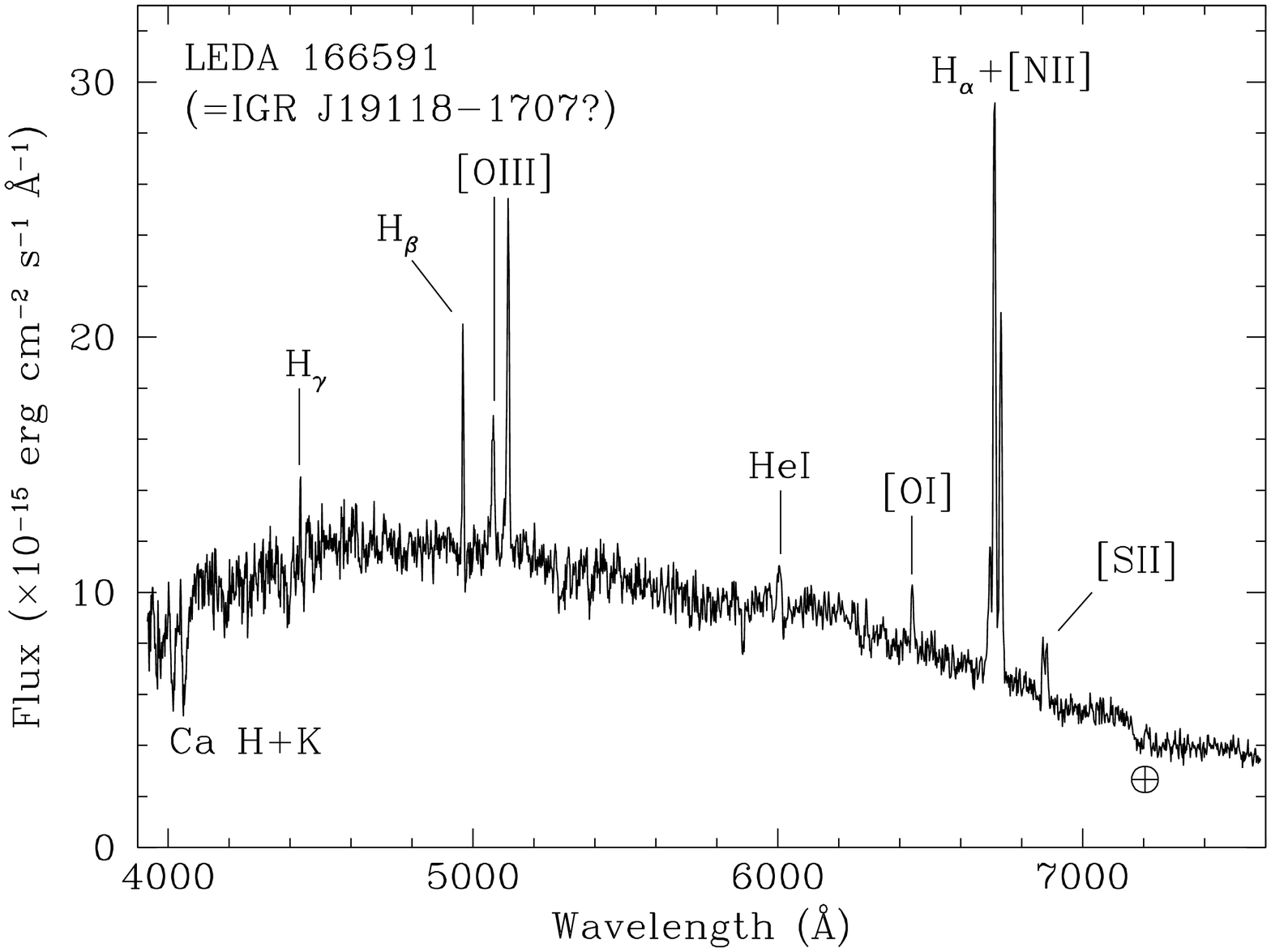,width=9cm,angle=270}}
\vspace{-.5cm}
\caption{Spectra (not corrected for the intervening Galactic absorption)
of the optical counterparts of the 2 LINERs belonging to the sample of 
{\it INTEGRAL} sources presented in this paper.
For each spectrum, the main spectral features are labeled. The
symbol $\oplus$ indicates atmospheric telluric absorption bands.}
\end{figure*}

It is found that 32 objects of our sample have optical spectra that allow 
us to classify them as AGNs (see Figs. 8-13). Thirty-one of them exhibit
strong, redshifted broad and/or narrow emission lines typical of nuclear 
galactic activity: 18 of them can be classified as Type 1 (broad-line) and 
13 as Type 2 (narrow-line) AGNs (see Tables 4 and 5, respectively, for 
their breakdown in terms of subclasses). In any case, we wish to draw 
the reader's attention to our identification, for the first time, of two 
definite low-ionization nuclear emission-line regions (LINERs; Heckman 
1980) among the unknown-nature hard X--ray sources detected with {\it 
INTEGRAL}: they are MCG +04$-$26$-$006 and IGR J19118$-$1707 (see Fig. 13).

In a single galaxy, IGR J17009+3559, the nuclear activity was discovered 
only thanks to soft X--ray emission detected with {\it Swift} (Krivonos et 
al. 2009), because no emission lines appear in its optical spectrum (Fig. 
12, lower right panel): we therefore classify it as an X--ray bright, 
optically normal galaxy (XBONG; see Comastri et al. 2002).
The ``normal galaxy" nature of this source is confirmed using the approach 
of Laurent-Muehleisen et al. (1998). The association of this galaxy with a 
NVSS radio source (Condon et al. 1998) also points to an AGN nature. We 
note that, as already mentioned and at variance with Krivonos et al. 
(2009), we do not detect any emission line in the optical spectrum of 
IGR J17009+3559.

The main observed and inferred parameters for each of these two broad 
classes of AGNs are reported in Tables 4 and 5, respectively; those
regarding the XBONG are shown in Table 5. In these tables, X--ray 
luminosities were computed from the fluxes reported in Voges et al. 
(1999, 2000), Saxton et al. (2008), Krivonos et al. (2009), Bird et 
al. (2010), Cusumano et al. (2010), Krivonos et al. (2010), Landi et 
al. (2010), Rodriguez et al. (2010), and Tueller et al. (2010).

For most of the AGNs in our sample (that is, 22 objects out of 32), the 
redshift value was determined or explictly published\footnote{We 
here define as ``explicitly published" the galaxy redshifts that appeared 
in published papers, rather than simply in online catalogs or databases 
(such as, for instance, the SIMBAD database).} in this work for the 
first time. The redshifts of the remaining 10 sources are consistent with 
those reported in the literature, i.e., in the Hyperleda archive 
(Prugniel 2005), Mescheryakov et al. (2009), Kniazev et al. 
(2010), Krivonos et al. (2009), and Butler et al. (2009).

We were also able to give a more accurate classification of the three 
sources (IGR J12107+3822, IGR J13168$-$7157, and IGR J22292+6647) 
independently identified by Mescheryakov et al. (2009), Kniazev et al. 
(2010), and Butler et al. (2009), respectively: in detail, all of them 
could be classified as Seyfert 1.5 galaxies.

In addition, we slightly corrected the classification of Swift 
J1513.8$-$8125 from Seyfert 1.9 (as given in Tueller et al. 2010) to 
Seyfert 1.2, based on the appearance of its optical spectrum (Fig. 9, 
lower left panel). At this stage, we are unable to say whether this 
classification mismatch is caused by intrinsic spectral variability of the 
source in the optical.

When we examined in detail the optical and X--ray properties
of the AGN sources of our sample, we found the following noteworthy 
issues for some selected cases. 
We note that IGR J00158+5605 could have been classified as a narrow-line 
Seyfert 1 AGN, given the narrowness of the full width at half maximum 
(FWHM) of its H$_\beta$ emission ($\sim$2000 km s$^{-1}$); however, owing 
to the absence of the Fe {\sc ii} bumps around 4600 \AA~and 5200 \AA, we 
opted for a Seyfert 1.5 galaxy classification.

It is also seen that, among all narrow-line AGNs (in Table 5) for 
which an estimate of the local absorption can be obtained, the LINER
MCG +04$-$26$-$006 appears to display no reddening local to the AGN host. 
This suggests that this source may be a ``naked" LINER.
While ``naked" Seyfert 2 galaxies, i.e. AGNs that lack the broad-line 
region (BLR) are not unheard of (see, e.g., Panessa \& Bassani 2002, 
Bianchi et  al. 2008 and Paper VI), the detection of a similar
situation in a LINER is quite peculiar, so it is potentially interesting 
and deserves further study.

A reliable assessment of the Compton nature of the narrow-line AGNs 
of our sample (see Table 5) can be obtained using the diagnostic of 
Bassani et al. (1999), that is, the ratio of the measured 2--10 keV X--ray 
flux to the unabsorbed flux of the [O {\sc iii}]$\lambda$5007 forbidden 
emission line. Because of its definition, however, this diagnostic could 
be explored only for the Seyfert 2 galaxy IGR J00465$-$4005 and the LINER 
MCG +04$-$26$-$006, which are the only two sources of Table 5 for which 
the above fluxes are simultaneously known.
After correcting the [O {\sc iii}]$\lambda$5007 emission line flux 
of these objects for the absorption local to the AGN (see again Table 5),
we found that the parameter $T$, defined in Bassani et al. (1999), has 
values 5.7 and 130, respectively, indicating that both sources are in the 
Compton thin regime. This finding strengthens the result of Landi et al.
(2010) who classified these sources as Compton thin (albeit considerably
absorbed) AGNs on the basis of their local hydrogen column, determined
by means of X--ray spectral fitting and larger than 10$^{23}$ cm$^{-2}$ in
both cases.

Following an independent method, we also verified this result by applying 
the diagnostic of Malizia et al. (2007), which uses the ratio of the flux 
measurement in the 2--10 keV band to that in the 20--100 keV band. 
Because of its definition, among the objects listed in
Table 5, this method can be used with the presently available information 
only for the two sources above, in addition to the XBONG IGR J17009+3559. 
We found that this diagnostic yields values 0.033, 0.069, and 0.15, 
respectively, for the three sources considered.
We caution the reader that, despite the definition of the diagnostic of 
Malizia et al. (2007), at present only the (1--10 keV)/(17--60 keV) flux 
ratio is available for IGR J17009+3559 (see Table 5): this means that, for 
this source, the value of this diagnostic should be considered as a strict 
upper limit. Nevertheless, when comparing these numbers with those of the 
sample of Malizia et al. (2007, their Fig. 5), we found that none of these 
sources fall in the locus in which possible Compton thick AGNs dwell. 
This result therefore independently confirms that these three AGNs are in 
the Compton thin regime.

Finally, we applied the prescriptions of Wu et al. (2004) and Kaspi et al. 
(2000), which use the width and the strength of the broad component of the 
H$_\beta$ emission as a probe of the orbital velocity and the size of the 
BLR. With them we calculated an estimate of the mass of the central black 
hole in 16 of the 18 Type 1 AGNs of our sample (a procedure that could not 
be applied to IGR J1248.2$-$5828 and IGR J19077$-$3925 as no broad 
H$_\beta$ emission component was detected in their spectra). The 
corresponding black hole masses for these 16 cases are reported in Table 
6. Here we assumed a null local absorption for all Type 1 AGNs. The main 
sources of error in these mass estimates generally come from the 
determination of the H$_\beta$ emission flux, which spans from 3\% to 30\% 
in our sample (see Table 4), and from the scatter in the $R_{\rm BLR} - 
L_{\rm H_\beta}$ scaling relation, which introduces typical uncertainties 
of 0.4--0.5 dex in the black hole mass estimate (Vestergaard 2004).

\begin{table}
\caption{BLR gas velocities (in km s$^{-1}$) and 
central black hole masses (in units of 10$^7$ $M_\odot$) for 16
broad line AGNs belonging to the sample presented in this paper.}
\begin{center}
\begin{tabular}{lrc}
\noalign{\smallskip}
\hline
\hline
\noalign{\smallskip}
\multicolumn{1}{c}{Object} & \multicolumn{1}{c}{$v_{\rm BLR}$} & $M_{\rm BH}$ \\
\noalign{\smallskip}
\hline
\noalign{\smallskip}

IGR J00158+5605         &  1900 & 5.3 \\
IGR J02086$-$1742       & 12100 & 180 \\
IGR J06058$-$2755       &  4100 & 7.7 \\
Swift J0845.0$-$3531    &  7400 &  69 \\
IGR J09094+2735         &  1800 & 1.4 \\
PKS 1143$-$696          &  3500 &  42 \\
IGR J12107+3822         &  5900 & 3.5 \\
IGR J13168$-$7157       & 11500 &  59 \\
IGR J13187+0322         &  7500 & 200 \\
Swift J1513.8$-$8125    & 10300 &  79 \\
IGR J15311$-$3737       &  2800 & 9.2 \\
IGR J18078+1123         & 11000 &  58 \\
1RXS J191928.5$-$295808 &  3800 &  16 \\
1RXS J211928.4+333259   &  2800 & 0.7 \\
1RXS J213944.3+595016   &  2300 & 4.9 \\
IGR J22292+6647         &  4700 &  12 \\

\noalign{\smallskip}
\hline
\hline
\noalign{\smallskip}
\end{tabular}
\end{center}
\end{table}

\subsection{Statistical considerations}

\begin{figure}[h!]
\hspace{-0.5cm}
\psfig{file=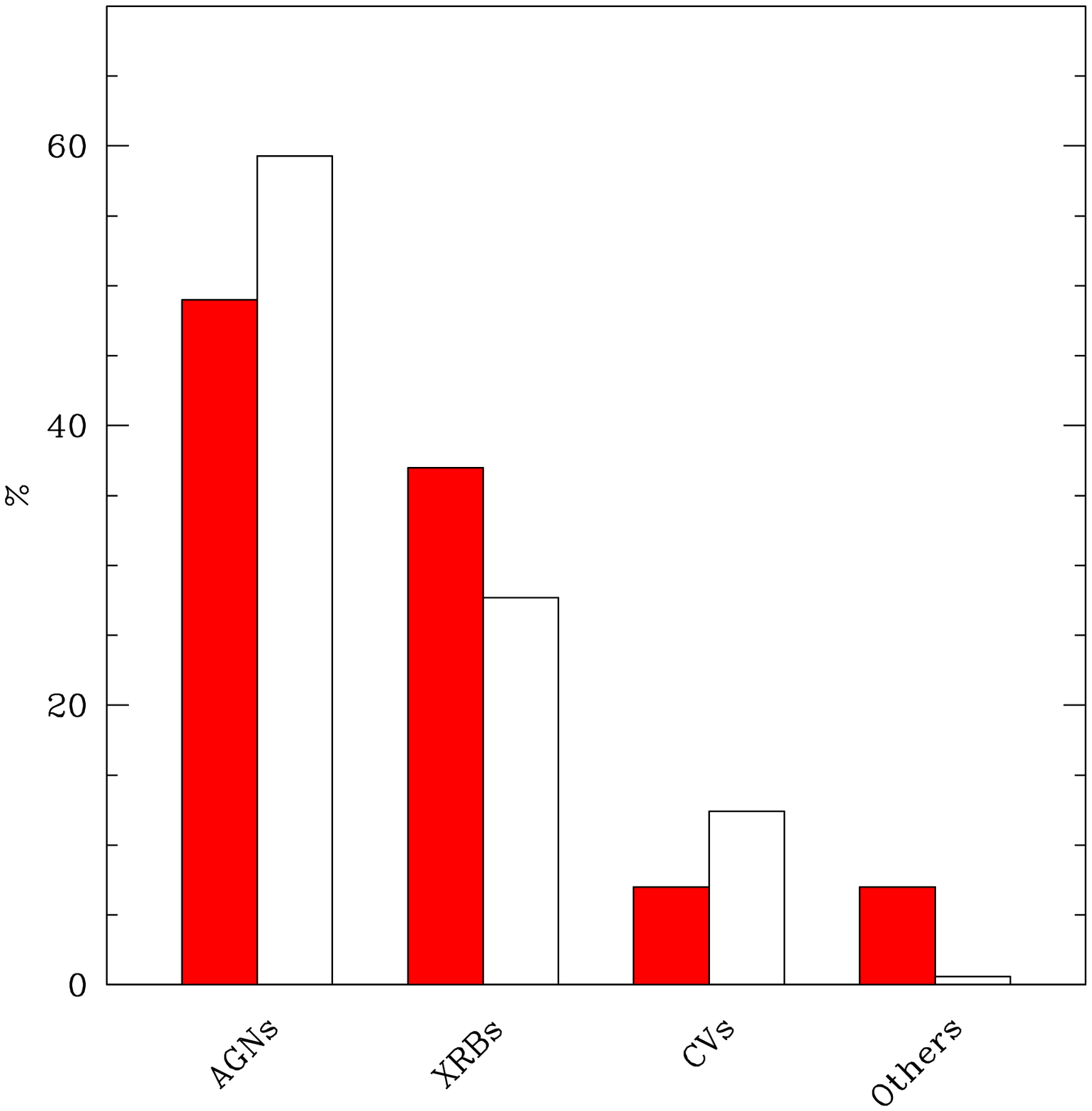,width=9.5cm}
\vspace{-0.7cm}
\caption{Histogram, subdivided into source types, showing the percentage 
of {\it INTEGRAL} objects of known nature in the 4$^{\rm th}$ IBIS survey
(Bird et al. 2010; left-side, darker columns), and {\it INTEGRAL} sources 
from various surveys and identified through optical or NIR spectroscopy 
(right-side, lighter columns).}
\end{figure}

We again update the statistics of Papers V-VII with the results from the 
sample presented here and other spectroscopic optical and NIR 
identifications of {\it INTEGRAL} sources (Zurita Heras et al. 2009; 
Mescheryakov et al. 2009; Burenin et al. 2009a,b; Degenaar et al. 2010).

We find that the 177 {\it INTEGRAL} sources identified until now on the 
basis of their optical or NIR spectroscopy are distributed into the main 
broad classes discussed in this paper as follows: 105 (59.3\%) are AGNs, 
49 (27.7\%) are X--ray binaries, and 22 (12.4\%) are CVs. Only 1 case 
(0.6\%) does not belong to any of the above categories, the identification 
of IGR J08023$-$6954 being a chromospherically active star (see Paper VI 
and Rodriguez et al. 2010).

Studying the AGN subcategories in greater detail, one can see that 52 
sources (i.e., 49\% of the AGN identifications) are broad-line AGNs, 42 
(40\%) are narrow-line AGNs, while the QSO, XBONG, and BL Lac subclasses 
are populated by 4, 4, and 3 objects (4\%, 4\%, and 3\%), respectively.

For the Galactic objects, it is found that 38 and 11 objects (76\% and 
24\% of the X--ray binary identifications) are HMXBs and LMXBs, 
respectively; in addition, 22 sources are classified as CVs, most of which 
(18, that is 82\% of them) are definite or likely dwarf novae (mostly of 
magnetic type), and the remaining 4 are symbiotic stars.

In this paper, we can provide a general overview of these statistics and 
compare them with those of our previous works (Papers V-VII). In 
particular, and as already briefly discussed in our previous papers, we 
focus on the evolution of the source populations. In terms of percentage, 
we see that AGNs still constitute the vast majority of the 
identifications, and continue to increase relative to the numbers of Paper 
VII; in contrast, the total percentage of objects identified as X--ray 
binaries decreased, while those identified as CVs remain basically 
unchanged but still represent the third largest group of {\it INTEGRAL} 
sources identified from their optical/NIR spectroscopy.

The majority of AGNs in the sample of optical identifications of {\it 
INTEGRAL} sources can indeed be partly explained by an observational bias: 
for instance, we have already emphasized in Paper VII that NIR 
spectroscopy more easily identifies Galactic X--ray binaries, probably 
because of their larger obscuration in optical bands due to their 
location in the Galactic plane; likewise, X--ray timing studies of course 
are more effective in discovering Galactic compact objects (see e.g. 
Walter et al. 2006). The increasing percentage of AGN discoveries may also 
be due to the change in the {\it INTEGRAL} observational strategy during 
the spacecraft lifetime, as it now maps more and more frequently the sky 
outside the Galactic plane (see Bird et al. 2010).

Among the X--ray binaries, the percentage of LMXB discoveries continues to 
slightly increase with time, possibly thanks to high-precision positional 
information afforded at soft X--rays coupled with deep optical/NIR imaging 
and spectroscopy acquired at large (8-m size) telescopes (see e.g. 
Degenaar et al. 2010).

One can compare the latest percentages (see Fig. 14) with those of the 513 
identified objects belonging to the deepest and most complete survey of 
the hard X--ray sky performed by {\it INTEGRAL} to date, that is, the 
4$^{\rm th}$ IBIS survey of Bird et al. (2010). This catalog contains, 
among the securely identified sources, 251 (49\%) AGNs, 190 (37\%) X--ray 
binaries (nearly equally divided into LMXBs and HMXBs), and 36 (7\%) CVs, 
almost all of them of magnetic nature.

Comparisons of the relative percentages of these subclasses within the 
identified objects in the 4$^{\rm th}$ IBIS survey of Bird et al. (2010) 
with the sources identified from optical/NIR spectroscopy until now has 
confirmed, as noted in Papers V-VII, that this latter method is very 
effective in detecting AGNs and, to a lesser extent, CVs.

In a similar energy range as IBIS, the {\it Swift}/BAT instrument provides
a comparable distribution of the hard X--ray source population in the
different classes (Cusumano et al. 2010; Tueller et al. 2010), with a
slightly larger fraction of AGNs ($\sim$60\%) and a smaller fraction of
X--ray binaries ($\sim$25\%) than in the 4$^{\rm th}$ IBIS survey. 
These differences between the IBIS and BAT surveys can be attributed 
to different (and, to some extent, complementary) spacecraft observational 
strategies, {\it INTEGRAL} being mainly dedicated to Galactic plane scans 
and {\it Swift} covering more of the ecliptic pole regions and the sky 
outside the Galactic plane (see Bird et al. 2010, Cusumano et al. 2010, 
and Tueller et al. 2010 for details).

Returning to the sources identified in the present work, we found that for 
the first time two definite LINER-type AGNs are identified among the group 
of {\it INTEGRAL} sources of unknown nature. We also identified a 
non-magnetic CV within the sample selected in the present paper: to our 
knowledge, this is the first CV belonging to the lot of unidentified {\it 
INTEGRAL} objects that does not have properties that are indicative of 
accretion onto a WD with magnetic activity. These discoveries may be the 
result of the latest IBIS surveys (Bird et al. 2010; Krivonos et al. 2010) 
being the deepest obtained with this instrument and among the deepest and 
most complete investigations of the whole hard X--ray sky above 20 keV, 
hence permitting the detection of intrinsically faint objects at these 
energies.

\section{Conclusions}

As part of our ongoing identification program of {\it INTEGRAL} sources by 
means of optical spectroscopy (Papers I-VII) that has been employing various 
telescopes since 2004, we have identified and studied 44 additional 
objects of unknown or poorly explored nature belonging to surveys of the 
hard X--ray sky (Bird et al. 2010; Krivonos et al. 2010; Bozzo et al. 
2010; Coe et al. 2010; McBride et al. 2010). This has been made possible 
by using 7 different telescopes and archival data from 2 spectroscopic 
surveys.

We found that the selected sample consists of 32 AGNs (17 of which are 
Seyfert 1 galaxies, 11 are Seyfert 2 galaxies, 2 are LINERs, one is an 
XBONG, and one is a Type 1 QSO), 6 CVs (one of which is very likely 
of non-magnetic nature), 5 HMXBs (two of which belong to the SMC, two being 
Galactic Be/X binaries, and one being possibly a supergiant B[e]/X--ray 
binary), and one LMXB. In the present sample, as in our past papers within 
the framework of this research, we note that the absolute majority of 
identified sources belongs to the AGN class, and we detect a 
non-negligible presence of CVs.

We note that 148 of the 177 optical and NIR spectroscopic identifications 
considered in this subsection were obtained or refined within the 
framework of our spectroscopic follow-up program originally set up in 2004 
(Papers I-VII, the present work, and references therein).

The results presented in this work once again demonstrate the high 
effectiveness of the method of catalog cross-correlation and/or 
follow-up observations (especially with soft X--ray satellites capable 
of providing arcsec-sized error boxes, such as {\it Chandra}, {\it 
XMM-Newton} or {\it Swift}), and optical spectroscopy to determine the 
actual nature of still unidentified {\it INTEGRAL} sources. We however 
recall that, for 11 objects of our present sample, only a putative albeit 
likely optical counterpart could be identified because of the lack of soft 
X--ray observations providing a definite arcsec-sized position at high 
energies. Thus, these {\it INTEGRAL} sources should be timely 
observed by soft X--ray satellites affording arcsec-sized 
localizations to confirm the proposed associations.

Present and future surveys at optical and NIR wavelengths, such 
as the ongoing Vista Variables in the Via Lactea (VVV; Minniti et al. 
2010) public NIR survey, will permit us to check and identify variable 
objects in the fields of the objects detected in published and forthcoming 
{\it INTEGRAL} catalogs, thus easing the search for putative counterparts 
to these high-energy sources.

\begin{acknowledgements}

We thank Silvia Galleti for Service Mode observations at the Loiano 
telescope; Hripsime Navasardyan for Service Mode observations at the
Asiago telescope; Antonio Magazz\`u for Service Mode observations at the 
TNG; Manuel Hern\'andez and Jos\'e Vel\'asquez for Service Mode 
observations at the CTIO telescope, and Fred Walter for relaying the
observing information to them; Roberto Gualandi and Ivan Bruni for night 
assistance at the Loiano telescope; Gaspare Lo Curto for support at the ESO 
NTT telescope; Claudia Reyes for night assistance at the ESO NTT telescope. 
We also thank the anonymous referee for useful remarks which helped us
to improve the quality of this paper. 
This research has made use of the ASI Science Data Center Multimission 
Archive; it also used the NASA Astrophysics Data System Abstract Service, 
the NASA/IPAC Extragalactic Database (NED), and the NASA/IPAC Infrared 
Science Archive, which are operated by the Jet Propulsion Laboratory, 
California Institute of Technology, under contract with the National 
Aeronautics and Space Administration. 
This publication made use of data products from the Two Micron All 
Sky Survey (2MASS), which is a joint project of the University of 
Massachusetts and the Infrared Processing and Analysis Center/California 
Institute of Technology, funded by the National Aeronautics and Space 
Administration and the National Science Foundation.
This research has also made use of data extracted from the Six-degree 
Field Galaxy Survey and the Sloan Digitized Sky Survey archives;
it has also made use of the SIMBAD database operated at CDS, Strasbourg, 
France, and of the HyperLeda catalog operated at the Observatoire de 
Lyon, France.
The authors acknowledge the ASI and INAF financial support via grant 
No. I/008/07.
LM is supported by the University of Padua through grant No. 
CPDR061795/06. 
VC is supported by the CONACYT research grant 54480-F (M\'exico).
DM is supported by the Basal CATA PFB 06/09, and FONDAP Center for 
Astrophysics grant No. 15010003. 
GG is supported by Fondecyt grant No. 1085267.
\end{acknowledgements}

\end{document}